\documentclass[12pt]{iopart}

\usepackage{iopams}
\usepackage{amssymb}
\usepackage{amsfonts}
\usepackage{bm}
\usepackage{epsfig}
\usepackage{epsf}
\usepackage[]{graphicx}

\usepackage{color}

\usepackage{cite}
\usepackage{xspace}

\renewcommand{\d}{{\rm d}}

\newcommand {\ee}{{\rm e}}

\newcommand {\bfa} {{\bf a}}

\newcommand {\bfk} {{\bf k}}

\newcommand {\bfp} {{\bf p}}

\newcommand {\bfr} {{\bf r}}
\newcommand {\bfv} {{\bf v}}

\newcommand {\bfR} {{\bf R}}

\newcommand {\calA} {{\cal A}}

\newcommand {\calF} {{\cal F}}

\newcommand {\calN} {{\cal N}}

\newcommand {\calT} {{\cal T}}

\newcommand {\E}  {{\varepsilon}}
\newcommand {\om}  {{\omega}}
\newcommand {\Om}  {{\Omega}}

\newcommand {\aTF}  {a_{\rm TF}}
\newcommand {\Uat}  {U_{\rm at}}
\newcommand {\tUat}  {\widetilde{U}_{\rm at}}

\newcommand {\dUmax}  {U_{\max}^{\prime}}

\newcommand {\Ld}  {L_{\rm d}}

\newcommand {\Lp}  {L_{\rm p}}

\newcommand{\Nacc}{{N_{\rm acc}}}

\newcommand{\Lch}{{L_{\rm ch}}}

\newcommand{\Kch}{{K_{\rm ch}}}
\newcommand{\lamch}{{\lambda_{\rm ch}}}

\newcommand{\lamu}{{\lambda_{\rm u}}}
\newcommand{\ku}{{k_{\rm u}}}
\newcommand{\Omu}{{\Omega_{\rm u}}}
\newcommand {\Nu}  {N_{\rm u}}
\newcommand{\Ku}{{K_{\rm u}}}

\newcommand{\calUpl} {{\cal U}_{\rm pl}}

\newcommand{\Eph} {{E_{\rm ph}}}

\newcommand{\MBNExplorer} {\textsc{MBN Explorer}\xspace}

\begin{document}


\title[Simulations of channeling and radiation processes in oriented crystals]
{All-atom relativistic molecular dynamics simulations of channeling and radiation processes in oriented crystals}

\author{Andrei V. Korol, Gennady B. Sushko, and Andrey V. Solov'yov$^*$}

\address{
MBN Research Center, Altenh\"{o}ferallee 3, 60438 
Frankfurt am Main, Germany}

\ead{solovyov@mbnresearch.com}

\begin{abstract}
We review achievements made in recent years in the field of numerical modeling of ultra-relativistic projectiles propagation in
oriented crystals, radiation emission and related phenomena.
This topic is highly relevant to the problem of designing novel gamma-ray light sources based on the exposure of oriented crystals 
to the beams of ultra-relativistic charged particles . 
The paper focuses on the approaches that allow for advanced 
computation exploration beyond the continuous potential and the binary collisions frameworks.
A comprehensive description of the  multiscale all-atom relativistic molecular dynamics approach implemented in the \MBNExplorer package
is given.
Several case studies related to modeling of ultra-relativistic projectiles 
(electrons, positron and pions) channeling and photon emission in oriented straight, bent and periodically bent crystals are presented.
%
%
In most cases, the input data used in the simulations, such as crystal orientation and thickness, the bending radii, periods and amplitudes,
as well as the energies of the projectiles, have been chosen to match the parameters used in the accomplished and the ongoing experiments. 
Wherever available the results of calculations are compared with the experimental data and/or the data obtained by other numerical means.
\end{abstract}

\section{Introduction \label{Introduction}}

Development of light sources (LS) for wavelengths $\lambda$ well below one angstrom (corresponding 
photon energies $\Eph \gg 10$ keV) is a challenging goal of modern physics.
Sub-angstrom wavelength powerful spontaneous and, especially, coherent radiation will
have many applications in the basic sciences, technology and medicine.
They may have a revolutionary impact on nuclear and solid-state physics,
as well as on the life sciences \cite{HighImpact}.
Modern X-ray Free-Electron-Laser (XFEL) can generate X-rays with wavelengths $\lambda\sim1$ \AA{}  
\cite{Seddon-EtAl_RepProgPhys_v80_115901_720_2017,Doerr-EtAl_NatureMethods_v15_p33_2018,
SwissFEL_ApplScie_v7_720_2017, Bostedt-EtAl_RMP_v88_015007_2016}.
Existing synchrotron facilities provide radiation of shorter wavelengths but orders of magnitude less intensive
\cite{YabashiTanaka_NaturePhotonics_v11_p12_2017,Ayvazyan_EtAl_EPJD_v20_p149_2002}.
Therefore, to create a powerful LS in the range $\lambda \ll 1$ \AA{} new approaches and technologies are needed.

One of the approaches to designing novel gamma-ray LSs relies on the exposure of oriented crystals 
to the beams of ultra-relativistic charged particles \cite{HighImpact}. 
In an exemplary case study, presented in the cited paper,
brilliance of radiation emitted in Crystalline Undulator (CU) based LSs have been estimated.
It was shown that in the range $\Eph = 10^{-1}-10^2$ MeV the brilliance is comparable to   
or higher than the synchrotrons brilliance at much lower photon energies.  
A rigorous computation of the brilliance \cite{PavlovBrilliance_2020} based on numerical simulations carried out by means of 
the \MBNExplorer software package \cite{MBN_Explorer_Paper,MBN_Explorer_Site} has confirmed the estimates.  
The brilliance can be boosted by orders of magnitude through the process of superradiance \cite{Gover-EtAl_RPM_v91_035003_2019}
by a pre-bunched beam \cite{ChannelingBook2014,Patent}. 
In this case the brilliance can be comparable with the values achievable 
at the XFEL facilities which operate in much lower photon energy range \cite{HighImpact}. 

Crystal-based LSs can generate radiation in the photon energy range where the technologies based 
on the fields of permanent magnets become inefficient or incapable. 
The limitations of conventional LS is overcome by exploiting very strong crystalline fields.
The orientation of a crystal along the beam enhances significantly 
the strength of the particles interaction with the crystal due to strongly correlated 
scattering from lattice atoms \cite{Lindhard}.
This allows for the guided motion of particles through crystals of different geometry and for the enhancement of radiation.
A comprehensive review of physics of channeling and radiation emission in periodically bent crystals
one finds in \cite{ChannelingBook2014}.

Construction of novel crystal-based LSs is an extremely challenging task, which constitutes a highly interdisciplinary and broad field. 
A roadmap review \cite{HighImpact} presents an overview of the field, 
outlines the achievements that have been made and discusses the goals which can be achieved.
It also provides an updated review of the issues relevant to practical realization of the novel LSs including, 
in particular, channeling experiments with bent and periodically bent crystals, 
technologies and methods developed or/and proposed to create bent crystalline structures.    
The results, data, supportive detailed description of the formalism involved, the discussion of the related phenomena 
as well as the outlook has been presented in a consistent and complete manner for the first time in the cited paper.

The goal of the current paper is to present recent achievements in one of the research directions within the aforementioned 
interdisciplinary field, namely, numerical modeling of the channeling and related phenomena.
Within this direction we focus on the approaches that allow for advanced 
computation exploration beyond  the continuous potential framework \cite{Lindhard}.
The computational tools, which are based on the continuous potential concept and on the binary collisions approach, 
developed in recent years are reviewed briefly in Section \ref{Overview}.
A detailed description of the  multiscale all-atom relativistic molecular dynamics approach implemented in the \MBNExplorer package
is given in Section \ref{MBNExplorer}.
In the last years the channeling module of the package \cite{MBN_ChannelingPaper_2013} has been used extensively to simulate
propagation and radiation emission by projectiles in 
in various crystalline media. 
In Section \ref{CaseStudies} we present several case studies related to modeling of ultra-relativistic projectiles 
channeling and calculation of the spectral intensities by means of \MBNExplorer.

\section{Overview of numerical approaches to simulate channeling phenomenon 
\label{Overview} }

Various approximations have been used to simulate channeling phenomenon in oriented crystals. 
Whilst most rigorous description relies can be achieved within the framework of quantum mechanics (see recent 
paper \cite{WistisenPiazza_PRD_v99_116010_2019} and references therein) quite often classical description in terms 
of particles trajectories is highly adequate and accurate.
Indeed, the number of quantum states $N$ of the transverse motion of a channeling 
electron and/or positron increases with its energy as $N\sim A \sqrt{\gamma}$
where $A\sim 1$  and $\gamma=\E/mc^2$ stands for the relativistic Lorentz factor of a projectile of energy $\E$ and mass $m$
\cite{AndersenEtAl_KDanVidensk_v39_p1_1977,KKSG_simulation_straight}. 
The classical description implies strong inequality $N \gg 1$, which becomes well fulfilled
for projectile energy in the hundred MeV range and above. 

Simulation of channeling and related phenomena has been implemented in several software packages
within frameworks of different theoretical approaches.
Below we briefly characterize the most recently developed ones.
         
\begin{itemize}
\item
The computer code {\it Basic Channeling with Mathematica} \cite{BogdanovEtAl_JPhysCS_v236_012029_2010} 
uses continuous potential for analytic solution  of the channeling related problems.
The code allows for computation of classical trajectories of channeled electrons and positrons
in continuous potential as well as for computation of wave functions and energy levels of
the particles.
Calculation of the spectral distribution of emitted radiation is also supported. 

\item
A toolkit for the simulation of coherent interactions between high-energy charged projectiles 
with complex crystalline structures called DYNECHARM++ has been developed 
\cite{BagliGuidi_NIMB_v309_p124_2013}. 
The code allows for calculation of electrostatic characteristics (charge densities, electrostatic
potential and field) in complex atomic structures and 
to simulate and track a particle's trajectory. 
Calculation of the characteristics is based on their expansion in the Fourier series through 
the ECHARM (Electrical CHARacteristics of Monocrystals) method \cite{BagliEtAl_PRE_v81_026708_2010}. 
Two different approaches to simulate the interaction have been adopted, 
relying on (i) the full integration of particle trajectories within the 
continuum potential approximation, and (ii) the definition of cross-sections of coherent processes.
Recently, this software package was supplemented with the RADCHARM+ module 
\cite{BandieraEtAl_NIMB_v355_p44_2015} which allows for the computation of the emission spectrum 
by direct integration of the quasi-classical formula of Baier and Katkov \cite{Baier}.

\item
The CRYSTALRAD simulation code, presented in Ref. \cite{Sytov_PR_AB_v22_064601_2019}
is an unification of the CRYSTAL simulation code \cite{SytovTikhomirov_NIMB_v355_p383_2015} 
and the RADCHARM++ routine \cite{BandieraEtAl_NIMB_v355_p44_2015}. 
The former code is designed for trajectory calculations taking into account various coherent
effects of the interaction of relativistic and ultrarelativistic charged particles with straight or bent single 
crystals and different types of scattering. 
The program contains one- and two-dimensional models that allow for modeling of classical
trajectories of relativistic and ultrarelativistic charged particles in the field of atomic 
planes and strings, respectively. 
The algorithm for simulation of motion of particles in presence of multiple
Coulomb scattering is modeled accounting for the suppression of incoherent scattering
\cite{Tikhomirov_PR_AB_v22_064601_2019}.
In addition to this, nuclear elastic, diffractive, and inelastic
scattering are also simulated.
   
\item
\textcolor{black}{
In Ref. \cite{Guidi-EtAl:PRA_v86_042903_2012} the algorithm based on the Fourier transform method for 
planar radiation has been presented and implemented to compute the emission spectra of ultra-relativistic 
electrons and positrons within the Baier-Katkov quasiclassical formalism.
Special attention has been given to treat the radiation emission in the planar channeling regime in bent 
crystals with account for the contributions of both volume reflection and multiple volume reflection events.
The simulation presented took into consideration both the nondipole
nature and arbitrary multiplicity of radiation accompanying volume reflection. 
A large axial contribution to the hard part of the radiative energy loss spectrum as well as the strengthening of planar
radiation, with respect to the single volume reflection case, in the soft part of the spectrum have been demonstrated.
}
   
\item
The codes described in Ref. \cite{Dechan01} (see also \cite{ChannelingBook2014}) 
allow for simulation of classical trajectories of ultra-relativistic projectiles in straight and 
periodically bent crystals as well as for computing spectra-angular distribution of the radiated
energy within the quasi-classical formalism \cite{Baier}.
The trajectories are calculated by solving three-dimensional equations
of motion with account for: (i) the continuous interplanar potential; 
(ii) the centrifugal potential due to the crystal bending; 
(iii) the radiative damping force; 
(iv) the stochastic force due to the random 
scattering of the projectile by lattice electrons and nuclei.

\item
Recently presented code \cite{Nielsen_arXiv2019} allows one to determine the trajectory
of particles traversing oriented single crystals and to evaluate the radiation
spectra within the quasi-classical approximation.
To calculate the electrostatic field of the crystal lattice the code uses thermally averaged
Doyle-Turner continuous potential \cite{DoyleTurner1968} (see also \ref{AtomicPotential}). 
Beyond this framework, included are multiple Coulomb scattering and energy loss due to 
radiation emission. 
It is shown that the use of Graphics Processing Units (GPU) instead of the CPU processors
speeds up calculations by several orders of magnitude.

\item
In Ref. \cite{KKSG_simulation_straight} a Monte Carlo code was 
described which allows one to simulate the electron and positron channeling.
The code did not use the continuous potential concept but utilized the algorithm
of binary collisions of the projectile with the crystal constituents.
However, as it has been argued \cite{MBN_ChannelingPaper_2013,ChannelingBook2014,Reply2018},
the code was based on the peculiar model of the elastic scattering of the projectile 
from the crystal atoms. 
Namely, atomic electrons are treated as point-like charges placed at fixed positions around the nucleus. 
The model implies also that the interaction of an projectile with each atomic constituent, electrons included,
is treated as the classical Rutherford scattering from a static, infinitely massive point charge.        
It was demonstrated in the cited papers that in practical simulations, non-zero statistical weight 
of hard collisions with spatially fixed electrons overestimates the increase of the root-mean square 
scattering angle with increasing the propagation distance of the channeling particle. 
As a result, the model over-counts dechanneling-channeling events resulting from the hard collisions. 

\end{itemize}

\section{Atomistic molecular dynamics approach \label{MBNExplorer}}

Numerical modeling of the channeling and related phenomena beyond the continuous potential framework
can be carried out by means of the multi-purpose computer package \MBNExplorer 
\cite{MBN_Explorer_Paper,MBN_Explorer_Site}.
The \MBNExplorer package was originally developed as a universal computer program to allow 
investigation of structure and dynamics of molecular systems of different
origin on spatial scales ranging from nanometers and beyond. 

 In order to address the channeling phenomena, an additional module has been incorporated 
 into \MBNExplorer to compute the motion for relativistic projectiles along with 
 dynamical simulations of the propagation environments, including the
crystalline structures, in the course of the projectile’s motion  \cite{MBN_ChannelingPaper_2013}.
The computation accounts for the interaction of projectiles with separate atoms of the 
environments, whereas a variety of interatomic potentials implemented in \MBNExplorer supports 
rigorous simulations of various media. 
The software package can be regarded as a powerful numerical tool to reveal the dynamics of 
relativistic projectiles in crystals, amorphous bodies, as well as in biological environments. 
Its efficiency and reliability has been benchmarked for the channeling channeling of 
ultra-relativistic projectiles (within the sub-GeV to tens of GeV energy range) in straight, bent and 
periodically bent crystals \cite{Polozkov-EtAl:EPJD_v68_268_2014,Sushko-EtAl:NIMB_v355_p39_2015,%
MBN_ChannelingPaper_2013,Sushko-EtAl:JPConfSer_v438_012019_2013,%
Sushko-EtAl:JPConfSer_v438_012018_2013,Bezchastnov_AK_AS:JPB_v47_195401_2014,Sushko-EtAl:NTV_v1_p341_2015,%
Sushko-EtAl:NTV_v1_p332_2015,%
Sushko:Thesis_2015,Korol-EtAl:NIMB_v387_p41_2016,Korol-EtAl:EPJD_v71_174_2017,Korol-EtAl:NIMB_v424_26_2018,%
Pavlov-EtAl:JPB_v52_11LT01_2019,Pavlov-EtAl:EPJD_2020,HaurylavetsEtAl-in-preparation-2020}.
In these papers verification of the code against available experimental data and predictions of 
other theoretical models was carried out.

\subsection{Methodology \label{MBNExplorer:Methodology}}

The description of the simulation procedure is sketched below.

Within the framework of classical relativistic dynamics propagation of an ultra-relativistic 
projectile of the charge $q$ and mass $m$ 
through a crystalline medium implies integration of the following two coupled equations of motion
(EM):
\begin{eqnarray}
\partial \bfr / \partial t = \bfv,
\qquad
\partial \bfp / \partial t = - q \, \partial U/\partial \bfr
\label{Methodology:eq.01} 
\end{eqnarray}
where $U=U(\bfr)$ is the electrostatic potential due to the crystal constituents, 
$\bfr(t), \bfv(t)$, and $\bfp(t) = m\gamma\bfv(t)$ stand, respectively, 
for the position vector, velocity and momentum of the particle at instant $t$,   
$\gamma = \left(1-v^2/c^2\right)^{-1/2} = \E/mc^2$ is the relativistic Lorentz factor, 
$\E$ is the particle's energy, and $c$ is the speed of light. 

In \MBNExplorer, the EM are integrated using the forth-order Runge-Kutta scheme with variable time step. 
At each step, the potential $U=U({\bf r})$ is calculated as the sum of potentials $U_{\mathrm{at}}(\bfr)$ 
of individual atoms 
\begin{eqnarray} 
U(\bfr) = 
\sum_{j} U_{\mathrm{at}}\left(\left|\bfr - \bfR_j\right|\right)
\label{MC_Simulations.02}
\end{eqnarray} 
where $\bfR_j$ is the position vector of the $j$th atom. 
The code allows one to evaluate the atomic potential using the approximations due to 
Moli\`{e}re \cite{Moliere} and Pacios \cite{Pacios1993} (see also Section \ref{AtomicPotential}).
A rapid decrease of these potentials with increasing the distances from the atoms 
allows the sum~(\ref{MC_Simulations.02}) to be truncated in practical calculations. 
Only atoms located inside a sphere of the (specified) cut-off radius 
$\rho_{\max}$ with the center at the instant location of the projectile.
The value $\rho_{\max}$ is chosen large enough to ensure negligible contribution
to the sum from the atoms located at $r>\rho_{\max}$. 
The search for such atoms is facilitated by using the linked cell algorithm 
implemented in \MBNExplorer \cite{MBN_Explorer_Paper,MBNExplorer_Book}.

As a first step in simulating the motion along a particular direction, 
a crystalline lattice is generated inside the rectangular simulation box
of the size $L_x\times L_y \times L_z$.
The $z$-axis is oriented along the beam direction.
To simulate the axial channeling the $z$-axis is directed along a chosen crystallographic direction 
$\langle klm \rangle$ (here integers $k,l,m$ stand for the Miller). 
In the case planar channeling, the $z$-axis is parallel to
the $(klm)$ plane, and the $y$ axis is perpendicular to the plane.
The position vectors of the nodes $\bfR_j^{(0)}$ ($j=1,2,\dots, N$) 
within the simulation box are generated in accordance with the 
type of the Bravais cell of the crystal and using the predefined values of 
the lattice vectors.
The simulation box can be cut along specified faces, thus allowing tailoring the 
generated crystalline sample to achieve the desired form of the sample.

Several build-in options, characterized below, allow one to further  modify
the generated crystalline structure \cite{Sushko:Thesis_2015}.

\begin{itemize}
\item
The sample can be rotated around a specified axis thus allowing for the 
construction of the crystalline structure along any desired direction.
In particular, this option allows one to choose the direction of the $z$-axis well away
from major crystallographic axes, thus avoiding the axial channeling (when not desired).

\item
The nodes can be displaced in the transverse direction:
$y \to y + R(1-\cos\phi)$ where $\phi = \textrm{arcsin}(z/R)$.
As a result, a crystal bent with constant radius $R$ is generated.

\item
Periodic harmonic displacement of the nodes is achieved 
by means of the transformation
 $\bfr \to  \bfr + \bfa \sin(\bfk\cdot\bfr+\varphi)$.
The vector $\bfa$ and its modulus, $a$, determine the direction and amplitude of 
 the displacement,
 the wave-vector $\bfk$ specifies the axis along which the displacement to be propagated,
 and $\lamu=2\pi/k$ defines the wave-length of the periodic bending.
The parameter $\varphi$ allows one to change the phase-shift of periodic bending.

 In a special case $\bfa \perp \bfk$, this options provides simulation of
 linearly polarized periodically bent crystalline structure which is 
 an important element of a crystalline undulator.

\end{itemize}

\begin{figure} [h]
\centering
\hspace*{1.5cm}
\includegraphics[scale=0.195,clip]{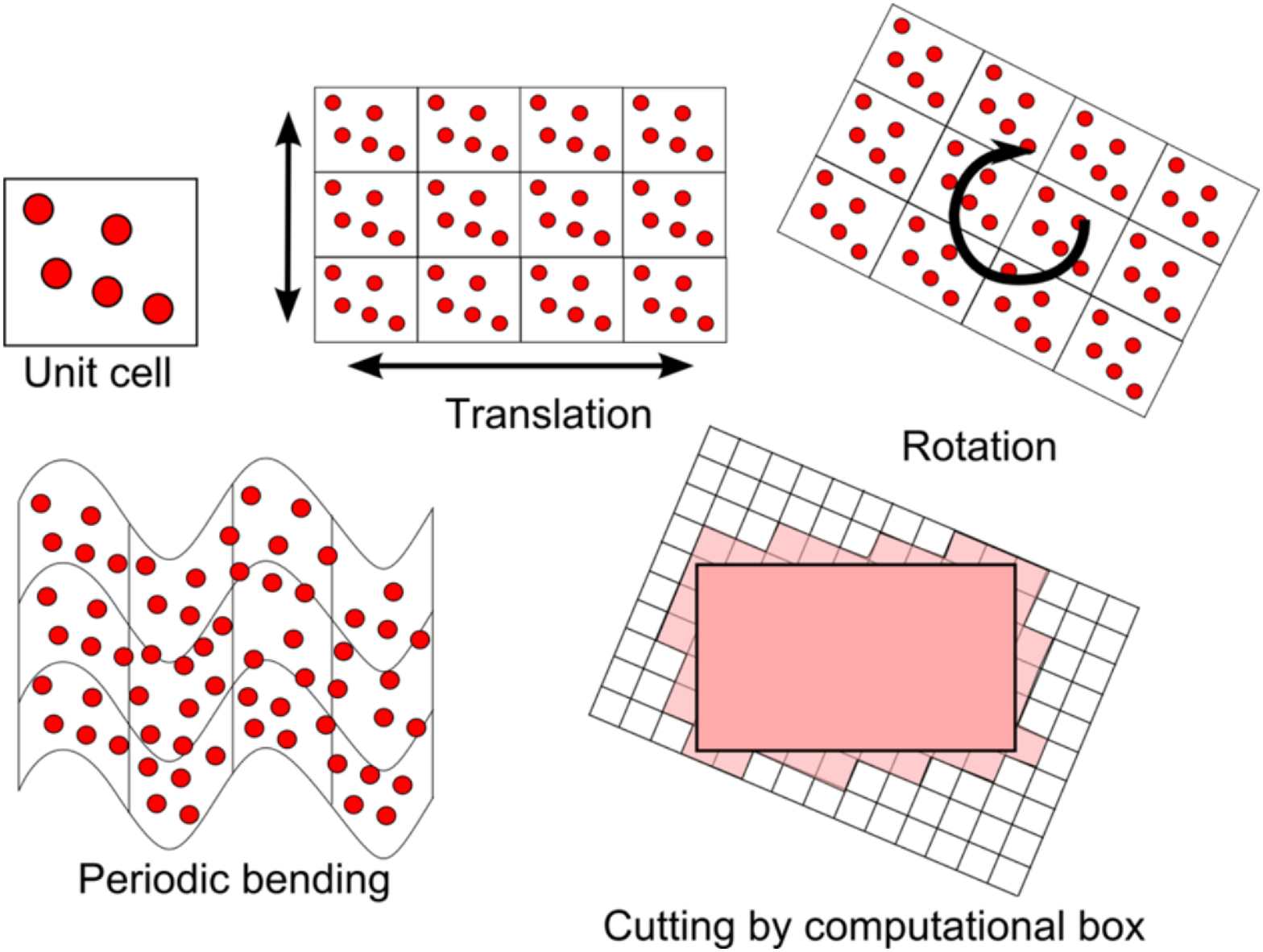}
\hspace*{0.5cm}
\includegraphics[scale=0.25,clip]{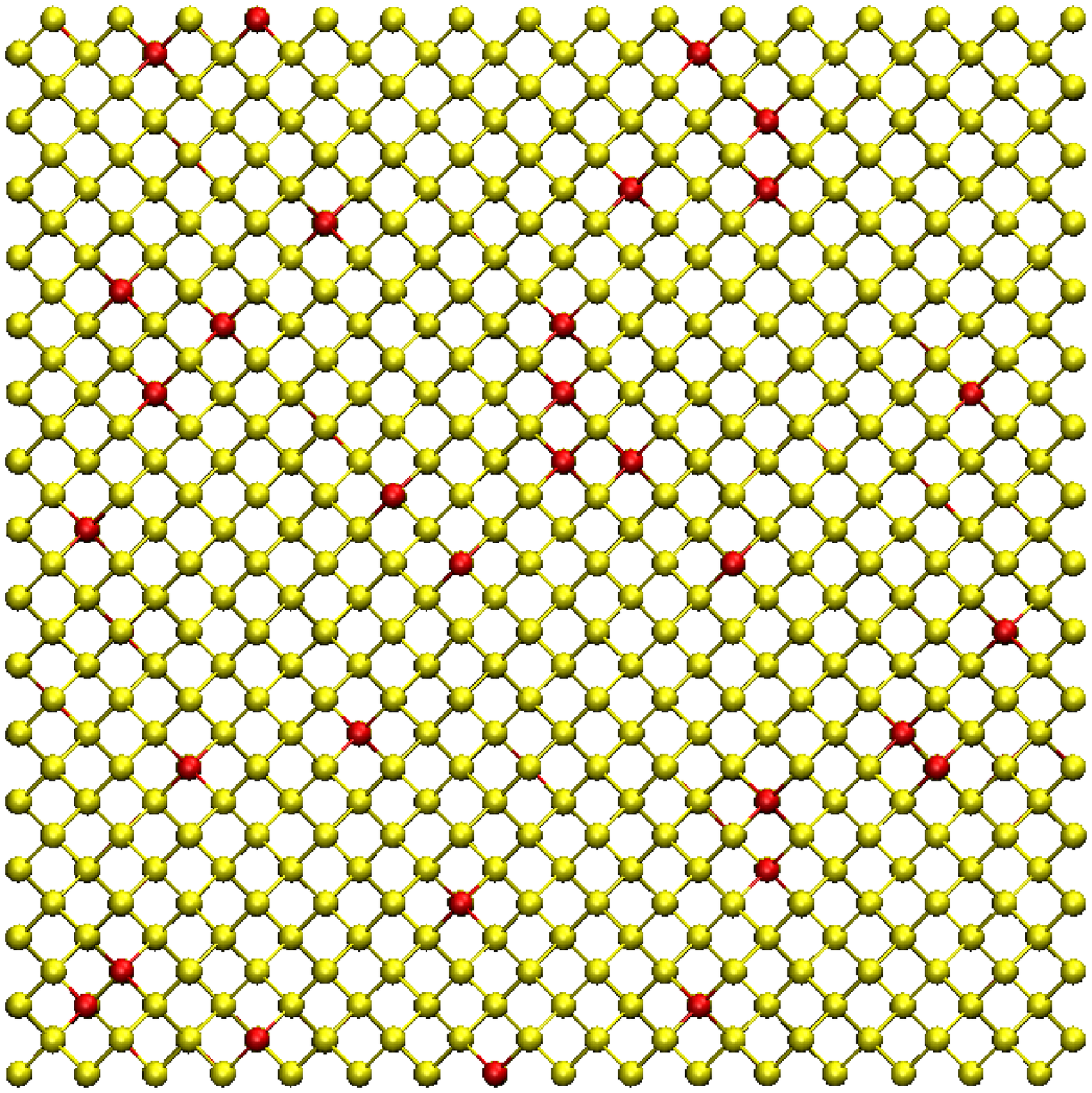}
\vspace*{0.3cm}
\caption{\textit{Left.} 
Construction of a crystalline medium is done through the definition of a unit cell of the crystal 
and a set of reversible transformations. 
This approach provides efficient mapping of the coordinates into crystalline reference frame 
i.e. cutting the periodic crystalline structure by a simulation box.
\textit{Right.}
Illustrative Si$_{1-x}$Ge$_x$ binary crystal sample with concentration $x = 0.05$ of germanium atoms. 
The silicon and germanium atoms are shown in yellow and red, respectively. 
The Ge atoms occupy positions in the same grid structure as Si but the difference in the lattice 
 constant leads to a deformation of the whole sample.
}
\label{CrystalTransformations.fig}
\end{figure}

These transformations, illustrated schematically in Fig. \ref{CrystalTransformations.fig}\textit{left}, 
are reversible and, therefore, 
allow for efficient construction of a crystalline structure in an arbitrary spatial area. 
Also, the simulation box can be cut along specified faces, thus allowing
tailoring the generated structure to achieve the desired form of the sample.
In addition to the aforementioned options, \MBNExplorer  allows one
to model binary structures 
(for example, Si$_{1-x}$Ge$_x$ or diamond-boron superlattices, see Fig. \ref{CrystalTransformations.fig}\textit{right}) 
can generated by introducing random or regular substitution of atoms in the 
initial structure with the dopant atoms.

Once the nodes are defined, the position vectors of the atomic nuclei are generated with account for random displacement 
from the nodes due to thermal vibrations corresponding to a given temperature $T$.
For each atom the displacement vector $\bDelta$ is generated by means of the normal distribution
\begin{eqnarray}
w(\Delta)
=
{1\over (2\pi u_T^2)^{3/2}}
\exp\left(-{\Delta^2 / 2u_T^2}\right)
\label{PlanarPot:eq.02}
\end{eqnarray}
where  $u_T$ denotes the root-mean-square amplitude of the thermal vibrations.
The values of $u_T$ for a number of crystals are summarized in \cite{Gemmell}.

By introducing unrealistically large value of $u_T$
(for example, exceeding the lattice constants)
it is possible to consider large random displacements.
As a result, the amorphous medium can be generated.

The trajectory of a particle entering the initially constructed crystal at the instant $t=0$
is calculated by integrating equations (\ref{Methodology:eq.01}).
Initial transverse coordinates, $(x_0, y_0)$, and velocities, $(v_{x,0}, v_{y,0})$, 
are generated randomly accounting for the conditions at the crystal entrance 
(i.e., the crystal orientation and beam emittance).
A particular feature of \MBNExplorer is in simulating the crystalline environment "on the fly", 
i.e. in the course of propagating the projectile.
This is achieved by introducing a \textit{dynamic simulation box} which moves following 
the particle (see Refs.~\cite{MBN_ChannelingPaper_2013,ChannelingBook2014} for the details). 

\begin{figure} [h]
\centering
\includegraphics[scale=0.3,clip]{Figure_02a.eps}
\hspace*{0.5cm}
\includegraphics[scale=0.25,clip]{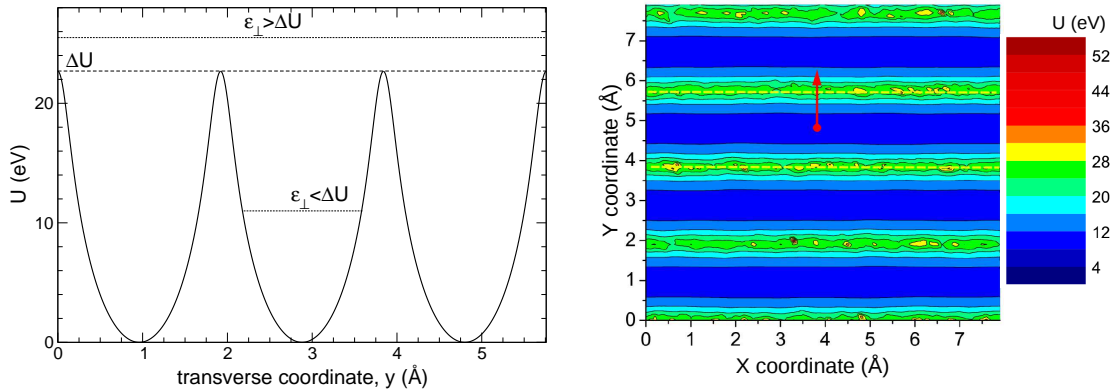}
\caption{\textit{Left.} 
Continuous interplanar  Si(110) potential $U(y)$ for a positron calculated in the Moli\`{e}re approximation at $T=300$ K.
The coordinate $y$ is measured along the $\langle 110 \rangle$ axial direction.
The channeling regime corresponds to the transverse energies $\E_{\perp}<\Delta U$.
\textit{Right.}
The positron potential in Si(110) calculated following Eq. (\ref{MC_Simulations.02}) as a sum of individual atomic potentials
with account for thermal vibrations of the atoms. 
The red arrow is aligned with  the $\langle 110 \rangle$ direction.
}
\label{Si110_Potentials.fig}
\end{figure}

The important methodological issue concerns formulation of a criterion for distinguishing between channeling and
non-channeling regimes of projectiles' motion.
Depending on theoretical approach used to describe interaction of a projectile with
a crystalline environment, the criterion can be introduced in different ways. 
For example, within the continuous potential framework \cite{Lindhard} 
the transverse and longitudinal motions of the projectile are decoupled. 
As a result, it is straightforward to define the channeling projectiles as 
those with transverse energies $\E_{\perp}$ less than the height $\Delta U$ of the inter-planar 
(or, inter-axial) potential barrier, see Fig. \ref{Si110_Potentials.fig}\textit{left}). 
 Within this framework, the \textit{acceptance} $\calA$ is determined at the entrance to the
crystal and can be defined as the ratio of the number of particles with $\E_{\perp} < \Delta U$ to the 
total number of particles. 

Within the framework of molecular dynamics, the simulations are based on solving the EM
(\ref{Methodology:eq.01}) accounting, as in reality, for the interaction of a projectile 
with individual atoms of the crystal. 
The potential $U(\bfr)$ (\ref{MC_Simulations.02}) experienced by the projectile varies rapidly in the course of the motion 
(see Fig. \ref{Si110_Potentials.fig} \textit{right})
coupling the transverse and longitudinal degrees of freedom.
Therefore, other criterion must be provided to identify the channeling segments in the projectile’s trajectory.
\textcolor{black}{
To distinguish the channeling segments we employ a kinematic criterion introduced in Refs. 
\cite{KKSG_simulation_straight,MBN_ChannelingPaper_2013}.
For a particular case of planar channeling the criterion as follows:
a particle is considered to be moving in the channeling mode if the sign of its transverse velocity $v_y$
(normal to the planar direction) has been changed at least two times inside the same channel.
This corresponds to at least one full oscillation executed by the particle.
The criterion must be supplemented by geometrical definition of the channels. 
In the case of straight crystals the positron channels are the volume areas restricted by the neighboring 
planes, whereas the electron channels are the areas between the
corresponding neighboring mid-planes. 
For bent and periodically bent crystals this definition is modified accordingly.
}
It was noted however \cite{Backe-ChanCriterion} that this criterion is not perfect in a sense that 
it leads to some artificial features in the evolution of the fraction of channeling particles 
with the penetration distance, see Section \ref{MBNExplorer:StatisticalAnalysis}.

\subsection{Statistical analysis of trajectories\label{MBNExplorer:StatisticalAnalysis}}

Taking into account randomness in sampling the incoming projectiles and 
in positions of the lattice atoms due to the thermal fluctuations, 
one concludes that each simulated trajectory corresponds to a unique crystalline
environment.
Thus, all simulated trajectories are statistically independent and can be
analyzed further to quantify the channeling process as well as the emitted radiation.

\begin{figure} [h]
\centering
\includegraphics[clip,width=10cm]{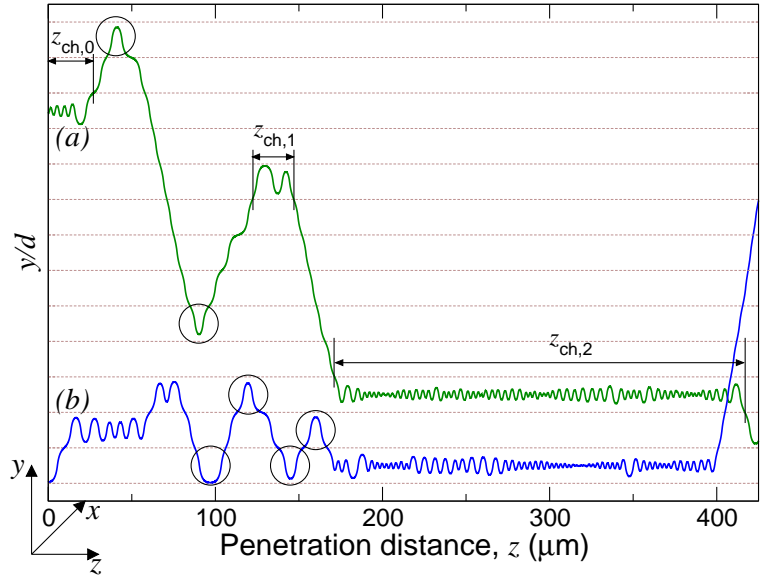}
\caption{
Selected simulated trajectories of 10 GeV electrons propagating in straight oriented 
diamond(110) crystal.
The trajectories illustrate the channeling and the over-barrier motion as well as
the dechanneling and rechanneling effects. 
The z-axis of the reference frame is directed along the incoming projectiles, 
the $(xz)$-plane is parallel to the (110) crystallographic planes (dashed lines) and 
the $y$-axis is perpendicular to the planes. 
The diamond(110) interplanar distance is $d = 1.26$ \AA. 
For the accepted trajectory (a) the characteristic lengths of the channeling motion are indicated: 
the initial channeling segment, $z_{\rm ch,0}$, and \textcolor{black}{channeling} segments in the bulk,
$z_{\rm ch,1}$ and $z_{\rm ch,2}$. 
The non-accepted trajectory (b) corresponds to $z_{\rm ch,0} = 0$. 
Encircled are the parts of trajectories that do not satisfy the criterion adopted for the definition of the channeling mode:
the projectile stays in the same channel but changes the direction of the transverse motion only once.  
}
\label{e-10GeV-Si110_traj.fig}
\end{figure}

Figure \ref{e-10GeV-Si110_traj.fig} shows two simulated trajectories of 10 GeV electrons 
which enter straight oriented diamond crystal along the (110) crystallographic planes. 
Dashed horizontal lines in Fig. \ref{e-10GeV-Si110_traj.fig} mark the cross section of the (110) 
crystallographic planes separated by the distance $d=1.26$ \AA{}.
Thus, the $y$-axis is aligned with the $\langle 110 \rangle$ crystallographic axis.
The horizontal $z$-axis corresponds to the direction of the incoming particles.
To avoid the axial channeling, the $z$-axis was chosen along the $[10, -10, 1]$ crystallographic direction.
A projectile enters the crystal at $z=0$ and exits at $z=L$.
The crystal is considered infinitely large in the $x$ and $y$ directions.
In the simulations, the integration of the equations of motion (\ref{Methodology:eq.01})
produces a 3D trajectory. 
The curves shown in the figure represent the projections of the corresponding 
3D trajectories on the $(yz)$ plane.  

These exemplary trajectories illustrate a variety of features which
characterize the motion of a charged projectile in an oriented crystal:
the channeling motion, the over-barrier motion, the dechanneling and
the rechanneling processes, rare events of hard collisions.
Apart from providing the possibility of illustrative comparison, the simulated trajectories
allow one to quantify the channeling process in terms of several parameters and 
functional dependencies which can be generated on the basis of statistical
analysis of the trajectories 
\cite{MBN_ChannelingPaper_2013,ChannelingBook2014,Sushko-EtAl:JPConfSer_v438_012018_2013,Sushko-EtAl:JPConfSer_v438_012019_2013,
Polozkov-EtAl:EPJD_v68_268_2014,Sushko-EtAl:NIMB_v355_p39_2015,Sushko-EtAl:NTV_v1_p341_2015,Korol-EtAl:NIMB_v387_p41_2016}.

Randomization of the "entrance conditions" leads to different chain of scattering 
events for the different projectiles at the entrance to the bulk. 
As a result, not all trajectories start with the channeling segments.
In Fig. \ref{e-10GeV-Si110_traj.fig},  trajectory (a) refers to an accepted projectile that
changes the direction of transverse motion more than two times while moving in the same channel.
In contrast, trajectory (b) corresponds to the non-accepted projectile.
To quantify this feature we define acceptance as the ratio
\begin{eqnarray}
\calA = {N_{\rm acc} \over N_0}
\label{Acceptance}
\end{eqnarray}
where
$N_{\rm acc}$ stands for the number of accepted particles and $N_0$ is the total number of the incident particles.
The non-accepted particles experience unrestricted over-barrier motion at the entrance 
but can rechannel at some distance $z$.

As defined, acceptance depends on the beam emittance, on the type of the crystal and the channel, and on the bending radius. 
In a bent crystal, the channeling condition \cite{Tsyganov1976} implies that 
the centrifugal force $F_{\rm cf}= pv/R \approx \E/R$\textcolor{black}{, acting on the particle
in the co-moving frame ($R$ stands for the bending radius)}
is smaller than the maximum interplanar force $F_{\max}$.
It is convenient to quantify this statement by introducing the dimensionless 
\textit{bending parameter} $C$:
\begin{equation}
C = {F_{\rm cf} \over F_{\max}} = {\E \over R F_{\max}} = {R_{\rm c} \over R}.
\label{Results:eq.01}
\end{equation}
The case $C = 0$ ($R=\infty$) characterizes the straight crystal whereas
$C = 1$ corresponds to the critical (minimum) bending radius $R_{\rm c}=\E/F_{\max}$
\cite{Tsyganov1976}. 
Within the framework of the continuous interplanar potential model,
one can calculate $F_{\max}$ by means of the formulae presented in \ref{InterPlanarPot}.

\begin{figure}
\centering
\includegraphics[width=10.0cm,clip]{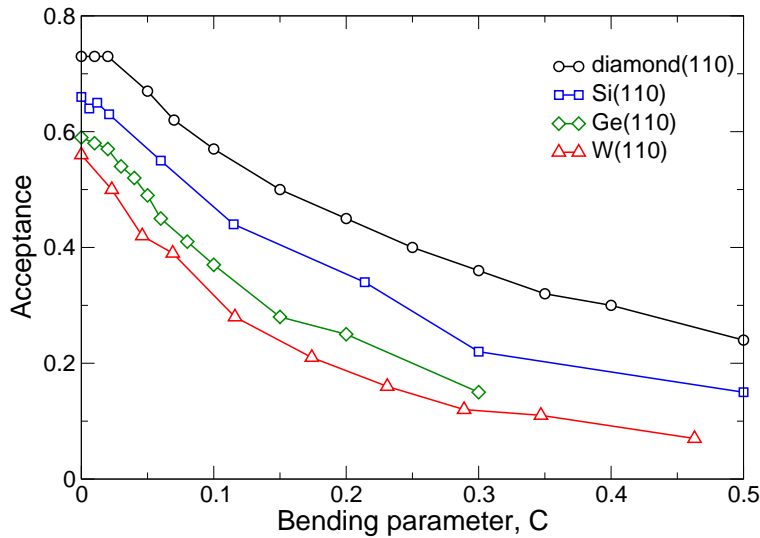}
\caption{
Acceptance versus bending parameter $C$ (see Eq. (\ref{Results:eq.01}))
for 855 MeV electrons.
The presented data, obtained by means of means of \MBNExplorer, are take from:
Ref. \cite{Sushko-EtAl:NTV_v1_p341_2015} for diamond(110),
Refs. \cite{Polozkov-EtAl:EPJD_v68_268_2014,Sushko-EtAl:JPConfSer_v438_012019_2013} for Si(110),
Ref. \cite{Sushko:Thesis_2015} for Ge(110),
and \cite{Korol-EtAl:NIMB_v424_26_2018} for W(110).
}
\label{Figure.02} 
\end{figure}

Figure \ref{Figure.02} shows acceptance as a function of the bending parameter $C$ for 855 MeV electrons 
channeling in several oriented crystals as indicated.
The symbols mark the data that were obtained by statistical analysis of the simulated trajectories.
For each crystal, the corresponding values of bending radius can be calculated from Eq. (\ref{Results:eq.01}) 
using the following values 
of $F_{\max}$ calculated in the Moli\`{e}re approximation and at room temperature:
7.0, 5.7, 10.0, and 42.9 GeV/cm for diamond, silicon, germanium, and tungsten, respectively.

An accepted projectile, stays in the channeling mode of motion over some interval $z_{ch,0}$ until 
an event of the dechanneling (if it happens).
The initial channeling segment is explicitly indicated for trajectory (a). 
For the non-accepted particle this segment is absent, $z_{ch,0}=0$.
To quantify the dechanneling effect for the accepted particles, one can introduce \textit{the penetration length} $\Lp$ 
\cite{MBN_ChannelingPaper_2013} 
defined as the arithmetic mean of the initial channeling segments $z_{\rm ch,0}$ calculated with respect 
to all accepted trajectories:
 \begin{eqnarray}
\Lp 
=
{\sum_{j=1}^{N_{\rm acc}} z_{\rm ch,0}^{(j)} \over N_{\rm acc} }.
\label{Eq.Lp}
\end{eqnarray}
For sufficiently thick crystals  the penetration length approaches the so-called dechanneling length $\Ld$ 
that characterizes the decrease of the fraction of channeling particles in terms of the exponential decay 
law, $\propto \exp(-z /\Ld)$
 \cite{Beloshitsky-EtAl:RadEff_v20_p95_1973}. 
The concept of exponential decay has been widely exploited to estimate the dechanneling-channeling lengths for various 
ultra-relativistic projectiles in straight and bent crystals 
\cite{Backe_EtAl_2008,BogdanovDabagov2012,Scandale_etal:PL_B719_p70_2013,BiryukovChesnokovKotovBook,
Mazzolari_etal:PRL_v112_135503_2014,Wienands_EtAl_PRL_v114_074801_2015,Wistisen_EtAl_PR-AB_v19_071001_2016,%
BackeLauth:NIMB-v355-p24-2015,Sytov-EtAl_EPJC_v77_901_2017}.
To derive the \textit{approximate} relationship between the dechanneling length, which does not 
depend on the crystal thickness, and
the penetration length, the value of which depends on $L$,
one assumes that the exponential decay law is applicable at any distance $z$ in the crystal.
Then, calculating the penetration length as the mean value of the channeling segment in the 
 crystal of thickness $L$ one obtains:
 \begin{eqnarray}
\Lp = \Ld\left(1 - \ee^{-L/\Ld}\right)\,.
\label{Eq.02}
\end{eqnarray}
It is seen, that for any finite thickness $L$ the value of $\Lp$, calculated from statistical analysis 
of the simulated trajectories, is smaller the dechanneling lengths defined within the framework of the 
diffusion approximation to the dechanneling process 
\cite{Beloshitsky-EtAl:RadEff_v20_p95_1973}. 
The transcendent Eq. (\ref{Eq.02}) can be easily solved numerically with respect to $\Ld$ for any given 
values $\Lp$ and $L$ used in the simulations.
The solution becomes trivial in the limit of a thick crystal,  $L \gg \Lp, \Ld$, when the exponential 
term on the right-hand side can be omitted producing $\Ld \approx \Lp$.

Random scattering of the projectiles on the crystal atoms can result in the rechanneling, 
i.e. the process of capturing the particles into the channeling mode of motion. 
In a sufficiently long crystal, the projectiles can experience dechanneling and 
rechanneling several times in the course of propagation, as it is illustrated by both trajectories in 
Fig. \ref{e-10GeV-Si110_traj.fig}. 
These multiple events can be quantified by introducing the \textit{total channeling length} $\Lch$, which
characterizes the channeling process in the whole crystal. 
The total channeling length $\Lch$ per particle is calculated  by averaging the sums of all channeling 
segments $z_{{\rm ch}0} + z_{{\rm ch}1} + z_{{\rm ch}2} + \dots$, calculated for each trajectory, 
over all trajectories.

\Table{\label{Germanium-Sushko.Table}
Acceptance $\calA$, penetration length $\Lp$, total channeling length $\Lch$ (both in microns) for 855 MeV electrons
in straight ($C=0$) and bent ($C>0$) germanium (110) and (111) channels \cite{Sushko:Thesis_2015}.
The last column shows the \textcolor{black}{dechanneling lengths measured experimentally at the MAinz MIcrotron (MAMI) facility}
\cite{Sytov-EtAl_EPJC_v77_901_2017}. 
 }
 \br
      &  \multicolumn{3}{c}{(110)}                &&  \multicolumn{4}{c}{(111)} \\
 $C$  & $\calA$&      $\Lp$      &    $\Lch$      &&  $\calA$  &     $\Lp$         &    $\Lch$     &  $\Ld$ (exp.)\\
 \mr
 0.00 &  0.59  & 6.6 $\pm$ 0.3 & 18.1 $\pm$ 0.7 &&  0.45     &  10.6 $\pm$ 0.5 & 20.6 $\pm$ 0.8 &               \\
 0.01 &  0.58  & 6.5 $\pm$ 0.3 & 16.0 $\pm$ 0.7 &&  0.42     &  10.3 $\pm$ 0.6 & 18.4 $\pm$ 0.9 &              \\
 0.02 &  0.57  & 6.3 $\pm$ 0.3 & 11.6 $\pm$ 0.6 &&  0.42     &  10.1 $\pm$ 0.5 & 14.9 $\pm$ 0.9 &              \\
 0.03 &  0.54  & 6.1 $\pm$ 0.3 &  9.1 $\pm$ 0.5 &&  0.38     &   9.9 $\pm$ 0.5 & 12.4 $\pm$ 0.9 &              \\
 0.04 &  0.52  & 5.9 $\pm$ 0.3 &  7.6 $\pm$ 0.5 &&  0.36     &   9.7 $\pm$ 0.6 & 11.0 $\pm$ 1.0 &              \\
 0.05 &  0.49  & 5.7 $\pm$ 0.3 &  6.9 $\pm$ 0.4 &&  0.32     &   9.5 $\pm$ 0.5 & 10.4 $\pm$ 1.1 &   9 $\pm$ 5    \\
 0.06 &  0.45  & 5.4 $\pm$ 0.3 &  6.2 $\pm$ 0.6 &&  0.29     &   9.2 $\pm$ 0.5 &  9.6 $\pm$ 1.1 &              \\
 0.08 &  0.41  & 5.2 $\pm$ 0.3 &  5.5 $\pm$ 0.5 &&  0.22     &   9.0 $\pm$ 0.6 &  9.1 $\pm$ 1.9 &  7.3 $\pm$ 1.2 \\
 0.10 &  0.37  & 4.8 $\pm$ 0.3 &  5.0 $\pm$ 0.5 &&  0.14     &   7.9 $\pm$ 0.7 &  7.7 $\pm$ 2.8 &  5.9 $\pm$ 1.5 \\ 
 0.15 &  0.28  & 4.0 $\pm$ 0.3 &  4.1 $\pm$ 0.7 &&           &                   &                &                 \\
 0.20 &  0.25  & 3.5 $\pm$ 0.2 &  3.5 $\pm$ 0.7 &&           &                   &                &                  \\
 \br
 \end{tabular}
 \end{indented}
 \end{table}

The results on the acceptance and the characteristic lengths are presented in Table  \ref{Germanium-Sushko.Table} \cite{Sushko:Thesis_2015}. 
All the data refer to zero emittance 855 MeV electrons beams entering a $L = 75$ microns thick germanium (110) and (111) crystals. 
Statistical uncertainties due to the finite numbers of the simulated trajectories correspond to the
99.9\% confidence level.
Also shown are the data on $\Ld$ measured in the recent experiment \cite{Sytov-EtAl_EPJC_v77_901_2017}.
As mentioned above, for $L \gg \Lp$ the value of $\Lp$ obtained in the simulations equals to the dechanneling length, which determines the exponential decay
of the channeling fraction.
Comparing the data from the last column with that for $\Lp$ we state very good agreement between our theory and the experiment. 

To further quantify the impact of the rechanneling effect basing on the information which can be extracted from the simulated trajectories,
one can compute the channeling fractions  $f_{\rm ch, 0}(z)=N_{\rm ch, 0}(z)/\Nacc$ and $f_{\rm ch}(z)=N_{\rm ch}(z)/\Nacc$ 
\cite{MBN_ChannelingPaper_2013,Sushko-EtAl:JPConfSer_v438_012019_2013}.
Here $N_{\rm ch, 0}(z)$ stands for the number of particles that propagate in the same channel
where they were accepted up to the distance $z$ where they dechannel.
The second quantity, $N_{\rm ch}$, is the total number of particles which are in the channeling mode at the distance $z$. 

\begin{figure} [h]
\centering
\includegraphics[clip,width=10cm]{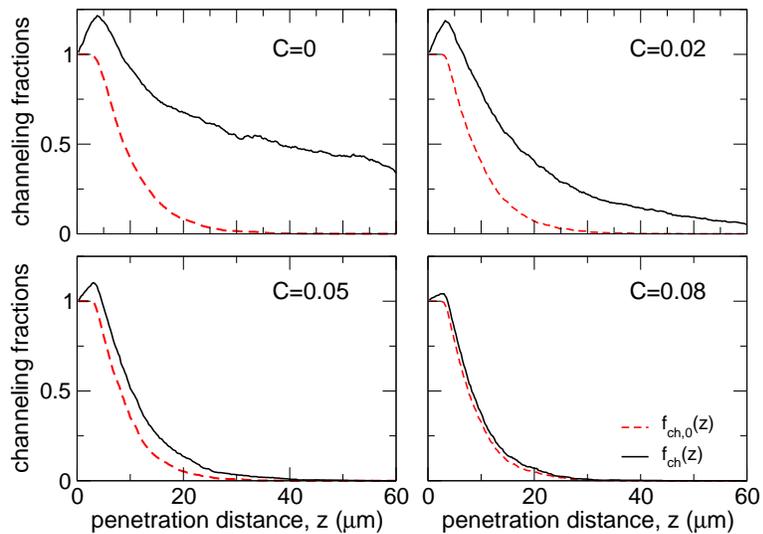}
\caption{
Channeling fractions $f_{\rm ch 0}(z)$ (dashed curves) and 
$f_{\rm ch}(z)$ (solid curves) 
calculated for 855 MeV electrons in straight ($C=0$) and bent ($C>0$) 
Ge(110) channels. Ref. \cite{Sushko:Thesis_2015}.
}
\label{e855-Ge111-fractions_v02.fig}
\end{figure}

As $z$ increases the fraction $\xi_{\rm ch0}(z)$ decreases due to the dechanneling of the accepted particles.
In the contrast, the fraction $\xi_{\rm ch}(z)$ 
can increase with $z$ when the particles, including those not accepted at the entrance, 
can be captured in the channeling mode in the course of the  {rechanneling}. 
These dependencies for a 855 MeV electron channeling in straight and bent germanium(110) channels are 
presented in Fig. \ref{e855-Ge111-fractions_v02.fig}.

A striking difference in the behaviour of the two fractions as functions of the 
penetration distance $z$ is mostly pronounced for the straight channel.
Away from the entrance point, the fraction $f_{\rm ch 0}(z)$ (dashed curve)
follows approximately the exponential decay law (see discussion below).
At large distances, the fraction $f_{\mathrm ch}(z)$ (solid curve), which accounts for the 
rechanneling process, decreases much slower following the power law, 
$f_{\rm ch}(z)\propto z^{-1/2}$ \cite{KKSG_simulation_straight}.
As the bending curvature increases, $C\propto 1/R$, the rechanneling events become rarer, and
the difference between two fractions decreases.
For $C\gtrsim 0.1$ both curves virtually coincide.
\textcolor{black}{We note here that the impact of rechanneling was also highlighted by other authors
in connection with the experimental studies with 
both straight \cite{Backe_EtAl_2008} and bent \cite{Mazzolari_etal:PRL_v112_135503_2014} crystals.}

Let us present simple analytical fits for both channeling fractions.
The parameters, which enter the fitting formulae, can be related to those which are used to quantify the
dechanneling process within the framework of the diffusion theory of dechanneling process (see, e.g., \cite{BiryukovChesnokovKotovBook}).

\begin{figure} [h]
\centering
\includegraphics[clip,width=10cm]{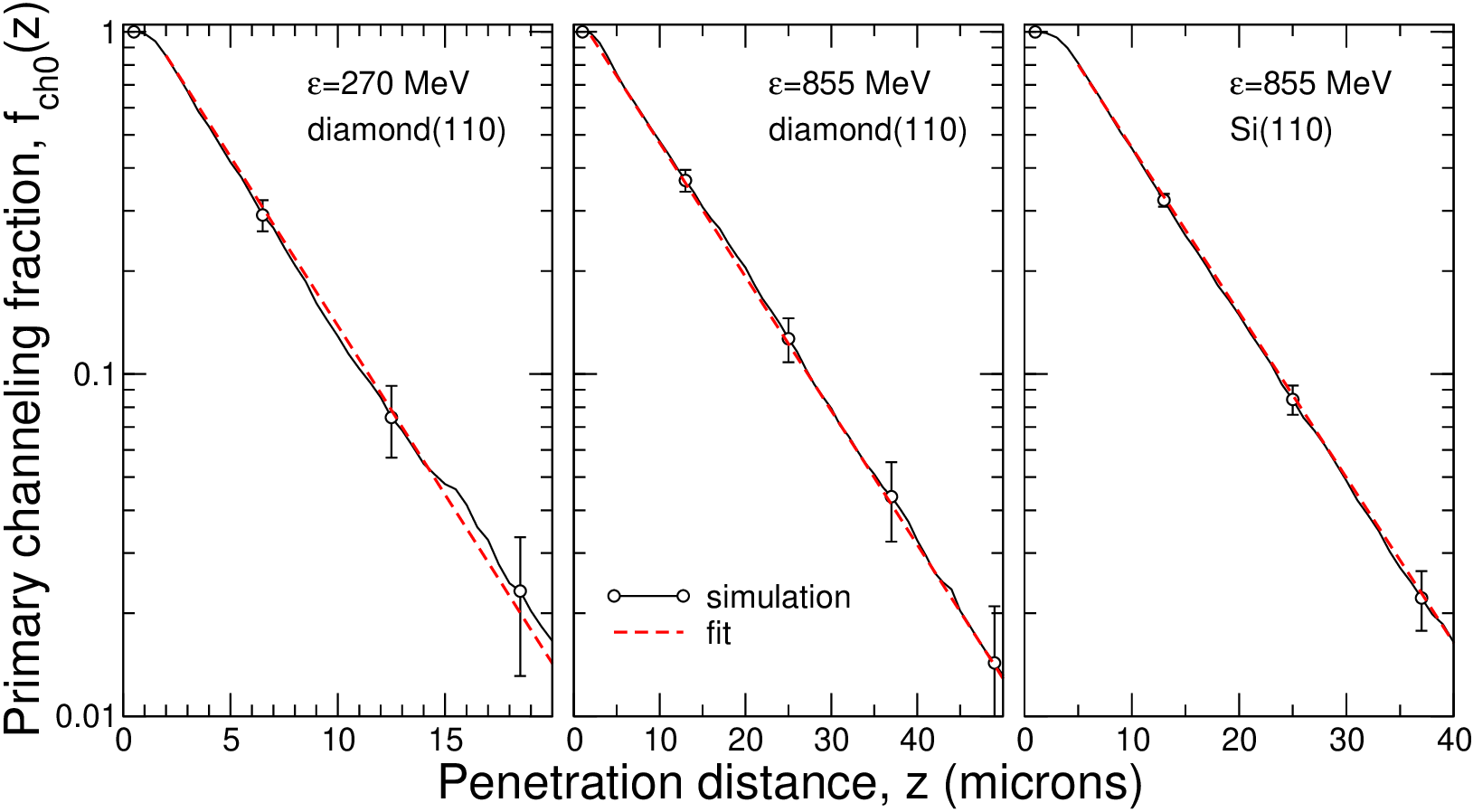}
\caption{
Primary channeling fraction $f_{\rm ch, 0}(z)$ versus penetration 
distance for 270 and 855 MeV electrons in, correspondingly,
300 and 140 microns thick diamond(110) crystal (left and middle graphs),
and for 855 MeV electrons in 500 microns thick Si(110) (right).
Solid curves with open circles and error bars stand for the simulated data.
Red dashed curves correspond to the fit (\ref{Nch0_Fit:eq.01}).
With account for statistical uncertainties the intervals for 
the fitting parameter $z_0$ and $\lambda$ are (from left to right):
$z_0=(1.42 \pm 0.05)$, $(1.7\pm 0.1)$, $(3.0\pm 0.1)$ microns
and 
$\Ld=(4.4 \pm 0.1)$, $(11.14\pm 0.14)$, $(9.0\pm 0.1)$ microns.
}
\label{e-C100-Si110-Nch0.fig}
\end{figure}

Apart from the some interval $[0, z_0]$ at the crystal entrance, 
the primary channeling fraction, $f_{\rm ch, 0}(z)$, can be approximated by 
the exponential decay law:
\begin{eqnarray}
\tilde{f}_{\rm ch, 0}(z) = \ee^{-(z-z_0)/\Ld}\,,
\label{Nch0_Fit:eq.01}
\end{eqnarray}
where $z_0$ and $\Ld$ are fitting parameters.
The parameter $z_0>0$ appears due to the aforementioned criterion used to identify the accepted particles: the initial segment of each trajectories is
analyzed to decide whether the particle moves in the channeling mode. 
Then, the penetration distance $L_{\rm p}$ is written as follows:
\begin{eqnarray*}
L_{\rm p} = z_0 + \int_{z_0}^L (z-z_0) \tilde{f}_{\rm ch, 0}(z) {\d z \over \Ld} + (L-z_0) \tilde{f}_{\rm ch, 0}(L)\,.
\label{Nch0_Fit:eq.01a}
\end{eqnarray*}
Here,  the integral term is due to the particles that dechannel at some distance $z\in[z_0,L]$
inside the crystal while the last, non-integral term accounts for the particles that travel the whole distance $L$ in the channeling regime.
Carrying out the integration one derives    
 \begin{eqnarray}
\Lp 
=
z_0 + \Ld\left(1 - \ee^{-(L-z_0)/\Ld}\right)\,.
\label{Nch0_Fit:eq.01b}
\end{eqnarray}

Figure \ref{e-C100-Si110-Nch0.fig} compares the simulated $f_{\rm ch, 0}(z)$ 
(solid curves with symbols) and
approximate $\tilde{f}_{\rm ch, 0}(z)$ (dashed curves) dependences 
for $\E=270$ and 855 MeV electron channeling in diamond(110) and silicon(110) 
crystals.
The values of the fitting parameters are indicated in the caption.

To derive the fitting formula for the channeling fraction $f_{\rm ch}(z)$, which accounts
for the re-channeling, one can utilize the approximation based on the one-dimension diffusion equation 
for the density of channeling particles, $\rho(\theta,z)$, as a function
of penetration distance and the angle $\theta$ with respect to the planar channel centerline:
\begin{eqnarray}
{\partial \rho \over \partial z} 
=
D {\partial^2 \rho \over \partial \theta^2}\,.
\label{Nch_Fits:eq.01}
 \end{eqnarray}
The diffusion coefficient $D$, which is assumed to be independent on $\theta$ and $z$.
Imposing the initial and boundary conditions as
$\rho(\theta,0) = \delta(\theta)$ and $\rho(\pm\infty,z) = 0$
one writes the solution of Eq. (\ref{Nch_Fits:eq.01}) \footnote{The initial condition and 
the solution are written for the case of ideally collimated beam. 
For a non-collimated beam one carries out the substitution $2 D z \to 2 D z + \theta_0^2$,
where $\theta_0^2$ is the beam divergence at the crystal entrance}
\begin{eqnarray}
\rho(\theta,z) 
=
{1\over \sqrt{4\pi D z}} \exp\left(-{\theta^2 \over 4 D z}\right)\,.
\label{Nch_Fits:eq.02}
\end{eqnarray}
The channeling fraction is calculated as follows
\begin{eqnarray}
f_{\rm ch}(z)
=
\int_{-\theta_{\rm L}}^{\theta_{\rm L}}
\rho(\theta,z) \d \theta
=
{\rm erf}
\left({\theta_{\rm L} \over \sqrt{4 D z } }\right)
\label{Nch_Fits:eq.03}
\end{eqnarray}
where $\theta_{\rm L}$ is Lindhard's critical angle 
and ${\rm erf}(.)$ stands for the error function.

To get initial values of the parameters entering the right-hand-side of 
(\ref{Nch_Fits:eq.03}) one estimates the critical angle as
$\Theta_{\rm L} = \sqrt{2U_0/\E}$ with $U_0$ standing for the depth of the continuous
interplanar potential well.
The diffusion coefficient $D$ can be related the rms angle of multiple scattering 
$\langle \theta^2 \rangle(z)$:
\begin{eqnarray}
D = {\langle \theta^2 \rangle \over 2z}\,.
\label{Nch_Fits:eq.04}
\end{eqnarray}
For estimation purposes one can use the following formula  \cite{Backe_JINST_v13_C02046_2018}:
\begin{eqnarray}
\langle \theta^2 \rangle 
=
\left({10.6 (\mbox{MeV}) \over \E\, (\mbox{MeV})}\right)^2
{z \over L_{\rm rad} }
\label{Nch_Fits:eq.05}
\end{eqnarray}
where $L_{\rm rad}$ is the radiation length.
It was argued in the cited paper that the factor 10.6 (MeV) provides a more accurate 
approximation in the limit of small penetration distances, $z\ll L_{\rm rad}$, 
whereas a conventional value of 13.6 (MeV) (see, e.g., \cite{ParticleDataGroup2018})
is applicable for $z \gtrsim L_{\rm rad}$.
Utilizing Eq. (33.20) from Ref.\,\cite{ParticleDataGroup2018}) one obtains the 
following values of $L_{\rm rad}$ for amorphous carbon, silicon and germanium media:
12.2, 9.47 and 2.36 cm.
This values are much larger than thickness of crystals used in channeling experiments with
electrons and positrons. 

Using (\ref{Nch_Fits:eq.04}) and (\ref{Nch_Fits:eq.05}) one derives the following
fitting formula for the channeling fraction 
\begin{eqnarray}
\tilde{f}_{\rm ch}(z) 
= 
{\rm erf}\left(A\sqrt{\E \over z}\right)\,.
\label{Nch_Fits:eq.05a}
\end{eqnarray}
Within the model described above the parameter $A$ is written as follows
$A=(L_{\rm rad}U_0)^{1/2}/106$ (with $L_{\rm rad}$, $U_0$, $\E$ and $\E$ measured in
cm, eV, MeV, and microns, correspondingly). 
The calculated value can be used as the initial guess for the best fit to
the $f_{\rm ch}(z)$ obtained from numerical simulations.

Note that for sufficiently large penetration distances ${\rm erf}\left(A\sqrt{\E /z}\right) \propto z^{-1/2}$.

\begin{figure} [h]
\centering
\includegraphics[clip,width=12cm]{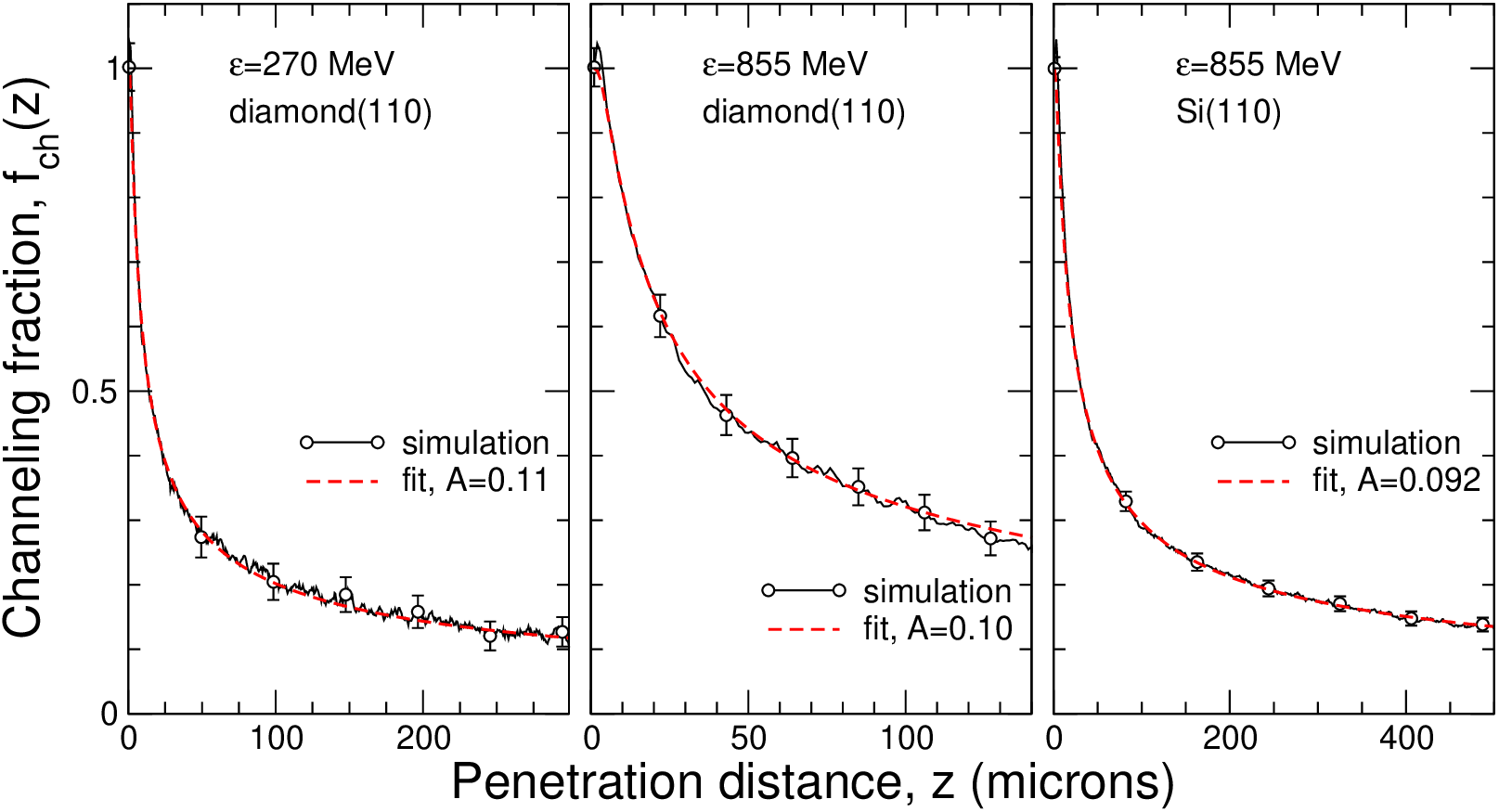}
\caption{
Channeling fraction $f_{\rm ch}(z)=N_{\rm ch}(z)/\Nacc$ versus penetration 
distance for 270 and 855 MeV electrons in, correspondingly,
300 and 140 microns thick diamond(110) crystal (left and middle graphs),
and for 855 MeV electrons in 500 microns thick Si(110) (right).
Solid curves with open circles and error bars stand for the simulated data.
Red dashed curves correspond to the fit (\ref{Nch_Fits:eq.05a}).
With account for statistical uncertainties the intervals for 
the fitting parameter $A$ are (from left to right):
$A=(0.11 \pm 0.07)$, $(0.10\pm 0.5)$, and $(0.092\pm 0.007)$ (micron/eV)$^{1/2}$.
}
\label{e-C100-Si110-Nch.fig}
\end{figure}

Figure \ref{e-C100-Si110-Nch.fig} shows three case studies which illustrate the applicability of 
the fitting formula (\ref{Nch_Fits:eq.05a}).
Solid curves with symbols present the dependences $f_{\rm ch}(z)$ obtained from simulation 
of $\E=270$ and 855 MeV electron channeling in diamond(110) and silicon(110) crystals.
The dashed curves stand for the fits with parameter $A$ (in (micron/eV)$^{1/2}$) 
indicated in the legend and in the caption as well.

\subsection{Calculation of spectral distribution of emitted radiation \label{MBNExplorer:Spectrum}}

Spectral distribution of the energy emitted within the cone $\theta\leq \theta_{0}\ll 1$ with 
respect to the incident beam is computed numerically using the following formula:
\begin{eqnarray}
{\d E(\theta\leq\theta_{0}) \over \d \om} 
=
{1 \over N_0}
\sum_{n=1}^{N_0} 
\int\limits_{0}^{2\pi}
\d \phi
\int\limits_{0}^{\theta_{0}}
\theta \d\theta\,
{\d^3 E_n \over \d \om\, \d\Om}.
\label{Methodology:eq.03}
\end{eqnarray}
Here, $\om$ is the radiation frequency, 
$\Om$ is the solid angle corresponding to the emission angles $\theta$ and $\phi$.
The quantity $\d^3 E_n/\d \om\, \d\Om$ stands for the spectral-angular distribution 
emitted by a particle that moves along the $n$th trajectory.
The sum is carried out over \textit{all} simulated trajectories, and thus its takes into account the 
contribution of the channeling segments of the trajectories as well as of those corresponding to the
non-channeling regime.

The numerical procedures implemented in \MBNExplorer to calculate the distributions 
$\d^3 E_n/\d \om\, \d\Om$ \cite{MBN_ChannelingPaper_2013} are based on the quasi-classical 
formalism \cite{Baier} that combines classical description of the particle's motion with 
the quantum corrections due to the radiative recoil quantified by the ratio $\hbar\om/\E$.
In the limit  $\hbar \om /\E \ll 1$ the quasi-classical formula reduces to that known in
classical electrodynamics (see, e.g., \cite{Jackson}).
The classical description the radiative process is adequate to characterize the 
emission spectra by electrons and positrons of the sub-GeV and GeV energy range 
(see, for example,~\cite{ChannelingBook2014} and references therein).  
The quantum corrections lead to strong modifications of the radiation spectra of multi-GeV 
projectiles channeling in bent and periodically bent crystals
\cite{Bezchastnov_AK_AS:JPB_v47_195401_2014,Sushko-EtAl:NTV_v1_p341_2015,Sushko-EtAl:NIMB_v355_p39_2015}.

The calculated spectral intensity can be normalized to the Bethe-Heitler value (see, for example, Ref.~\cite{Tsai:RMP_v46_p815_1974}) and, 
thus, can be presented in the form of an enhancement factor over the bremsstrahlung spectrum in the corresponding amorphous medium.

\begin{figure} [h]
\centering
\includegraphics[scale=0.5,clip]{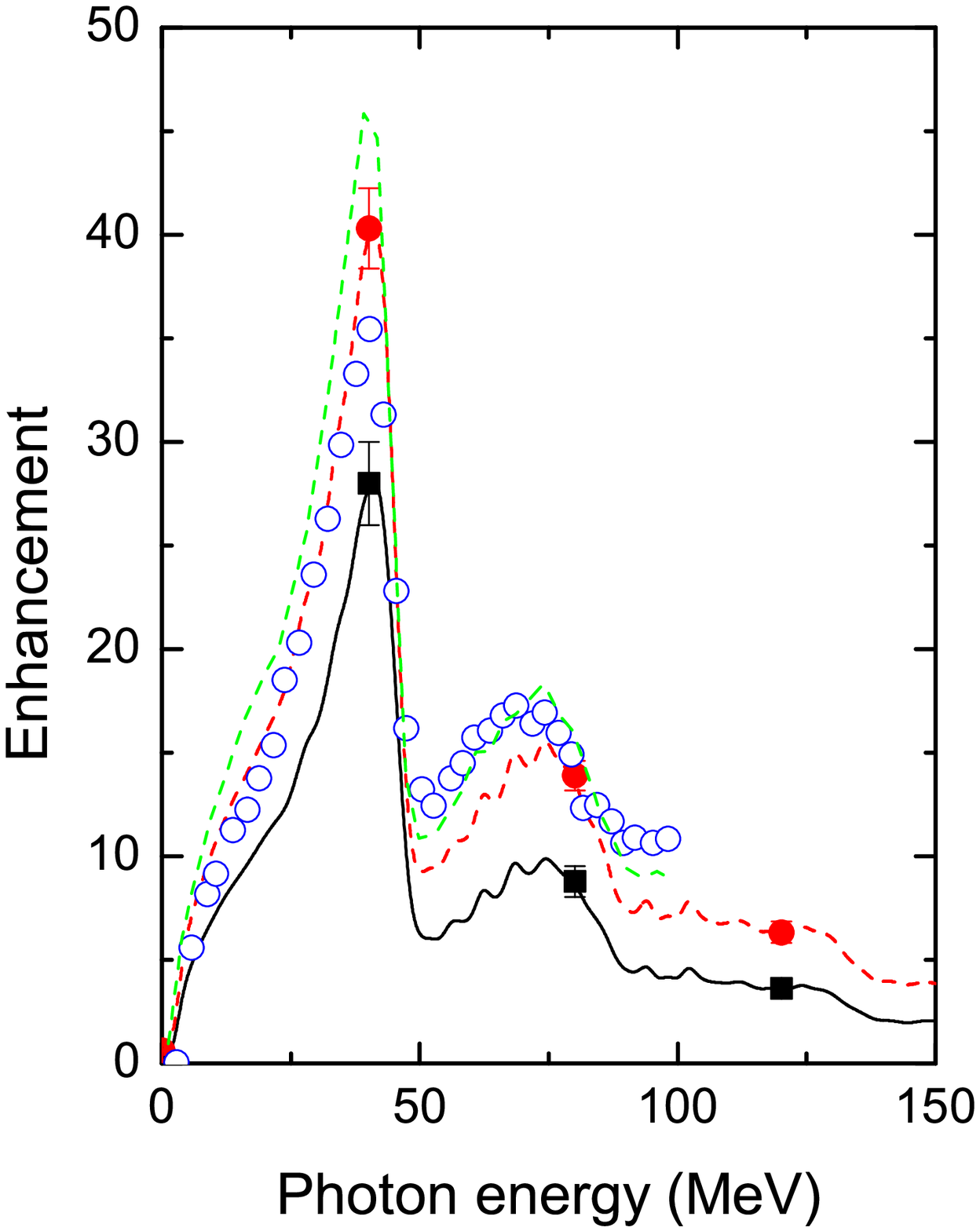}
\includegraphics[scale=0.455,clip]{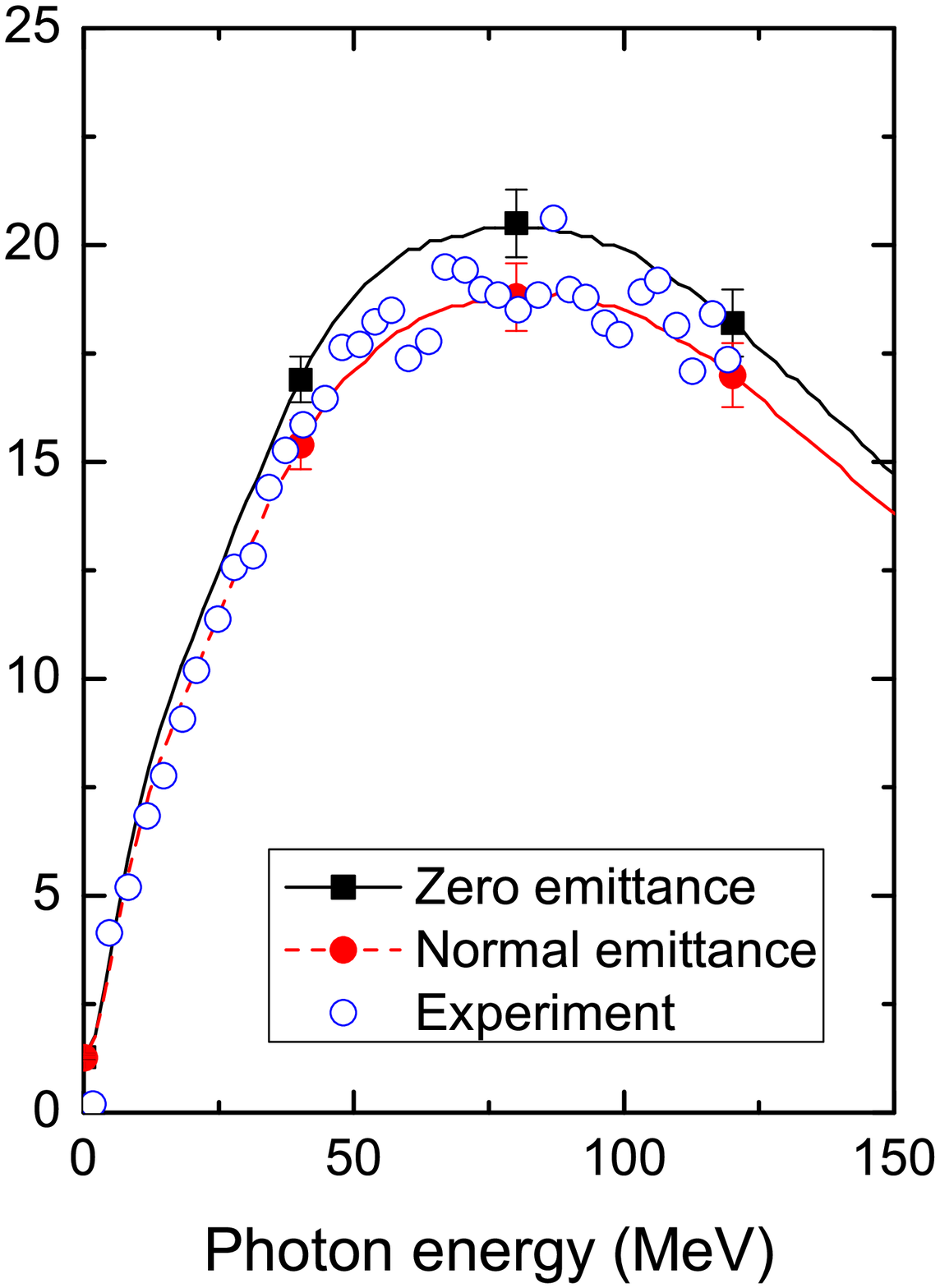}
\caption{
Enhancement factor of the channeling radiation over the Bethe-Heitler spectrum
for 6.7 GeV positrons (left) and electrons (right) in straight Si(110) crystal.
%
\textcolor{black}{
The experimental data, shown with open circles, are taken from Ref. \cite{Bak-EtAl:NuclPhysB_v251_p254_1985}
where the results of earlier experiments at CERN were quoted \cite{Atkinson-EtAl:PhysLettB_v110_p162_1982}.
}
The calculations performed with \MBNExplorer \cite{MBN_ChannelingPaper_2013,Sushko:Thesis_2015}
are shown with black solid curves, which present the results obtained for fully collimated beams (zero emittance),
and red dashed curves, which correspond to the emittance of $62$ $\mu$rad as in the experiment.
The symbols (closed circles and rectangles) mark a small fraction of the points
and are drawn to illustrate typical statistical errors (due to a finite number of the trajectories simulated) in different parts of the spectrum.
Green dashed curve, shown on the left figure, corresponds to the results presented in Ref. 
\cite{Tikhomirov:arXiv-1502.06588-2015}.
The data refer to the emission cone $\theta_0=0.4$ mrad. 
}
\label{ep-6_7GeV.fig}
\end{figure}

Figure \ref{ep-6_7GeV.fig} presents results of an exemplary case study of the emission spectra from 6.7 GeV positrons (left) and electrons (right)
channeled in $L=105$ $\mu$m thick oriented straight Si(110) crystal.
The spectra were computed for the emission cone $\theta_0= 0.4$ mrad in accordance with the experimental setup \cite{Bak-EtAl:NuclPhysB_v251_p254_1985}. 
This value exceeds the natural emission cone $\gamma^{-1}$ by a factor of about five. 
Therefore, the calculated curves account for nearly all radiation emitted.
Solid black and dashed red curves present the results of two sets of calculations carried out by means of \MBNExplorer.
The first set corresponds to the case of zero beam emittance, when the velocities of all projectiles at the crystal entrance are
tangent to Si(110) plane,  i.e. the incident angle $\psi$ is zero \cite{MBN_ChannelingPaper_2013}. 
The second set of trajectories was simulated allowing for the uniform distribution of the incident angle 
within the interval $\psi=[-\theta_{\rm L}, \theta_{\rm L}]$ with $\theta_{\rm L} = 62$ $\mu$rad being Lindhard’s critical angle. 
The calculated enhancement factors are compared  with the experimental results presented in Ref. \cite{Bak-EtAl:NuclPhysB_v251_p254_1985}
and the results of numerical simulations for positrons from Ref. \cite{Tikhomirov:arXiv-1502.06588-2015}.

Figure~\ref{ep-6_7GeV.fig} demonstrates that the simulated curves reproduce rather well the shape of the
spectra and, in the case of the positron channeling, the positions of the main and
the secondary peaks.
With respect to the absolute values both calculated spectra,  $\psi=0$ and $|\psi|\leq \psi_{\rm L}$, exhibit some deviations from the
experimental results.
For positrons, the curve with $\psi=0$ perfectly matches the experimental data in vicinity of the main peak but underestimates the measured yield of 
the higher (the second) harmonic.
Increase in the incident angle results in some overestimation of the main maximum but improves the agreement above $\hbar\om=40$~MeV.
For electrons, the $\psi=0$ curve exceeds the measured values, however, the increase in $\psi$ leads to a very good agreement if one takes into
account the statistical errors of the calculated dependence (indicated by symbols with error bars).

The aforementioned deviations can be due to several reasons.

\begin{figure} [h]
\centering
\includegraphics[clip,width=7.cm]{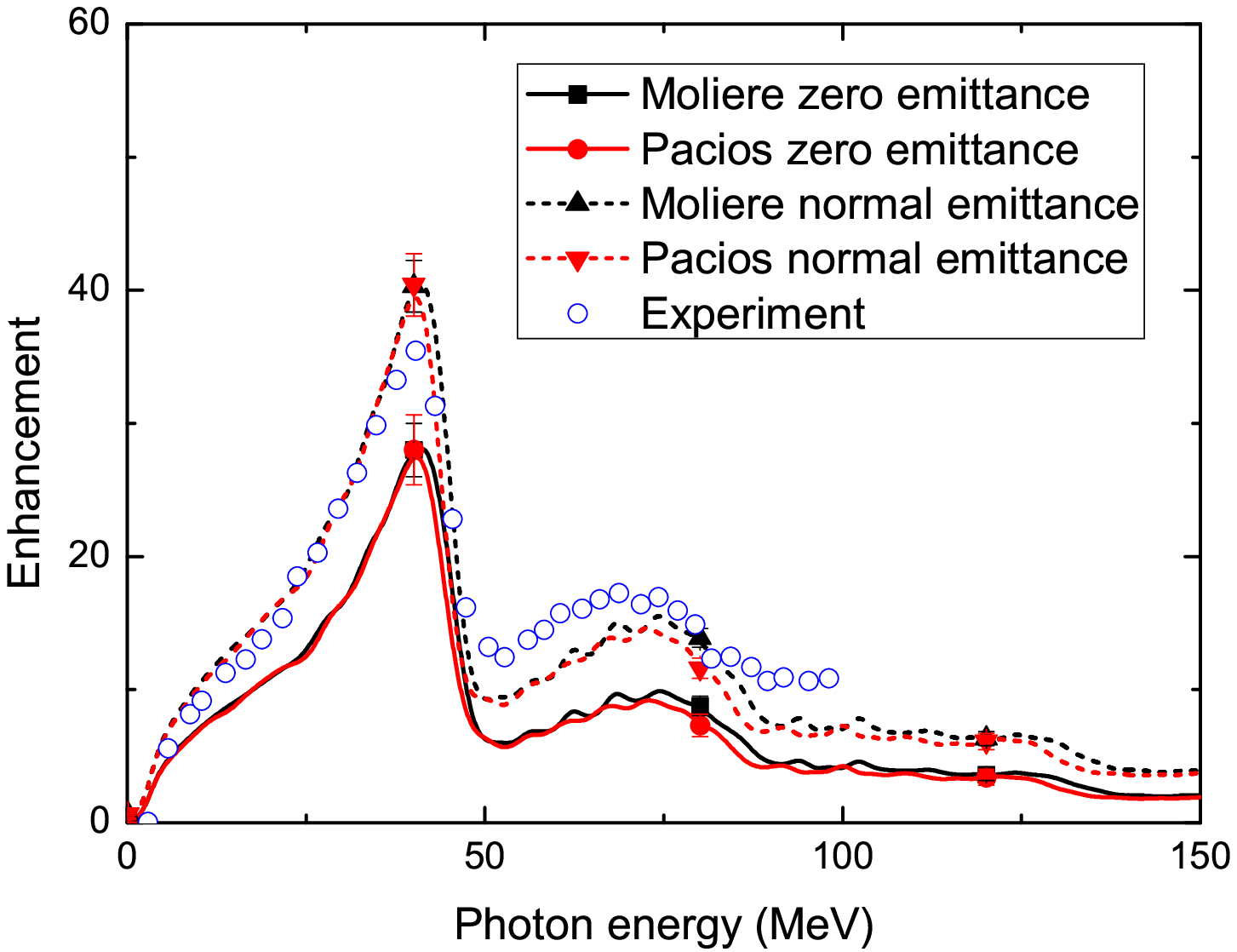}
\hspace*{0.5cm}
\includegraphics[clip,width=7.cm]{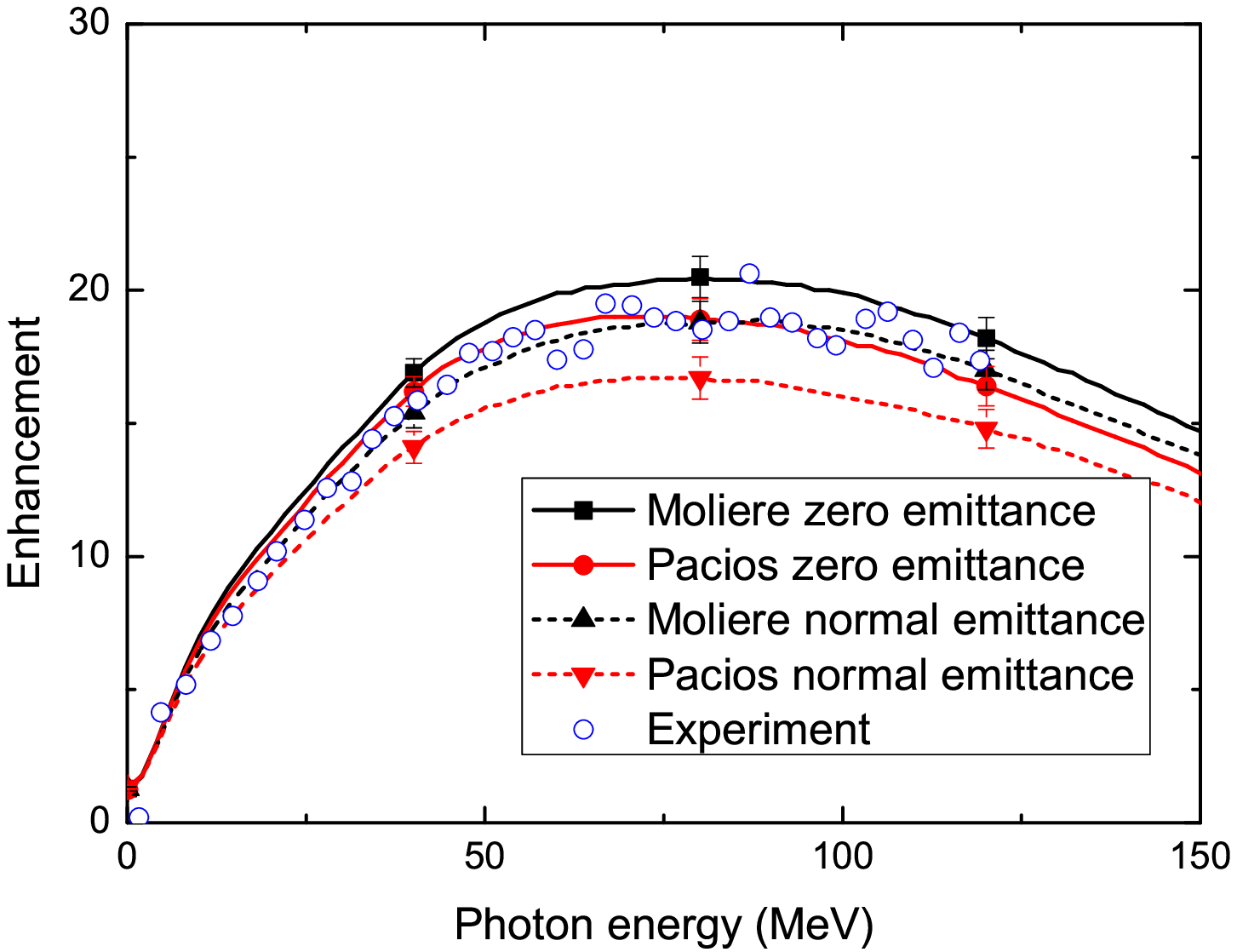}
\caption{
Simulated (lines) versus experimental (open circles, Refs. 
\textcolor{black}{\cite{Bak-EtAl:NuclPhysB_v251_p254_1985,Atkinson-EtAl:PhysLettB_v110_p162_1982}})
photon emission spectra for 6.7 GeV positrons (left) and electrons (right)
is straight Si(110) crystal.
The simulated spectra were obtained using 
(i) the  Moli\`{e}re approximation (black curves) and 
(ii) the Pacios approximation (red curves) for atomic potentials. 
The results for zero emittance of the beam are shown with solid lines, the results for beam emittance 
$\psi = \theta_L = 62$ $\mu$rad are shown with dashed lines.
Ref. \cite{Sushko:Thesis_2015}.
}
\label{ep-6_7GeV-Moliere-vs-Pacios.fig}
\end{figure}

First, the emission spectra can be sensitive to the choice of the approximation scheme used to describe the atomic potentials when constructing
the crystalline field as a superposition of the atomic fields, Eq. (\ref{MC_Simulations.02}).
The results presented in Fig.~\ref{ep-6_7GeV.fig} were obtained for the trajectories simulated within the Moli\`{e}re approximation framework.
Though this approximation is a well-established and efficient approach, more  realistic schemes for the crystalline fields, based, for example,
on X-ray scattering factors~\cite{DoyleTurner1968,ChouffaniUberall1999} or on accurate numerical approaches for 
calculation of the electron density in many-electron atoms \cite{Pacios1993}, can also be employed for the channeling simulations. 
Figure \ref{ep-6_7GeV-Moliere-vs-Pacios.fig} compares the experimentally measured spectra with those simulated numerically using the 
Moli\`{e}re and the Pacios approximations for atomic potentials \cite{Sushko:Thesis_2015}.
For positrons, both approximation result in virtually the same dependences. 
In the case of electrons, the spectra obtained with the Pacios potential are 5-10 per cent less intensive. 
Within the statistical errors both results are in a good
agreement with the experimental measurements.

Another source of the discrepancies can be attributed to some uncertainties in the experimental set-up described in 
\cite{Bak-EtAl:NuclPhysB_v251_p254_1985,Uggerhoj:RadEffDefSol_v25_p3_1993}.
In particular, it was indicated that the incident angles were in the interval
$[-\psi_{\rm L}, \psi_{\rm L}]$ with the value $\psi_{\rm L}=62$~$\mu$rad
for a 6.7~GeV projectile.
However, no clear details were provided on the beam emittance which becomes an
important factor for comparing theory vs experiment.
In the calculations a uniform distribution of the particles within
the indicated interval of $\psi$ was used, and this is also a source of the uncertainties.
The spectra was also simulated for larger cutoff angle equal to $2\psi_{\rm L}$
(these curves are not presented in the figure).
It resulted in a considerable ($\approx 30$ \%) decrease of
the positron spectrum in the vicinity of the first harmonic peak.

On the basis of the comparison with the experimental data it can be concluded that the code produces reliable results and can be 
further used to simulate the propagation of ultra-relativistic projectiles along with the emitted radiation.
In the Paper below we present several case studies of the channeling phenomena and radiation emission from ultra-relativistic projectiles 
traveling in various crystalline environments, incl. linear, bent and periodically bent 
crystals as well as in crystals stacks.
In most cases, the parameters used in the simulations, such as crystal orientation and thickness, the bending radii $R$, periods $\lamu$ and amplitudes $a$,
as well as the energies of the projectiles, have been chosen to match those used in past and ongoing experiments. 
Wherever available we compare results of our simulations with available experimental data and/or those obtained by means 
of other numerical calculations.

\section{\textcolor{black}{Case studies} \label{CaseStudies}}

In this section we present several case studies to modeling of ultra-relativistic projectiles 
(electrons, positron and negative pions) channeling and radiation emission by means of \MBNExplorer.
The case studies refer to the channeling in linear, bent and periodically bent oriented crystals. 
Wherever available the results of numerical simulations are compared with the experimental data or/and the results of 
calculations carried out by accompanying propagation of ultra-relativistic projectiles.

\subsection{\textcolor{black}{Channeling and radiation emission spectra in linear crystals} \label{LC}}

A case study presented in this section concerns comparative analysis of the channeling 
properties and spectral distribution of Channeling Radiation (ChR) of electrons and positrons in oriented
graphite(002), diamond(110), silicon(110) and tungsten(110) $L=1$ mm thick linear crystals.

The choice of the crystalline targets is explained as follows.
A silicon crystal, due to its availability and high degree of purity of a crystalline structure, has been extensively used in 
the experiments carried out at various
accelerator facilities starting from early days of the activity in the field up to nowadays, see e.g. 
\cite{RelCha,BiryukovChesnokovKotovBook,BackeLauth:NIMB-v355-p24-2015,Wistisen_EtAl_PR-AB_v19_071001_2016,ScandaleEtAl-NIMB-v446-p15-2019,
Takabayashi-EtAl:PLB_v785_p347_2018,Backe_EtAl:JINST_v13_C04022_2018,Mazzolari-EtAl:EPJC_v80_p63_2020}.
Diamond crystals (natural or/and synthesized \cite{ThuNhiTranThi_JApplCryst_v50_p561_2017}) has also been used in channeling experiments
(e.g. \cite{Sellschop-EtAl:PRL_v72_p2411_1994,Boshoff-EtAl_SAIP2016,Brau_EtAl-SynchrRadNews_v25_p20_2012,Backe_EtAl:JINST_v13_C04022_2018}).
The use of a diamond crystal is preferential in the experiments with high-intensity particle beams (such as the FACET beam at SLAC  \cite{FACET})
since it bears no visible influence from being irradiated \cite{UggerhojRPM}.

The technologies, available currently for preparing periodically bent crystals, do not immediately allow for lowering the values of bending period down to 
tens of microns range or even smaller keeping, simultaneously, the bending amplitude in the range of several angstroms. 
These ranges of $a$ and $\lambda_{\rm u}$ are most favourable to achieve high intensity of radiation in a CU \cite{ChannelingBook2014}.
One of the potential options to lower the bending period is related to using crystals heavier than diamond and silicon to propagate ultra-relativistic 
electrons and positrons.
In heavier crystals, both the depth, $\Delta U\propto Z^{2/3}$, of the 
interplanar potential and its the maximum gradient, $U^{\prime}_{\max}\propto Z^{2/3}$, 
attain larger values, resulting in the enhancement of the critical channeling angle and reduction of the critical radius 
$R_{\rm c} \propto 1/U^{\prime}_{\max}$ \cite{Tsyganov1976}.  
From this end, the tungsten crystal ($Z=74$) is a good candidate for the study
since the critical radius for W(110) is 0.16 cm which is seven times smaller than that in Si(110).
This allows one, at least in theory, to consider periodic bending with 
$\lambda_{\rm u}\lesssim 10$ $\mu$m.
This crystal was used in channeling experiments with both heavy \cite{Kovalenko_EtAl:JINR_RC-v72-p9-1995,BiryukovChesnokovKotovBook} 
and light \cite{Backe_EtAl:SPIE-6634-2007,Yoshida_EtAl:PRL_v80_p1437_1998} ultra-relativistic projectiles. 
In Ref. \cite{Kovalenko_EtAl:JINR_RC-v72-p9-1995} it was noted that the straight tungsten crystals show high structure perfection. 
This feature is also of a great importance for successful experimental realization of the CU idea \cite{Imperfectness2008}. 

Finally, we have simulated electron and positron channeling in graphite (002).
The channel is very wide (the interplanar distance $d=3.348$ \AA{}) and this, together with low electron density inside the channel, provide
excellent conditions for projectile to channel over very large distances.
Although it is known that crystalline structure of graphite is spoiled due to high degree of mosaicity \cite{BrianRand:Graphite} the channeling experiments
have been carried out with this material (see discussion in Ch. 4.7 in Ref. \cite{DresselhausKalis-Springer-2011}).
 
Figure \ref{Graphite-C-Si-Ge-W110_PlanarPot.fig} in Section \ref{InterPlanarPot} compares the continuous planar potentials for electrons and positrons 
in the four channels indicated. 
To be note is that cases of W(110) and graphite(002) can be considered as two extreme cases: of a very deep potential well ($130$ eV) for the former
and of a very wide one for the latter.

\MBNExplorer was used to simulate trajectories of 10 GeV projectiles and to compute the corresponding emission spectra.
For each crystal, the number of simulated trajectories was $N_0\approx 4000$.
The choice of the incident beam energies was motivated by the planned (though not carried out) channeling experiments with  4\dots20 GeV electron and positron beams
at the SLAC facility (USA).

\begin{figure} [h]
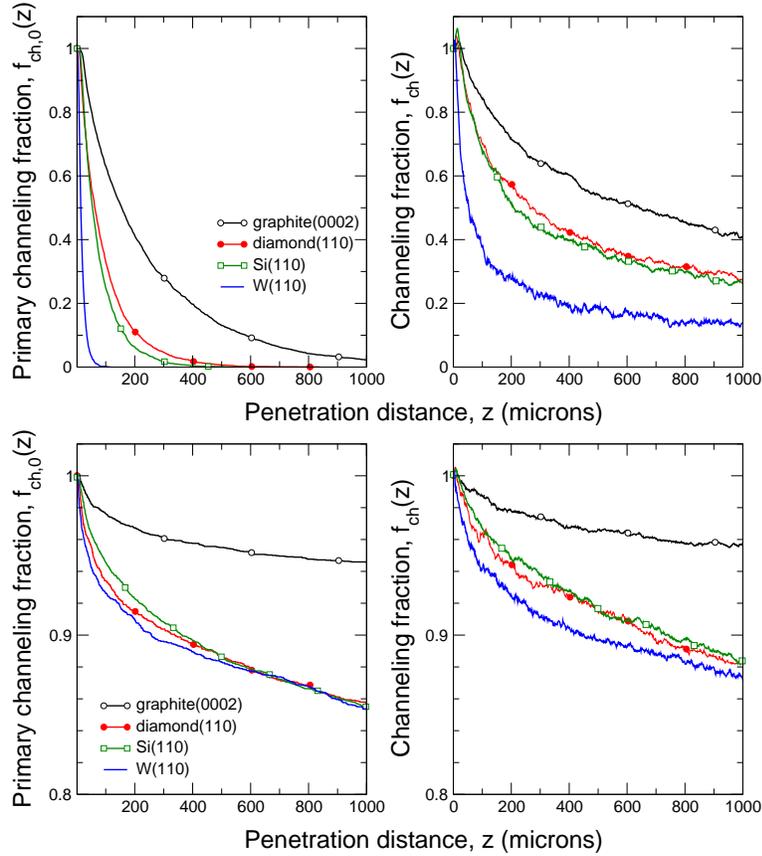

\centering
\includegraphics[scale=0.36,clip]{Figure_10a.eps}\\
\includegraphics[scale=0.36,clip]{Figure_10b.eps}
\caption{
Channeling fractions $f_{\rm ch, 0}(z)=N_{\mathrm ch, 0}(z)/N_{\mathrm{acc}}$ (left graphs)
and 
$f_{\rm ch}(z)=N_{\mathrm ch}(z)/N_{\mathrm{acc}}$ (right graphs)
of 10 GeV electrons (upper row) and positrons (lower row) 
versus the penetration distance $z$ in different oriented crystals as indicated.
}
\label{ep-10GeV-Graphite-C-Si-W-Nch0-Nch.fig}
\end{figure}
Figure \ref{ep-10GeV-Graphite-C-Si-W-Nch0-Nch.fig} shows the computed channeling fractions as functions of the penetration distance $z$ 
measured along the incident beam.
Let us point out that for both negatively and positively charged projectiles the channeling fractions calculated for graphite(002) 
are notably higher than for other channels presented.
The values of the acceptance factor, Eq. (\ref{Acceptance}), penetration length $\Lp$, Eq. (\ref{Eq.Lp}), and total channeling length $\Lch$,
calculated from statistical analysis of the trajectories, are listed in Table  \ref{ep-10GeV-Graphite-C-Si-W_acceptance_and_lengths}.
The quantity $L_{\rm d, est}$, also indicated in the table, stands for the estimated value of the dechanneling length $\Ld$ calculated by approximating 
the fraction $f_{\rm ch, 0}(z)$ of accepted projectiles with $\exp(-z/\Ld)$ at the penetration distances well away from the entrance
point.

\Table{\label{ep-10GeV-Graphite-C-Si-W_acceptance_and_lengths}
Values of the acceptance $\calA$, penetration length $L_{\rm p}$, total length of the channeling segments $L_{\rm ch}$
calculated for $\E = 10$ GeV electrons and positrons
in $L=1000$ $\mu$m thick  Graphite(002), C(110), Si(110) and  W(110).
$L_{\rm d, est}$ indicates the estimated values of the dechanneling length. 
}
 \br
             &  \multicolumn{4}{c}{Electrons}                    &  \multicolumn{3}{c}{Positrons} \\
Crystal      &$\calA$&$L_{\rm p}$ &$L_{\rm d, est}$&$L_{\rm ch}$ &$\calA$ &$L_{\rm p}$   &$L_{\rm d, est}$ \\
             &       & ($\mu$m)   & ($\mu$m)       &  ($\mu$m)   &        & ($\mu$m)     & ($\mu$m)        \\
\mr
Graphite(002)&  0.84 & 239$\pm$14 & 1090           & 508$\pm$16  &   0.99 &  993$\pm$9   & 13$\times10^3$  \\
C(110)       &  0.77 &  94$\pm$6  & 210            & 359$\pm$14  &   0.97 &  925$\pm$18  & 4970            \\
Si(110)      &  0.72 &  77$\pm$5  & 190            & 311$\pm$12  &   0.98 &  928$\pm$13  & 6060            \\
W(110)       &  0.62 &  18$\pm$1  & 43             & 144$\pm$ 6  &   0.97 &  921$\pm$18  & 4620            \\
 \br
 \end{tabular}
 \end{indented}
 \end{table}

\begin{figure} [h]
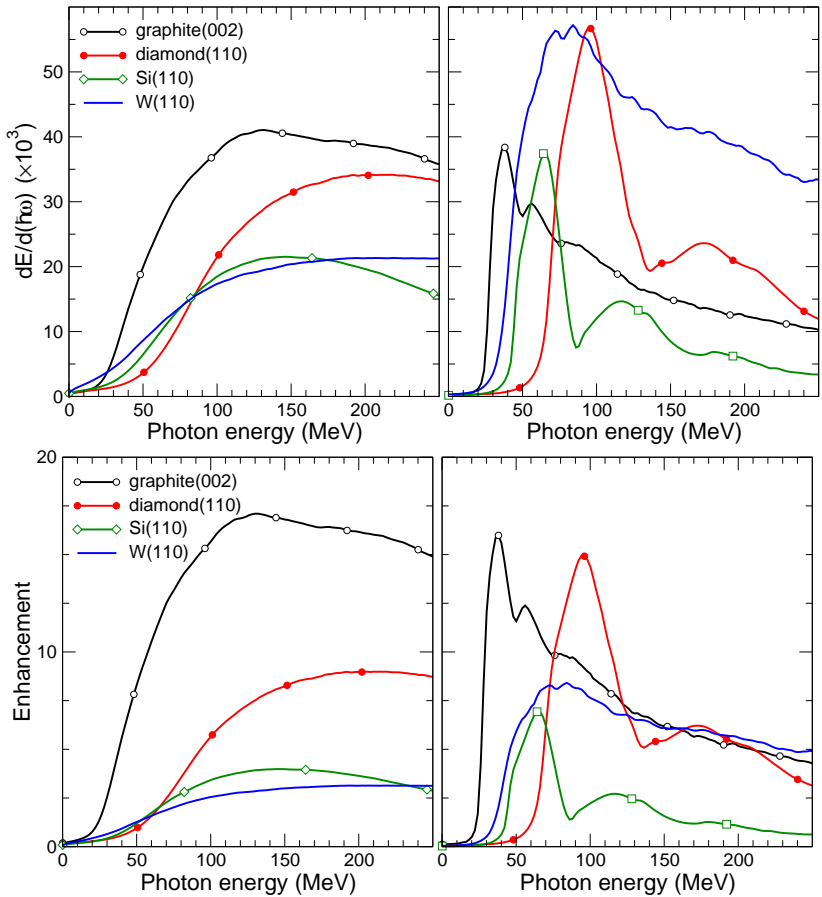

\centering
\includegraphics[scale=0.4,clip]{Figure_11a.eps}\\
\includegraphics[scale=0.4,clip]{Figure_11b.eps}
\caption{
Spectral distribution of channeling radiation (upper row) and enhancement factor over the emission in the amorphous medium 
(lower row) calculated 10 GeV electrons (left column) and positrons (right column)
channeled in different oriented crystals as indicated.
The data refer to the emission angle $\theta_0=1/\gamma\approx0.05$ mrad. 
}
\label{ep-10GeV-Graphite-C-Si-W_dE-Enhanc.fig}
\end{figure}

Spectral distributions of the emitted radiation are presented in the upper row in Fig. \ref{ep-10GeV-Graphite-C-Si-W_dE-Enhanc.fig}. 
For the sake of convenience, in the lower row of the figure we plot the spectral distribution of the enhancement factors over the Bether-Heitler spectra in 
corresponding amorphous media.


\subsection{Channeling of negative pions in bent Si(110) \label{Pion}}

\begin{figure}
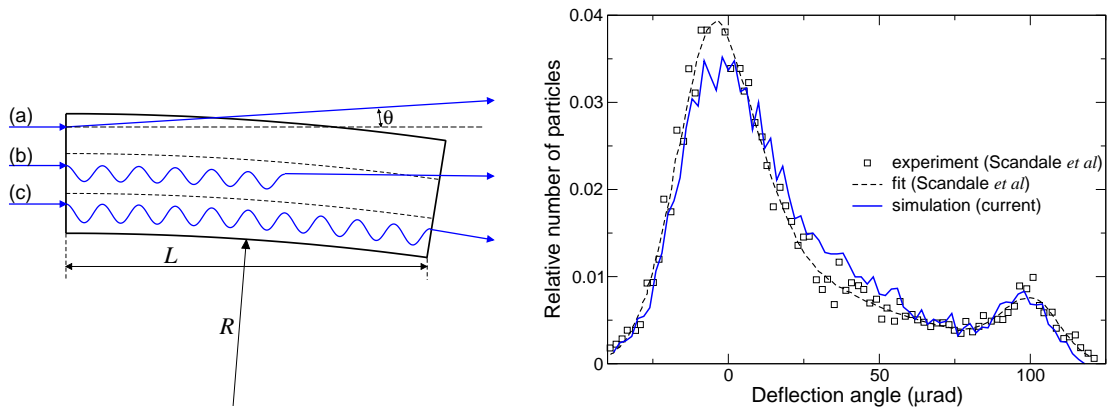

\centering
\includegraphics[width=6.5cm,clip]{Figure_12a.eps}
\hspace*{0.5cm}
\includegraphics[width=7.5cm,clip]{Figure_12b.eps}
\caption{
\textit{Left.} 
Schematic picture of particles deflection by a bent crystal. 
Thick (blue) solid curves illustrate trajectories of the particles which dechannel at the
crystal entrance, (a), dechannel somewhere inside the crystal, (b), propagate through the
whole crystal, (c).
The distribution of particles with respect to the deflection angle $\theta$ can be
measured  in experiment.
\textit{Right.} Distribution of 150 GeV pions in bent $L=1.91$ mm oriented Si(110) 
crystal with respect to $\theta$.
Symbols and dashed curve stand for the experimental data and its fit, correspondingly
\cite{Scandale_etal:PL_B719_p70_2013}.
Solid curve - the current simulation.
}
\label{pion-experiment.fig}
\end{figure}

In Ref. \cite{Scandale_etal:PL_B719_p70_2013} the dechanneling phenomenon of 
$\E=150$ GeV negative pions $\pi^{-}$ in Si(110) oriented crystal has been experimentally
investigated.
In the experiment, the beam of pions was deflected due to channeling effect
in the bent crystal, see illustrative left panel of Figure \ref{pion-experiment.fig}.
The crystal thickness in the beam direction was $L=1.91$ mm, the bending radius 
$R=19.2$ m.
The distribution of the beam particles with respect to the deflection angle $\theta$ 
was measured, see open squares in Fig. \ref{pion-experiment.fig} \textit{right}. 
The dashed curve shows the fit to the experimental data \cite{Scandale_etal:PL_B719_p70_2013}.
The fit was obtained as follows. 
Two peaks in the distribution, the one at $\theta=0$ corresponding to the particles scattered at the 
crystal entrance and another one at $\theta=L/R\approx 100$ $\mu$rad due to the particles channeled through 
the whole crystal, were approximated with two gaussians. 
The data in between two peaks was fitted with the exponential decay function which provided
a measure of the dechanneling length $\Ld$.

Thick blue curve in Fig. \ref{pion-experiment.fig} \textit{right} 
shows the angular distribution derived from the analysis of ca 12$\times 10^3$ trajectories 
simulated with the \MBNExplorer software.
Our result corresponds well to the experimental data. 
Some discrepancy seen can be attributed to 
(i) comparatively low statistics in both the simulation and
the experiment (total number of counts is about 5000, as one can calculate using the data 
presented in Fig. 3 in \cite{Scandale_etal:PL_B719_p70_2013}),
(ii) uncertainly in the beam direction along the (110) plane. 
Our recent study shows the sensitivity of the angular distribution to this parameter
\cite{HaurylavetsEtAl-in-preparation-2020}.

The bending radius quoted above is much smaller than the critical radius $R_{\rm c}$
\cite{Tsyganov1976}.
The latter can be estimated as $R_{\rm c}=\E/\dUmax\approx 0.26$ m where 
$\dUmax=5.7$ GeV/cm is the strength of the interplanar field in Si(110) calculated 
within the Moli\`{e}re approximation at $T=300$ K \cite{BiryukovChesnokovKotovBook,ChannelingBook2014}.
Thus, the bending parameter is small, $C=R_{\rm c}/R=0.013$.
As indicated in Ref. \cite{Scandale_etal:PL_B719_p70_2013}
"A slightly bent crystal ($R\gg R_{\rm c}$ \dots) is the ideal choice to perform 
 measurements of the dechanneling length". 

In what follows we demonstrate that the value $\Ld=0.92\pm 0.05$ reported in 
\cite{Scandale_etal:PL_B719_p70_2013}
overestimates the dechanneling length defined as the factor in the exponential decay, 
$\exp(-z/\Ld)$, of primarily channeled particles. 
In an experiment, the re-channeled particles cannot be separated from the primarily channeled
ones. 
As a result, the dechanneling rate deduced from the experimental data is intrinsically increased
due to the account for the re-channeled 
fraction.\footnote{\textcolor{black}{We note here that that the rechanneling
contribution in the deflection efficiency of negative particles has been accounted for 
in an earlier experiment with electrons, Ref. \cite{Mazzolari_etal:PRL_v112_135503_2014}.}}

\begin{figure}
\centering
\includegraphics[width=10cm,clip]{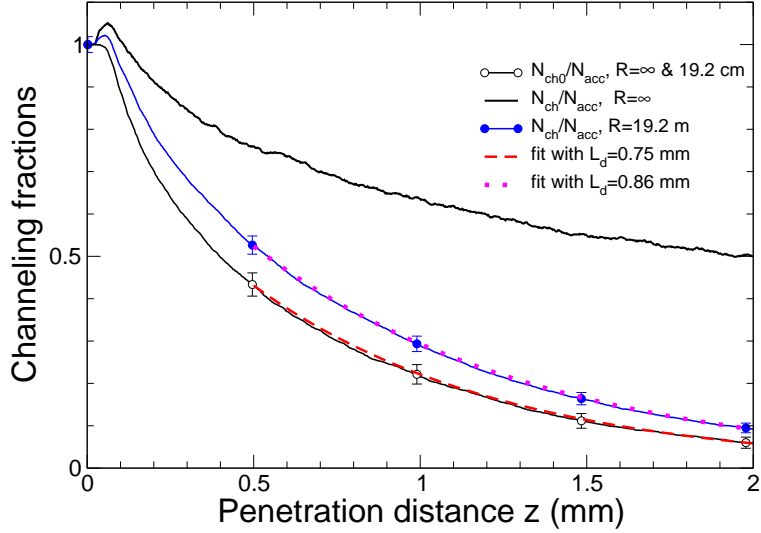}
\caption{ 
Fractions of the channeling negative pions in straight ($R=\infty$) and bent ($R=19.2$ m) 
Si(110) as a function of the penetration distance $z$. 
The curve with open circles shows the primary fraction
$N_{\rm ch, 0}(z)/\Nacc$ which is virtually identical for both straight and bent crystals.
The curve with filled circles correspond to the 
fraction $N_{\rm ch}(z)/\Nacc$ accounting for the re-channeling in the bent crystal. 
Error bars illustrate the statistical uncertainties due to the finite number of the trajectories
simulated.
For the sake of comparison, the fraction $N_{\rm ch}(z)/\Nacc$ calculated for the straight
crystal is also presented.
Dashed and dotted curves show the exponential decay fits with the $\Ld$ values as indicated in the
legend.}
\label{pion-Nch0_Nch.fig}
\end{figure}

To quantify the decay rates of channeled negative pions with and without dechanneling,
we have analyzed the fraction of primary channeled particles, $N_{\rm ch, 0}(z)/\Nacc$, and
the fraction $N_{\rm ch}(z)/\Nacc$ which accounts for the re-channeling effect ($\Nacc$ 
stands for the number of accepted particles), see Fig. \ref{pion-Nch0_Nch.fig}.
Because of the large bending radius the primary fraction in the bent crystal coincides
(within the statistical errors) with that calculated for the straight crystal ($R=\infty$).
Therefore, the curve $N_{\rm ch, 0}(z)/\Nacc$ in the figure refers to both cases.
As it has been already mentioned, the excess due to re-channeling 
of the total number of channeling particles $N_{\rm ch}(z)$ over the 
the number the primary channeling ones $N_{\rm ch, 0}(z)$ is the largest for a straight channel
and rapidly decreases with $R$. 
This decrease is seen explicitly when one compares the curve without symbols (straight channel)
and the curve with filled circles ($R=1.91$ m).
Even so, the difference between the fractions in the bent channel 
results in the different values of the decay rates.
In the region far away from the entrance point, both curves can be approximated with $A\exp(-z/\Ld)$.
For the primary fraction this fit (see the dashed curve) produces 
$\Ld = 0.75\pm 0.02$ mm which is the dechanneling length associated with the decay rate
of $N_{\rm ch, 0}(z)$ in both the straight and the bent Si(110).
The same approximation for $N_{\rm ch}(z)/\Nacc$ (the dotted curve) results in larger values
$(0.86\pm 0.02)$ mm.
This interval overlaps with the experimentally measured one 
\cite{Scandale_etal:PL_B719_p70_2013}.

\subsection{\textcolor{black}{Simulation of emission spectra in bent crystals} \label{Sec:BC-Radiation}}

\textcolor{black}{%
The motion of a channeling particle in a bent crystal contains two components:
the channeling oscillations and circular motion along the bent centerline.
The latter motion gives rise to the synchrotron-type radiation (SR) \cite{Jackson}.  
Therefore, the total spectrum of radiation formed by a ultra-relativistic 
projectile propagating in the channeling mode in a bent crystal bears features of both the ChR and SR
\cite{KaplinVorobev1978}.
The peculiarities which appear in spectral and spectra-angular distributions
of the emitted radiation due to the interference of the two mechanisms of radiation
were analyzed in Refs. \cite{SolovyovSchaeferGreiner1996,ArutyunovEtAl_NP_1991,Bashmakov1981,%
TaratinVorobiev1988,TaratinVorobiev1989} by analytical and numerical means.
}

\begin{figure}[hb]
\centering
\includegraphics[width=7cm,clip]{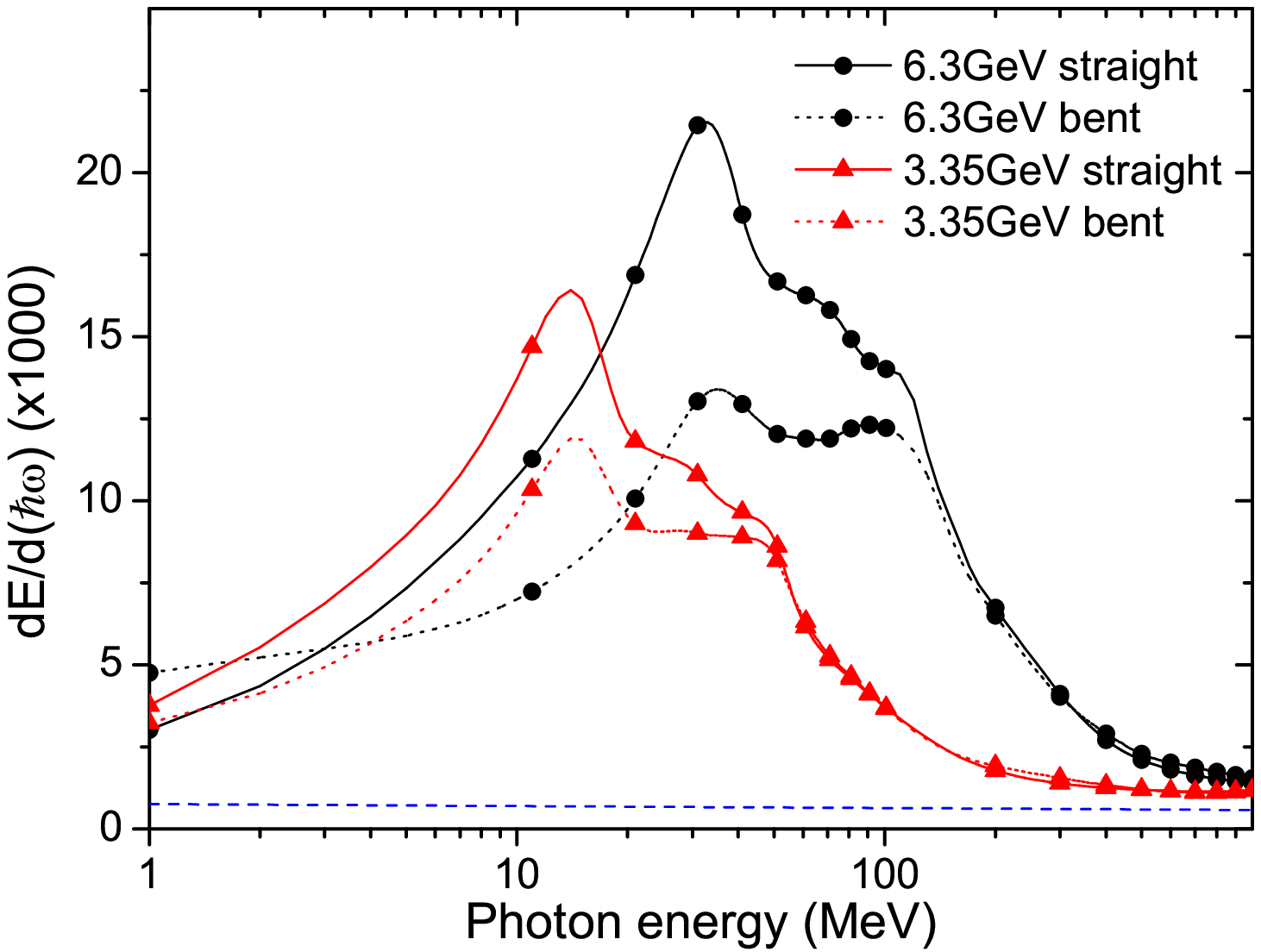}
\includegraphics[width=7cm,clip]{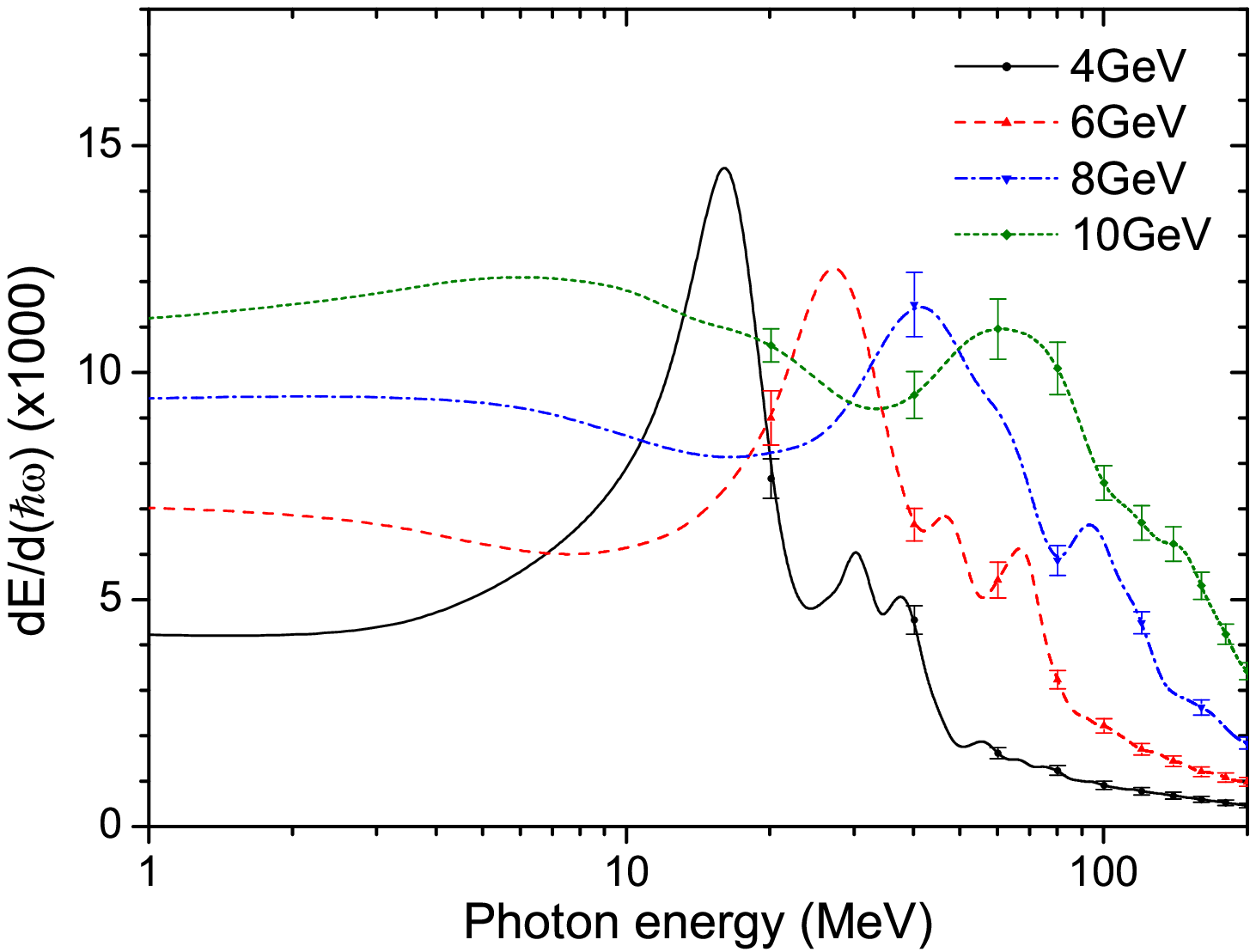}
\caption{ 
\textit{Left.} Radiation spectra by 3.35 and 6.3 GeV electrons in straight and bent 60 micron thick Si(111).
Broken line stands for the emission spectrum in amorphous silicon by the 3.35 GeV
projectile calculated within the Bethe-Heitler approximation.\\
Radiation spectra of 4\dots10 GeV positrons in bent Si(111).
Ref. \cite{Sushko-EtAl:NIMB_v355_p39_2015}.
}
\label{BC-spectra.fig}
\end{figure}

\textcolor{black}{%
A quantitative analysis of the emission spectra based on the simulation with the \MBNExplorer package
has been carried out for sub-GeV electrons and positrons channneling in 
silicon \cite{Polozkov-EtAl:EPJD_v68_268_2014}, diamond \cite{Polozkov_VKI_Sushko_AK_AS_SPB_Diamond_2015}
and tungsten \cite{Korol-EtAl:NIMB_v424_26_2018} crystals.
The emission spectra of 4\dots20 GeV projectiles in oriented bent Si(111) crystal have been presented 
in Ref. \cite{Sushko-EtAl:NIMB_v355_p39_2015} and the corresponding illustrative examples are shown in Fig. 
\ref{BC-spectra.fig}.
}

\textcolor{black}{%
In Fig. \ref{BC-spectra.fig}\textit{left}, comparing the solid curves, which show the intensity of radiation 
in the straight crystal,
with the dashed ones, corresponding to the bent crystal, one notes lowering of the channeling radiation peaks due to 
the crystal bending.
The synchrotron-type radiation, which is absent in the straight crystal, contributes to the low-energy part of the spectrum.
It is clearly seen in the spectral dependences for a  6.3 GeV electron: in the photon energy range 
$\hbar\om = 1\dots 3$ MeV the intensity in the bent crystal is higher than that in the straight one.
The contribution of the synchrotron radiation to the emission spectrum becomes more pronounced for 
more energetic projectile, as it is illustrated by the right panel of the figure.
}

\subsection{Radiation from diamond-based CU by multi-GeV  
electrons and positrons \label{Sec:Chineese2}}

In Refs. \cite{Sushko-EtAl:JPConfSer_v438_012018_2013} the results have been presented of the statistical
analysis of the channeling properties and of the spectral intensities of the radiation formed by 195...855 MeV 
electrons and positrons in the crystalline undulator with the parameters used in the experiments at the MAMI facility
\cite{Backe-EtAl:NuovCim_v34_p157_2011,Backe-EtAl:Channeling2010}.
The CUs were manufactured in Aarhus University (Denmark) using the molecular beam epitaxy technology to produce 
strained-layer Si$_{1-x}$Ge$_x$ superlattices with varying germanium content \cite{MikkelsenUggerhoj:NIMB_v160_p435_2000}.

Later, similar calculations have been extended to the range of  multi-GeV  projectile energies  
\cite{Chineese2,Sushko:Thesis_2015}.
To a great extent, this activity was inspired by the plans to carry out channeling experiments with  
diamond crystals at the SLAC facility (USA) 
\cite{FACET} using high-intensity 4...20 GeV electron and positron beams.  
The sets of simulations have been performed aiming at providing benchmark data for the emission spectra 
formed by projectile electrons and positrons 
in \textit{silicon-based} \cite{Sushko:Thesis_2015} and \textit{diamond-based} \cite{Chineese2} 
crystalline undulators with the parameters similar to those used in the experiments with sub-GeV electron beams
\cite{Backe-EtAl:NuovCim_v34_p157_2011,Backe-EtAl:Channeling2010}.

The parameters of the CU used in the simulations were as follows:
\begin{itemize}
 \item 
Bending period amplitude $\lamu=40$ microns.
 \item 
Number of periods $\Nu=8$ corresponds to the crystal thickness $L=\Nu\lamu = 320$ microns. 
 \item 
Bending period amplitude $a=2\dots6$  \AA{}.
\end{itemize}

 \begin{table}[h]
\hspace*{-1cm}
 \caption{
 Acceptance $\calA$ and penetration length $\Lp$ for 10~GeV positrons and electrons in straight ($a=0$) 
 and periodically bent ($\lamu=40$ microns)
planar channels Si(110). 
The bending parameter $C=a (2\pi/\lamu)^2 (\E/\dUmax)$ stands for the ratio of the centrifugal force
ti the interplanar force.
}
\begin{indented}%
\footnotesize\rm\item[]
 \begin{tabular}{@{}lllll}
 \br
    Projectile     & $a$ ({\AA})& $C$   & $\calA$ (\%)      & $\Lp$ ($\mu$m)  \\ 
\br
    positron       & 0           & 0    & 97.1 $\pm$ 0.9    & 302 $\pm$ 4     \\
                   & 2           & 0.08 & 89.8 $\pm$ 2.1    & 301 $\pm$ 5     \\
                   & 4           & 0.16 & 81.6 $\pm$ 2.6    & 287 $\pm$ 7     \\
                   & 6           & 0.24 & 71.9 $\pm$ 5.8    & 273 $\pm$ 15    \\ 
 \mr
    electron       & 0           & 0.0  & 65.8 $\pm$ 2.3    & 82  $\pm$ 4     \\
                   & 4           & 0.16 & 42.9 $\pm$ 3.3    & 52  $\pm$ 4     \\ 
 \br
 \end{tabular}
\end{indented}
 \label{ep-10GeV-320micron.table}
 \end{table}

Table~\ref{ep-10GeV-320micron.table} provides the values of acceptance and penetration length $\Lp$ obtained via statistical analysis of the 
trajectories simulated for 10 GeV electrons and positrons incident of straight ($a=0$) and periodically bent Si(110) crystal.
In the latter case, the bending parameter $C$ is defined as the ratio of the maximum values of the centrifugal force, 
$F_{\rm cf}\approx \E / R_{\rm min}$ with $R_{\min} = a^{-1}(\lamu/2\pi)^2$, and the interplanar force 
$\dUmax$. 
The latter was taken equal to $5.7$ GeV/c,  
which corresponds to the (110) interplanar potential calculated within the Moli\`{e}re approximation at $T=300$ K.
The data shown indicate that most of positrons travel in the channeling mode through the whole crystal. 
For electrons, both acceptance and channeling segments length are much lower.
These features reveal themselves in the emission spectra by the projectiles.

\begin{figure}[h]
\centering
\includegraphics[scale=0.26,clip]{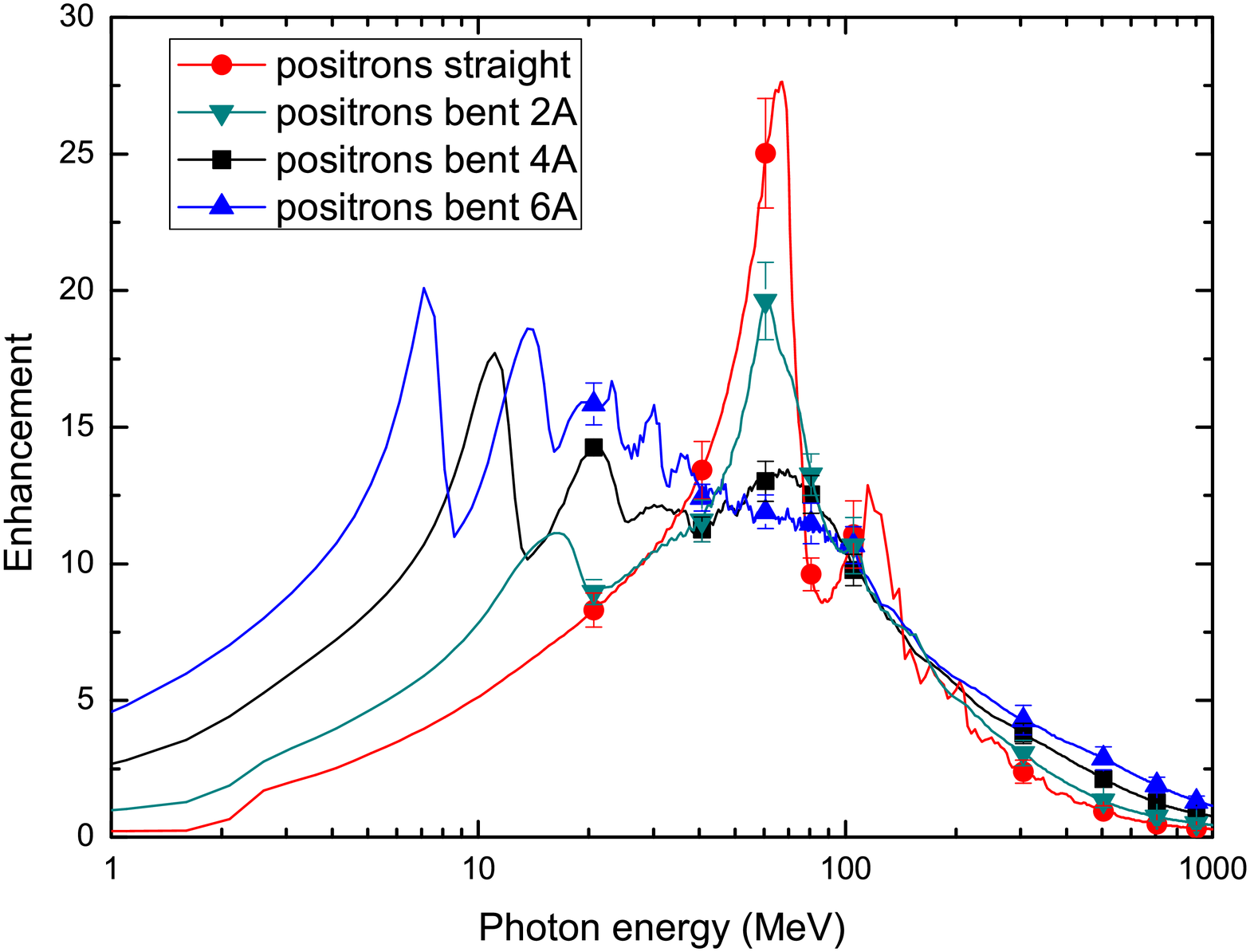}
\includegraphics[scale=0.26,clip]{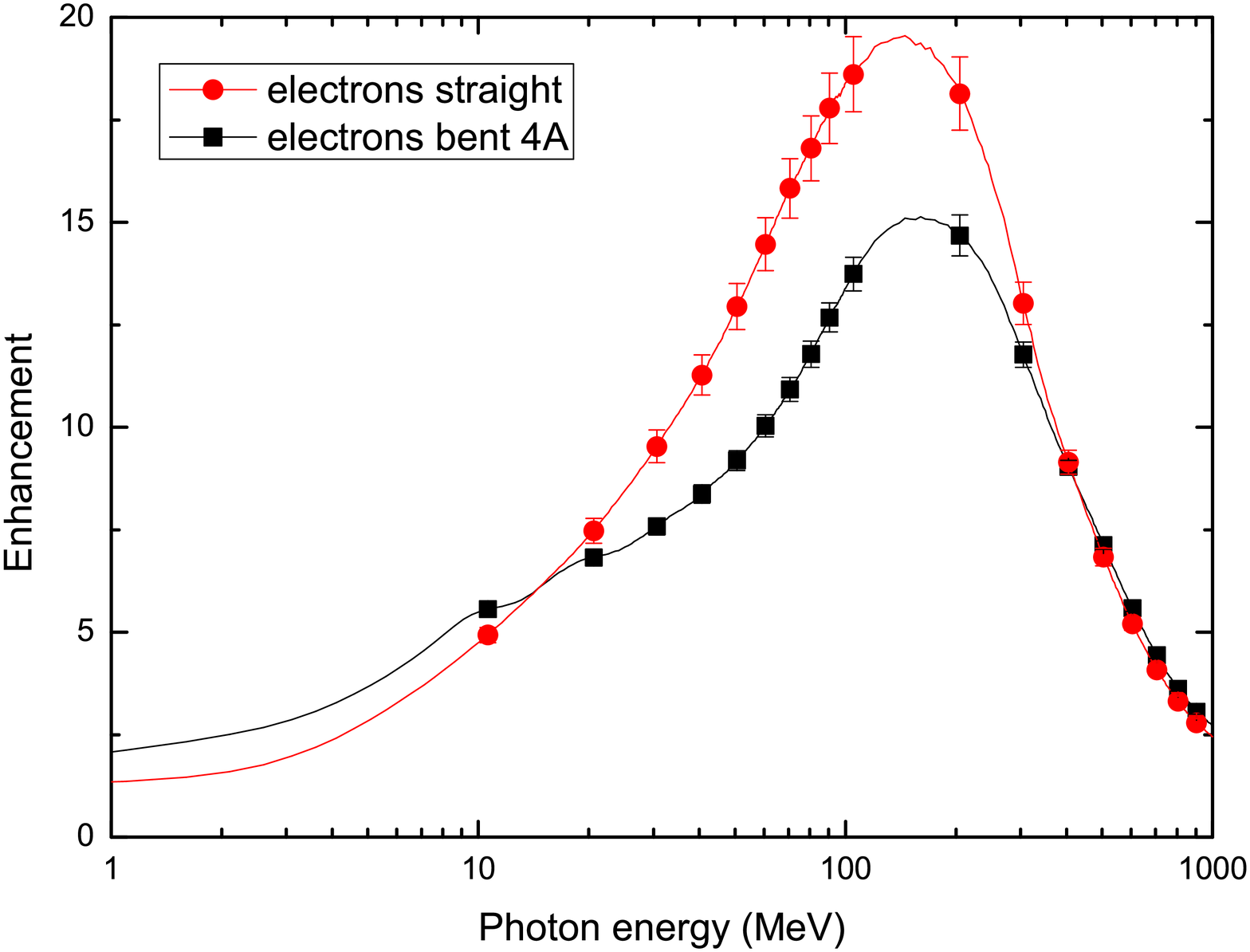}
\caption{
Enhancement factor of the radiation over the Bethe-Heitler spectrum for 10 GeV positrons (left) and electrons (right)
positrons in straight Si(110) and in Si(110)-based CU with different bending amplitudes as indicated.
The bending period is set to $\lamu=40$ microns.
All curves refer to the emission angle $\theta_{0}=7/\gamma\approx 0.36$ mrad.
}
\label{ep-10GeV-Enhance.fig}
\end{figure}

Figure~\ref{ep-10GeV-Enhance.fig} shows the enhancement factor (over the Bethe-Heitler background) of the radiation emitted 
by positrons (left panel) and electrons (right panel) \cite{Sushko:Thesis_2015}.
For both projectiles, the spectra in the straight channel (red curves) are dominated by powerful peaks due to the channeling radiation.
The peak is more pronounced for positrons since their channeling oscillations are quasi-harmonic resulting in the emission within
comparatively narrow bandwidth centered at $\hbar\om_{\rm ch}\approx 70$ MeV.
Strong anharmonicity of the electron channeling oscillation leads to the noticeable broadening of the peak with the maximum located at 
 $\hbar\om_{\rm ch}\approx 120$ MeV.
Periodical bending of the crystal planes gives rise to the CU Radiation (CUR). 
Since the dechanneling length of positrons greatly exceeds that of electrons, the CUR peaks in the positron spectra are much more pronounced.
The energy $\hbar\om_1$ of the first harmonic of CUR can be estimated from the relation
(see \cite{ChannelingBook2014}, Eq. (6.14)):
\begin{eqnarray}
\hbar\om_1\, [MeV]  = {9.5 \over 1 + K^2/2}\, {\E^2 \over \lamu}
 \label{Notations:eq.02}
\end{eqnarray}
where $\E$ is substituted in GeV and $\lamu$ in microns.
The quantity $K$ stands for the total undulator parameter due to both channeling oscillations and those due to the bending periodicity \cite{Dechan01}
\begin{eqnarray}
K = \sqrt{K_{\rm u}^2 + K_{\rm ch}^2}
 \label{Notations:eq.03}
\end{eqnarray}
 where  $\Ku=2\pi \gamma a/\lamu$ and $\Kch\propto 2\pi \gamma  a_{\rm ch}/\lamch $ with $a_{\rm ch} \leq d/2$ and $\lamch$ being the amplitude and period of 
 channeling oscillations.
In the case of positron channeling, assuming harmonicity of the oscillations one can derive the following expression for 
$K_{\rm ch}^2$ averaged over the allowed values of $a_{\rm ch}$ (see  \cite{ChannelingBook2014}, Eq. (6.14)):
\begin{eqnarray}
 \langle K_{\rm ch}^2 \rangle = { 2 \gamma U_0 \over 3mc^2}\,.
 \label{Notations:eq.05}
\end{eqnarray}
where $U_0$ is the depth of the interplanar potential well.
For $\E=10$ GeV in Si(110) ($U_0 \approx 22$ eV at $T=300$ K, see, e.g., Fig. \ref{C-Si-Ge110_PlanarPot.fig})
one obtains $\langle K_{\rm ch}^2 \rangle \approx 0.56$.

Using (\ref{Notations:eq.02})-(\ref{Notations:eq.05}) one estimates $\hbar\om_1$ for $a=2,4,6$ \AA{} as 16, 11.7 and 8 MeV, respectively.
These values correlate nicely with the positions of the first peaks of CUR seen in Fig. \ref{ep-10GeV-Enhance.fig} left.

\begin{figure}
\centering
\includegraphics[scale=0.45,clip]{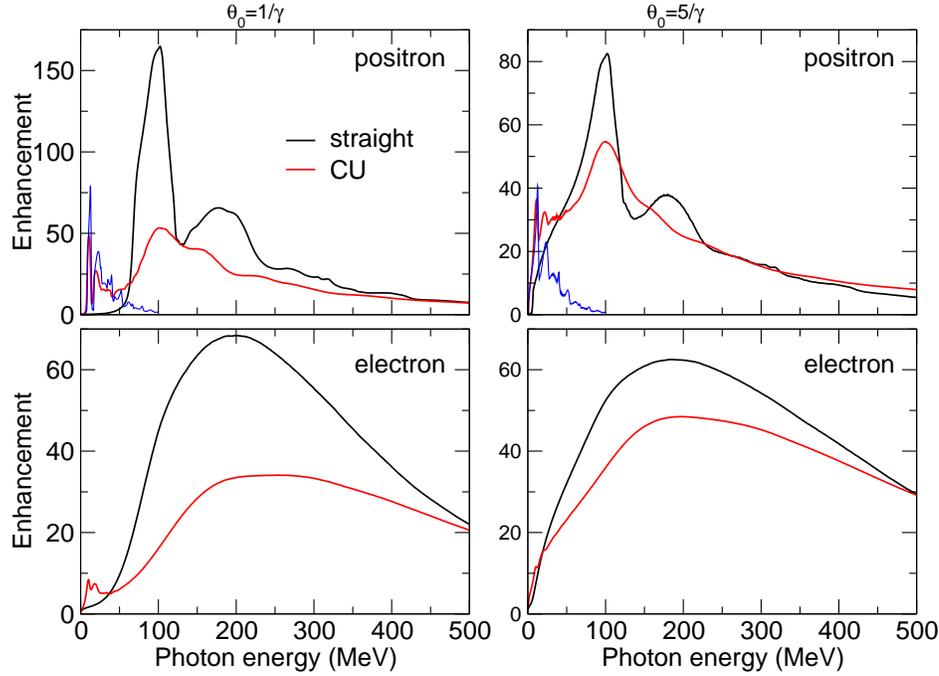}
\caption{ 
Enhancement factor of the radiation over the Bethe-Heitler spectrum for 10 GeV positrons (upper row) and electrons (lower row)
positrons in straight diamond (thick black lines) and in diamond-based CU (thick red lines) 
with  amplitude $a=4$ \AA{} and period $\lamu=40$ microns.
Thin blue solid lines show the emission spectra from ideal undulator with the same $a$ and $\lamu$.
Left column corresponds to the emission angle $\theta_{0}=1/\gamma=51.1$ $\mu$rad;
right column -- to  $\theta_{0}=5/\gamma \approx 256$ $\mu$rad. 
All spectra refer to the crystal thickness $L=320$ microns.
}
\label{ep-10GeV_1-5gamma-Enhance.fig}
\end{figure}

In Ref. \cite{Chineese2} the channeling of 4...20 GeV electron and positron beams in oriented diamond(110) crystal, both straight and periodically bent, was 
simulated and analyzed. 
As mentioned, this activity has been carried out to produce theoretical benchmarks for the experimental measurements planned to be carried out at 
the SLAC facility. 
From this viewpoint, the use of diamond crystals looked preferential since diamond bears no visible influence
from being irradiated by high-intensity beams available at SLAC.

The spectral distributions of radiated energy were computed for two values of the emission cone $\theta_0$ (see Eq. (\ref{Methodology:eq.03})):
(i) a 'narrow' cone $\theta_0=1/\gamma$, and (ii) a 'wide' cone $\theta_0=5/\gamma$, which collects virtually all radiation emitted by ultra-relativistic particles.
The results of calculations for 10 GeV projectiles, presented in the form of the enhancement factor over the emission spectra in amorphous medium, are 
shown in Fig. \ref{ep-10GeV_1-5gamma-Enhance.fig}.
For the sake of comparison the spectra formed by a positron moving in an "ideal undulator" (i.e., along the sine trajectory with the given values of 
$a$ and $\lamu$) are also shown in the upper figures.

\subsection{Interplay and specific features of radiation mechanisms for electrons 
\textcolor{black}{in crystalline undulators}\label{LALP-electrons}}

In recent papers \cite{Korol-EtAl:EPJD_v71_174_2017,Pavlov-EtAl:JPB_v52_11LT01_2019,Pavlov-EtAl:EPJD_2020} 
an accurate numerical analysis has been
performed of the evolution of the channeling properties and the radiation spectra for diamond(110) based CUs. 
Drastic changes in the radiation spectra with variation of the bending amplitude $a$ have been observed for 
different projectile energies and their sensitivity to the projectile’s charge has been noted. 
Some of the predictions made can be verified in channeling experiments with electrons at the MAMI facility.

The calculations were performed for 270--855 MeV electrons and positrons propagating in the 20 microns thick
diamond crystal.
The periodic bending was assumed to have a harmonic shape $S(z) = a \cos(2\pi z/\lamu)$ with the coordinate $z$ 
measured along the incident beam direction.
The bending period was fixed at $\lamu=5$ microns whereas the bending amplitude was varied from 
$a$ = 0 (straight crystal) up to $a = 4.0$ \AA{} in accordance with the parameters of crystalline samples used in the 
experiments at MAMI \cite{BackeLauth-Dyson2016}. 
%
\begin{figure}
\centering
\includegraphics[scale=0.55,clip]{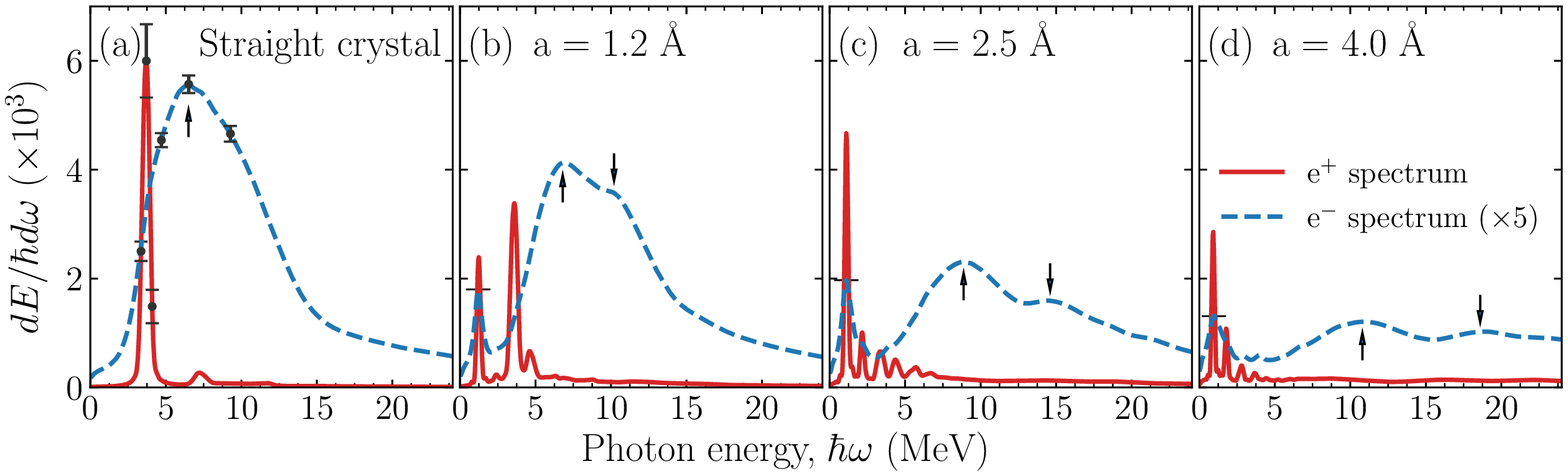}
\caption{ 
Spectral distributions of radiation by 855 MeV 
electrons (dashed blue curves, multiplied by a factor of 5) and positrons (solid red curves) 
in straight (a) and periodically bent (b)–(d) diamond (110) crystals. 
The upward arrows indicate the maxima of ChR for electrons, the downward arrows show the positions
of the additional maxima appearing in the bent crystals (see explanations in the text). 
The error bars shown in graph (a) illustrate the statistical errors due to the finite number of the
simulated trajectories.
The spectra correspond to the opening angle $\theta_0 = 0.24$ mrad. 
Figure from Ref. \cite{Pavlov-EtAl:JPB_v52_11LT01_2019}.
}
\label{Pavlov-Figure01.fig}
\end{figure}

Figure \ref{Pavlov-Figure01.fig} presents the emission spectra for the positrons and electrons with $\E = 855$ MeV 
calculated for the opening angle $\theta_0 = 0.24$ mrad, which is smaller than the natural
emission angle $\gamma^{-1}=0.59$ mrad.
For both types of projectiles the spectra formed in the straight crystal, graph (a), 
are dominated by the peaks of ChR, the spectral intensity of which by far exceeds that of the incoherent bremsstrahlung 
background  $2.5\times10^{-5}$ in the amorphous medium.
For positrons, nearly perfect harmonic channeling oscillations 
give rise to the narrow peak at $\hbar \om_{\rm ChR} \approx 3.6$ MeV.
Strong anharmonicity of the electron channeling oscillations makes the ChR peaks (marked with the
upward arrows) less pronounced and significantly broadened (note the scaling factor $\times5$ applied to the electron spectra). 

In periodically bent crystals, Figs. \ref{Pavlov-Figure01.fig}(b)-(d), the spectra exhibit 
additional features some of which evolve differently with increase in $a$.

\begin{itemize}

\item
\textcolor{black}{
For  both types of projectiles there are CUR peaks in the low-energy part of spectra.
The most powerful peaks correspond to the emission in the first harmonic at $\hbar\om_{\rm CUR}\approx 1$ MeV. 
To be noted is the non-monotonous dependence of the peak values on bending amplitude $a$.
This feature has been discussed in detail in Refs. \cite{Pavlov-EtAl:JPB_v52_11LT01_2019,Pavlov-EtAl:EPJD_2020}.
}

\item
\textcolor{black}{
For positrons, the intensity of ChR becomes strongly suppressed as bending amplitude increases:
for $a=1.2$ \AA{} the intensity is two times less than in the straight crystal.
For larger amplitudes, ChR virtually disappears \cite{Korol-EtAl:EPJD_v71_174_2017,Pavlov-EtAl:JPB_v52_11LT01_2019}.
This happens because the (mean) amplitude of channeling oscillations is a decreasing 
function of $a$. 
Indeed, as $a$ increases, the centrifugal force, especially in the points of maximum curvature, 
drives the projectiles oscillating with large amplitudes away from the channel resulting in
a strong quenching of the oscillations. 
A quantitative analysis of this feature one finds in Ref. \cite{Pavlov-EtAl:JPB_v52_11LT01_2019}.
}

\item
\textcolor{black}{
For electrons, the peak value of ChR does not fall off so dramatically. 
As $a$ increases, the peak (marked with the upward arrow) 
becomes blue-shifted and there appears additional structure (the downward arrow) on the right shoulder of the spectrum.
The analysis carried out in Ref. \cite{Pavlov-EtAl:EPJD_2020} has shown 
that both features are due to the emission by dechanneled electrons.
In a periodically bent crystal, a dechannel particle can experience 
(i) the volume reflection (VR) \cite{TaratinVorobiev:PLA_v119_p425_1987,TaratinVorobiev:NIMB_v26_p512_1987},
occurring mainly at the points of maximum curvature,
and 
(ii) the over-barrier motion in the regions with small curvature. 
These types of motion contribute to different parts of the radiation spectrum. 
The radiation, which accompanies VR, is emitted in same energy domain as the ChR. 
The over-barrier particles radiate at higher energies and this radiation reveals itself as an 
additional peak in the spectrum.
The radiation emission by over-barrier particles in the field of a periodically bent crystal 
was discussed qualitatively in Ref. \cite{Shulga-EtAl:PLA_v372_2065_2008} within the
continuous potential framework. 
More detailed quantitative analysis of the phenomena involved can be provided by means of all-atom molecular dynamics.
Below we present a brief overview of the results obtained and conclusions drawn in Ref. \cite{Pavlov-EtAl:EPJD_2020}.
}

\end{itemize}

To compare the contributions to the total emission spectrum coming from 
channeling and non-channeling particles the following procedure has been adopted.
Each simulated trajectory has been divided into segments corresponding to different types of motion.
Namely, we distinguished the following parts of the trajectory:
(i) the channeling motion segments, 
(ii) segments corresponding to the 
over-barrier motion across the periodically bent crystallographic planes,
(iii) segments corresponding to the motion in the vicinity of points of maximum curvature 
where a projectile experiences VR.
For each type of the motion, the spectrum of emitted radiation has been computed 
as a sum of emission spectra from different segments.
Thus, the interference of radiation emitted from different segments has been lost.

\begin{figure}
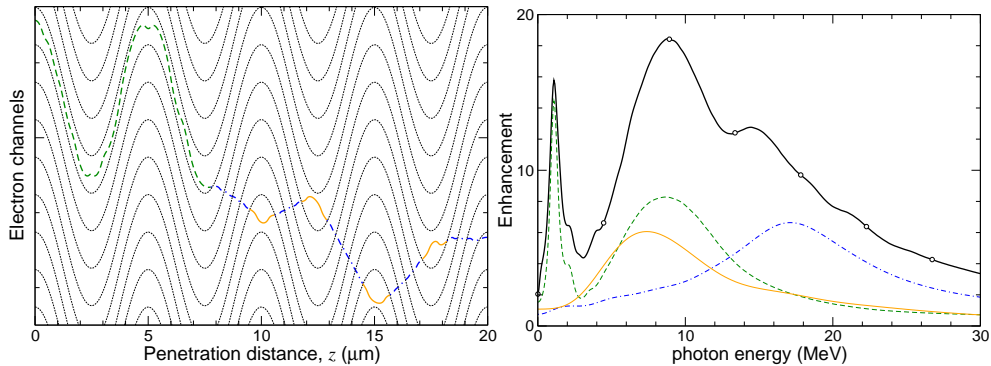

\centering
\includegraphics[width=6.5cm,clip]{Figure_17a.eps}
\includegraphics[width=6.55cm,clip]{Figure_17b.eps}
\caption{ 
\textit{Left.} An exemplary trajectory of a 855 MeV electron in a diamond (110) crystal  
               bent periodically with $a$ = 2.5 \AA{} and $\lamu=5$ $\mu$m.
               Highlighted are the segments corresponding to (i) the channeling regime (dashed green curve), 
             (ii) the over-barrier motion (dashed-dotted blue curves), (iii) to the VR events 
              (solid orange curves).
             Thin wavy lines mark the boundaries of the electron channels.\\
\textit{Right.} Solid black curve with open circles shows the enhancement factor of the total radiation 
                emitted by 855 MeV electrons in the diamond (110) crystal bent as described above.
                Dashed green, dashed-dotted blue and solid orange curves show the contributions coming from 
                the segments of the channeling and over-barrier motions and due to the VR, respectively.             
Ref. \cite{Pavlov-EtAl:EPJD_2020}.
}
\label{Pavlov_Figure07.fig}
\end{figure}

The aforementioned procedure is illustrated by Fig. \ref{Pavlov_Figure07.fig}.
Its left panel presents 
\textcolor{black}{a selected trajectory of a 855 MeV electron}
propagating in periodically bent crystal with bending amplitude 2.5 \AA.
Different types of segments are highlighted in different colour and type of the line as 
indicated in the caption.
The emission spectra corresponding to different types of motion (calculated accounting for
all simulated trajectories) are shown in the right panel.
The dependences presented allow one to associate the maxima in the total spectrum (black solid curve) 
with the corresponding type of motion. 
The radiation emitted from segments of channeling motion (dashed green curve) 
govern the spectrum in the vicinity of the CUR peak ($\hbar\om_{\rm CUR}\approx 1$ MeV)
and contributes greatly to the ChR at $\hbar\om_{\rm ch}\approx 6\dots12$ MeV.
Numerical analysis of the simulated trajectories has shown that the curvature of the trajectories
segments in the points of VR is close to that of the channeling trajectories.
As a result, the radiation from the VR segments is emitted in the same energy interval as ChR
so that the peak centered at $\approx 9$ MeV is due both to the channeling motion and to the 
VR events.
The over-barrier particles experience quasi-periodic modulation of the trajectory when crossing 
the periodically bent channels. 
The (average) period of these modulation is smaller than that of the channeling motion and
decreases with the increase of the bending amplitude.
For $a=2.5$ \AA{} this period is approximately two times less than the (average) 
period of channeling oscillations. 
As a result, radiation emitted from the over-barrier segments (dashed-dotted blue curve) 
is most intensive in the range $\hbar\om_{\rm ch}\approx 15\dots20$ MeV.
This contribution results in the additional structure in the total spectrum. 

\begin{figure}
\centering
\includegraphics[scale=0.5,clip]{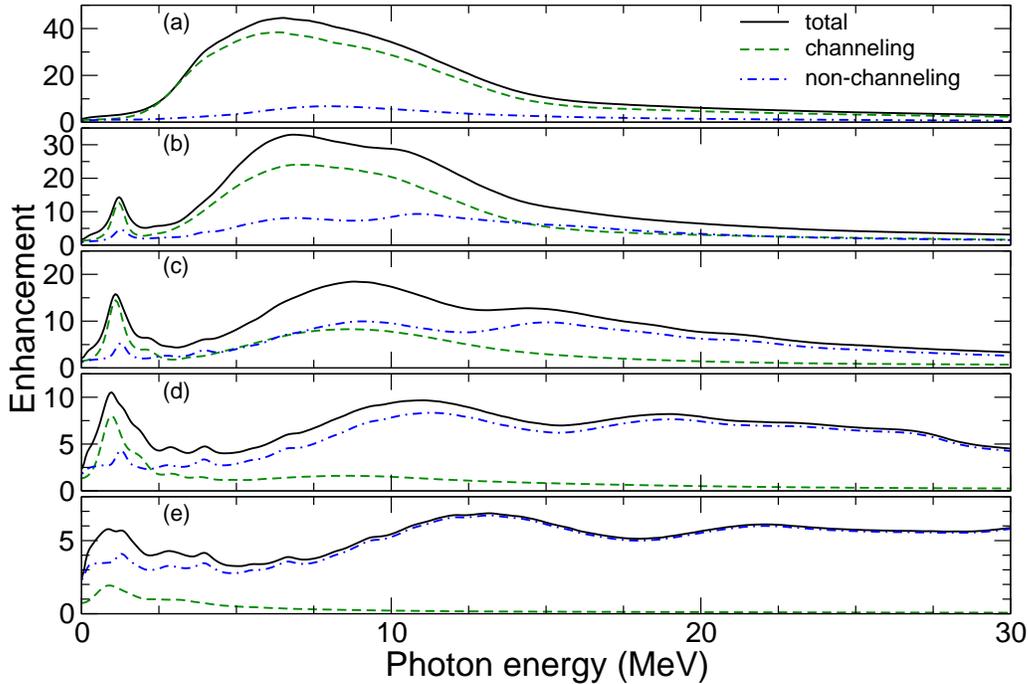}
\caption{Enhancement factor of the radiation over the
Bethe-Heitler spectrum for 855 MeV electrons in straight (a) and periodically bent
(graphs (b)-(e) correspond to $a = 1.2, 2.5, 4.0, 5.5$ \AA) diamond(110) crystals. 
Solid black curves show the total spectra, 
dashed green ones correspond to the radiation emitted from the channeling segments only,
and dashed-dotted blue curves present the spectra due to all non-channeling parts of the 
simulated trajectories.
Ref. \cite{Pavlov-EtAl:EPJD_2020}.
}
\label{Pavlov_Figure08_v02.fig}
\end{figure}

Figure \ref{Pavlov_Figure08_v02.fig} illustrates the evolution of the contributions from the channeling and
non-channeling particles to the emission spectrum with bending amplitude.
In the figure, each graph presents the total spectrum (solid curve) as well as the 
contributions of the channeling segments (dashed curve) and the non-channeling segments 
(both over-barrier and VR, dash-dotted curve).

In the straight crystal as well as in the periodically bent one with small bending amplitude ($a=1.2$ \AA)
the emission spectrum above 1 MeV is dominated by the channeling particles which provide main contributions
to the ChR peak.
As $a$ increases, the role of the non-channeling segments becomes more pronounced whereas
the channeling particles contribute less. 
The increase in $a$ leads to 
(i) increase of the curvature of a particle's trajectory in the vicinity of the VR points, 
(ii) decrease in the period of the quasi-periodic modulation of the trajectories of 
over-barrier particles.
As a result, two maxima seen in the graphs (b)-(e) become blue shifted as $a$ increases:
the maxima marked with upward arrows are due to the channeling motion and to the VR, 
those marked with downward arrows are associated with the over-barrier particles.
For large bending amplitudes, graphs (d)-(e), these maxima are virtually due to the emission
of the non-channeling particles only.

The low-energy part of the spectrum formed in periodically bent crystals is dominated by the peak 
located at $\hbar\om\approx 1$ MeV. 
For moderate amplitudes, $a\leq 2.5$ \AA, when the bending parameter (\ref{Results:eq.01}) is small,
this peak associated with CUR and is due to the motion of the accepted particles which cover a distance 
of at least one period $\lamu$ in a periodically bent channel.
For larger amplitude, $a=4.0$ \AA{} ($C=0.77$), the penetration length $\Lp$ of the accepted particles 
become less than half a period leading to noticeable broadening of the CUR peak.
For even larger amplitudes, there are further modifications of the peak related to 
the phenomenon different from the channeling.
Graph (e) shows the dependences for $a=5.5$ \AA{} which corresponds to the bending parameter larger than one,
$C=1.15$. 
As a result, only a small fraction of the incident electrons is accepted, 
and channels over the distance less than $\lamu/2$ having very small amplitude of channeling oscillations,
$a_{\rm ch}\ll d/2$.
Therefore, these particles virtually do not emit ChR but nevertheless contribute to the CUR part of the spectrum
(see the dashed curve in the graph). 
However, this contribution is not a dominant one. 
The main part of the peak intensity in the total spectrum comes from the non-channeling particles,
see the dash-dotted curve.
The explanation is as follows \cite{Pavlov-EtAl:EPJD_2020}.
As discussed above, a trajectory of a non-channeling particle consists of short segments corresponding 
to VR separated by segments $\Delta z \approx \lamu/2$ where it moves in the over-barrier mode.
In the course of two sequential VR the particle experiences 'kicks' in the opposite directions,
see the lower trajectory in Fig. \ref{Pavlov_Figure07.fig}(a)).
Therefore, the whole trajectory becomes modulated periodically with the period $2\Delta z \approx \lamu$.
This modulation gives rise to the emission in the same frequency as CUR. 

These effects, which are due to the interplay of different radiation mechanisms in periodically bent crystals, 
can be probed experimentally. 
In this connection one can mention recent successful experiments on 
\textcolor{black}{detecting the excess of
radiation emission due to VR in oriented bent Si(111) crystal by 855 MeV electrons 
\cite{Mazzolari_etal:PRL_v112_135503_2014} and 12.6 GeV electrons \cite{NielsenEtAl-PR-AB-v22-114701-2019}.}

\subsection{Channeling and radiation emission in SASP periodically bent crystals \label{SASP}}

The original concept of a CU assumes the projectiles channel in the crystal 
following the periodically bent planes or axes. 
For such motion, the undulator modulation frequencies $\Omu$ are smaller than the frequencies $\Om_{\rm ch}$
of the channeling oscillations. 
This regime implies periodic bending with large-amplitude, $a>d$, and large-period, $\lamu\gg a$. 
As a result, the CUR spectral lines appear at the energies below those of ChR 
\cite{KSG1998,KSG_review_1999,KSG_review_2004}.

Another regime of periodic bending, termed as Small-Amplitude Short-Period (SASP),
was suggested recently \cite{Kostyuk_PRL_2013}.
This regime implies bending with $a \ll d$ and $\lamu$ shorter than the period of 
channeling oscillations.
In contrast to the motion in a CU, the channeling trajectory in a SASP crystal
does not follow the short-period bent planes but acquires a short-period jitter-type modulations 
resulting from the bending.
These modulations lead to the radiation emission at the energies exceeding
the energies of the channeling peaks 
\cite{Kostyuk_PRL_2013,Korol-EtAl:NIMB_v387_p41_2016,Bezchastnov_AK_AS:JPB_v47_195401_2014,
Wistisen_etal:PRL_v112_254801_2014,UggerhojWistisen:NIMB_v355_p35_2015,Wistisen_etal:EPJD_v71_124_2017,
Wienands_EtAl:NIMB_v402_p11_2017}.
Interestingly, a similar radiative mechanism has recently been studied with 
respect to the radiation produced by relativistic particles in interstellar environments 
with turbulent small-scale fluctuations of the magnetic 
field~\cite{Medvedev_2000,Kellner_2013}.

In Ref. \cite{Korol-EtAl:NIMB_v387_p41_2016} results of a thorough study of channeling 
and radiation by 855 electrons and positrons passing through a SASP silicon crystal 
has been presented.
Comprehensive analysis of the channeling and radiation processes has been carried out on the grounds of 
numerical simulations.
Specific features which appear due to the SASP bending has been highlighted and elucidated within 
an analytically developed continuous potential approximation (see \ref{PotentialMotion-SASP}).
The parameters of the SASP bending were chosen to match those used in the experiment with 600 and
855 electrons carried out at MAMI \cite{Wistisen_etal:PRL_v112_254801_2014}.
A SASP crystal used in the experiment was produced by using Si$_{1-x}$Ge$_x$  graded composition 
with the Ge content $x$ varied from 0.3 \% to 1.3 \% to achieve a periodic bending of (110) planes
with the amplitude $a = 0.12 \pm 0.03$ \AA{} and period $\lamu = 0.43 \pm 0.004$ $\mu$m.
The number of periods quoted was 10.
No further details on the actual characterization of the profile of periodic bending were provided although
in a more recent paper \cite{Wienands_EtAl:NIMB_v402_p11_2017} it was noted that 
"\dots the shape is roughly sinusoidal".

\begin{figure} [ht]
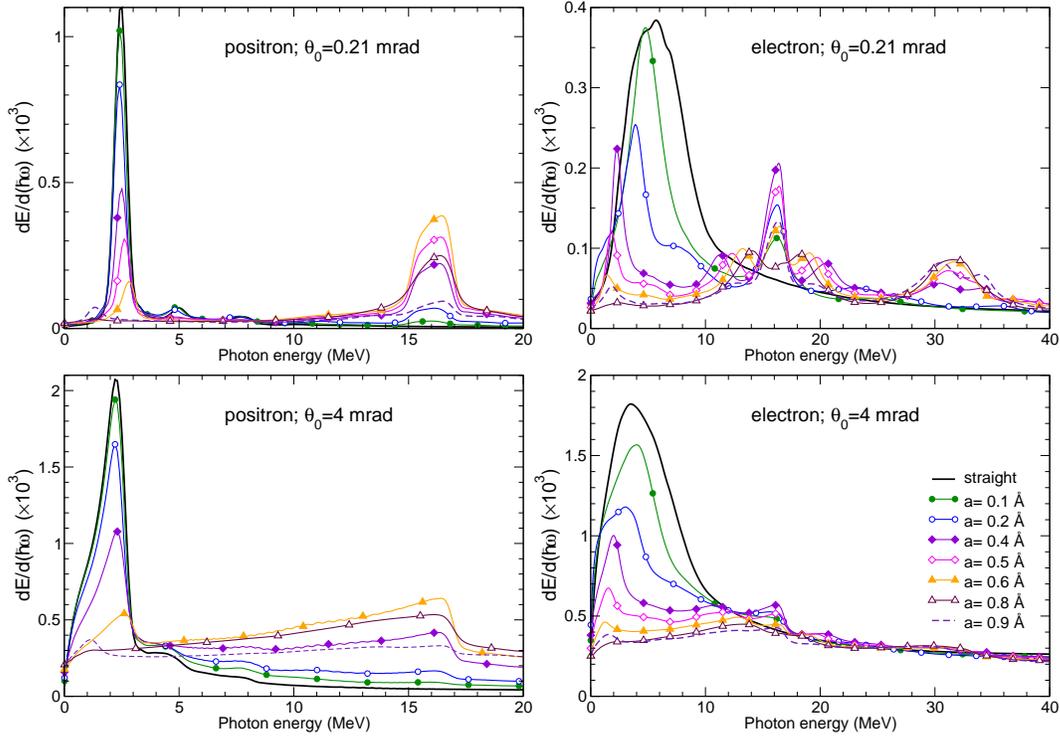

\centering
\includegraphics[width=7cm,clip]{Figure_19a.eps}
\includegraphics[width=7cm,clip]{Figure_19b.eps}
\\
\includegraphics[width=7cm,clip]{Figure_19c.eps}
\includegraphics[width=7cm,clip]{Figure_19d.eps}
\caption{
Spectral distribution of radiation emitted by 855 MeV positrons (left) 
and electrons (right) in straight (thick solid lines) and SASP bent Si(110) 
with period $\lamu=400$ nm and various amplitudes $a$ as indicated in the common
legend located in the right bottom graph. 
The upper and lower plots correspond to the aperture values 
$\theta_{0} = 0.21$ and $4$ mrad. 
All spectra refer to the crystal thickness $L=12~\mu$m.
The intensity of the bremsstrahlung radiation in amorphous silicon (calculated within the 
Bethe-Heitler approximation) is $0.016\times10^{-3}$ and  $0.15\times10^{-3}$ 
for $\theta_{0} = 0.21$ and $4$ mrad, respectively. 
}
\label{SASP-ep-855-400nm-12micron-Si.fig}
\end{figure}

In the simulations \cite{Korol-EtAl:NIMB_v387_p41_2016} a thicker crystalline sample, $L=12$ $\mu$m,
was probed assuming perfect cosine bending with period $400$ nm and amplitude varied
from $a=0$ (straight channel) up to $a=0.9$ \AA{}, which is close to the half of the (110) interplanar
spacing in silicon crystal ($d=1.92$ \AA).
The calculated emission spectra cover a wide range of the photon energies, 
from $\lesssim 1$~MeV up to $40$~MeV. 
The integration over the emission angle $\theta$ was carried out for two particular 
cones determined by the values $\theta_{0}=0.21$ and $4$~mrad. 
For a $855$ MeV projectile the natural emission angle is $\gamma^{-1} \approx 0.6$ mrad. 
Therefore, the smallest value of $\theta_{0}$ refers to a nearly forward emission, 
whereas the largest value, being significantly larger than $\gamma^{-1}$, 
provides the emission cone which collects almost all the radiation emitted.

The spectra computed display a variety of features seen in 
Fig. \ref{SASP-ep-855-400nm-12micron-Si.fig}.
To be noticed are the pronounced peaks of ChR in the spectra for 
the straight crystal (the black solid-line curves). 
Nearly perfectly harmonic channeling oscillations in the positron trajectories 
(the examples of the simulated 
trajectories can be found in \cite{MBN_ChannelingPaper_2013,Sushko-EtAl:JPConfSer_v438_012018_2013,ChannelingBook2014}) 
lead to the undulator-type spectra of radiation with small values of the undulator 
parameter, {$K^2 \ll 1$}. 
The positron spectra in straight Si(110) clearly display the fundamental peaks of ChR 
at the energy $\hbar\om\approx 2.5$~MeV, whereas the higher harmonics 
are strongly suppressed.
In particular, for the smaller emission cone the peak intensity in the fundamental harmonic
is an order of magnitude larger than that for the second harmonics at $\hbar\om\approx 5$~MeV, 
and only a tiny hump of the third harmonics can be recognized 
at about 7.5~MeV (see the top left plot in the figure). 
For electrons passing through the straight crystal, the ChR peaks 
are less intensive and much broader than these for the positrons, as a result of 
stronger anharmonicity of the channeling oscillations in the trajectories. 

The radiation spectra produced in the SASP bent crystals 
display additional peaks, which emerge from the short-period modulations of the 
projectile trajectories (see the discussion in Section \ref{SASP-Motion}). 
These peaks, more pronounced for the smaller emission cone, appear at the energies larger 
than the energies of the channeling peaks. 
For both types of the projectiles, the fundamental spectral peaks in the radiation 
emergent from the SASP bending correspond to the emission energy 
about $16$~MeV significantly above the ChR peaks. 
For positrons, the peaks of radiation due to the bending are displayed in the spectra for 
the amplitude values  $0.1\dots0.9$ \AA{}. 
For smaller values of $a$, the spectral peaks disappear because the positrons 
experience mainly "regular" channeling staying away from the crystalline 
atoms and being therefore less affected by the SASP bent planes 
(see Ref. \cite{Korol-EtAl:NIMB_v387_p41_2016} and Section \ref{SASP-Motion} below for 
the details). 
In contrast, the electrons experience the impact of the SASP bending at lower 
values of $a$. 
As seen in the right upper plot for the fundamental spectral 
peaks emergent from the bending, the peak for $a=0.1$ \AA{} is only two times 
lower than the maximal peak displayed for $a=0.4$ \AA{}.

To be noted are the spectral properties for smaller aperture value 
(upper plots in Fig. \ref{SASP-ep-855-400nm-12micron-Si.fig}). 
The electron spectra display peaks at the energies around 32 MeV for the values of $a$ exceeding $0.2$ \AA{}. 
These peaks are clearly the second harmonics of the radiation emergent from the SASP bending.  
In addition, the peaks of channeling radiation decrease in heights and shift towards the 
lower emission energies. 
The positron spectra, in contrast to the electron ones, 
exhibit less peculiarities and gradually converge to the Bethe-Heitler background with 
increasing radiation energies.
For the larger aperture value, $\theta_{0} = 4$~mrad, a sizable part of the 
energy is radiated at the angles $\theta > \gamma^{-1}$. 
The harmonics energies decrease with $\theta$ approximately as $(1+K^2/2 +(\gamma\theta)^2)^{-1}$.
As a result, the peaks of ChR and those of the radiation due to the SASP bending   
broaden and shift towards softer radiation energies. 

Recently, the impact of the radiation collimation on the intensity of the SASP peaks has been
measured in experiments with a 855 MeV electron beam at MAMI \cite{Wistisen_etal:EPJD_v71_124_2017}.
The crystal was produced by adding a fraction $x$ of germanium atoms to a silicon substrate. 
By alternating successively a linear increase of $x$ from 0.5 \% to 1.5 \%  with a linear decrease 
a sawtooth pattern of the SASP bending was achieved with 120 periods each of a $\lamu=0.44$ $\mu$m 
wavelength. 
It was indicated in the paper that "the expected oscillation amplitude" of the (110) planes
is $a\approx0.12$ \AA{}. 
In the experiment the radiation spectra enhancement over the amorphous silicon was measured.
The measurements were performed (i) with collimation to an emission angle 
$\theta_0\approx 0.24$ mrad,
and (ii) with no collimation.
It was noted that the latter case corresponded to the emission
cone $\theta_0 = 4\, \mbox{mrad}\, \gg \gamma^{-1}$ 
considered in \cite{Korol-EtAl:NIMB_v387_p41_2016}.

\begin{figure} [ht]
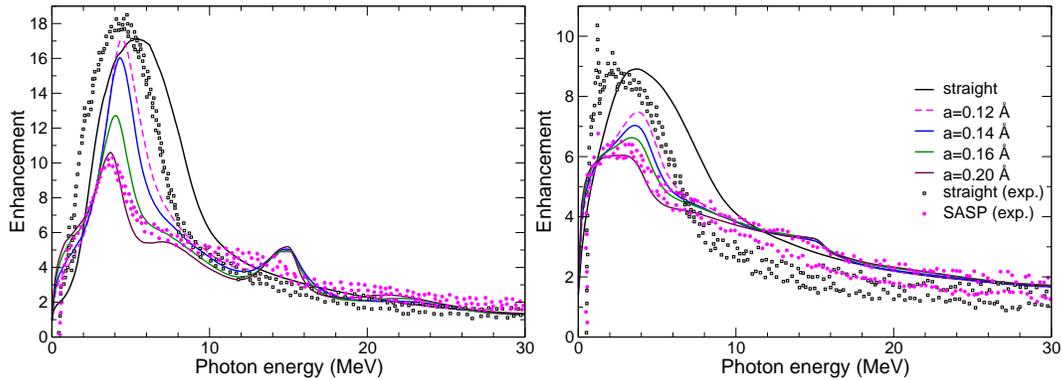

\centering
\includegraphics[width=7cm,clip]{Figure_20a.eps}
\includegraphics[width=7cm,clip]{Figure_20b.eps}
\caption{
Enhancement of radiation emitted by 855 MeV electrons in straight and SASP bent Si(110) 
with respect to the amorphous silicon.
The data refer to the crystal thickness $L=52.3$ $\mu$m and bending period $\lamu=436$ nm
(total number of periods equals to 120).
Left and right graphs refer to the emission cones with opening angle 
$\theta_{0} = 0.24$ and $4$ mrad, correspondingly. 
Symbols stand for the experimental data taken from Ref. \cite{Wistisen_etal:EPJD_v71_124_2017} where
the bending amplitude was assumed to be $a=0.12$ \AA.
Common legend is presented in the right graph.
}
\label{SASP-e-855-436nm-120period-SiGe.fig}
\end{figure}

In Fig. \ref{SASP-e-855-436nm-120period-SiGe.fig} we compare the experimental data (symbols)
with the results of simulations carried out with \MBNExplorer (solid lines). 
Shown are the dependences of the enhancement factor on the photon energy for straight and 
SASP bent Si(110).
Left graph corresponds to the narrow emission cone, $\theta_0= 0.24$ mrad, the right graph 
presents the data for the  wide cone, $\theta_0= 4$ mrad.
The simulations were performed for several values of the bending amplitude as indicated in the
common legend shown in the right graph.

The same SASP bent Si(110) with 120 undulations with period $\lamu=0.44$ $\mu$m was used in the 
experiment at the SLAC facility with a 16 GeV electron beam 
\cite{Wienands_EtAl:NIMB_v402_p11_2017}. 
In the experiment, the SASP signal can only be expected to appear when the crystal is
properly aligned and should reveal itself in a narrow emission angle and the presence of a peak in
the spectrum. 
Therefore, in Ref. \cite{Wienands_EtAl:NIMB_v402_p11_2017} the enhancement 
was looked for as the crystal was rotated in the beam, passing through the aligned condition, 
and a narrow radiation cone when scanning the
horizontal angular distribution with the SciFi detector, both measurements feasible at high beam
intensity.
However, it was mentioned in the cited paper that the measurements of the spectrum were not 
successful due to difficulties with the experimental setup and variations 
in beam energy that had not been expected.

\begin{figure} [h]
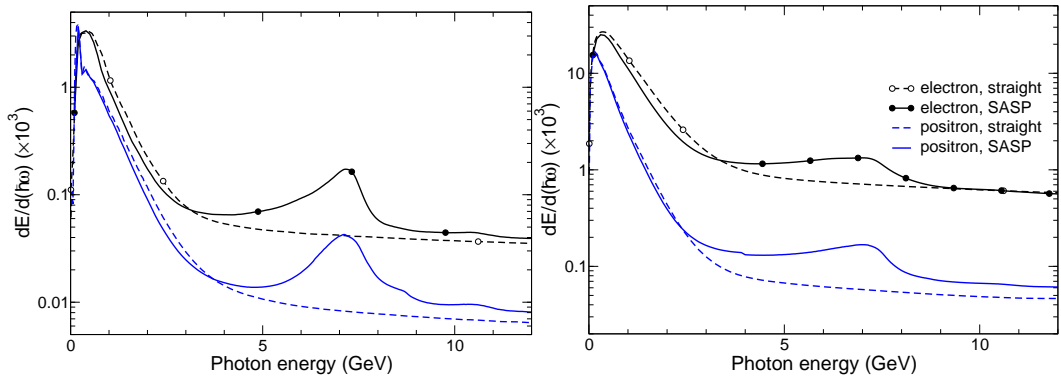

\centering
\includegraphics[width=7cm,clip]{Figure_21a.eps}
\includegraphics[width=7cm,clip]{Figure_21b.eps}
\caption{
Spectral distribution of radiation emitted by 20 GeV electrons and positrons
in straight and SASP bent Si(110).
The data refer to the crystal thickness $L=52.3$ $\mu$m, bending period $\lamu=436$ nm
and bending amplitude $a=0.12$ \AA.
Left and right graphs refer to the emission cones with opening angle 
$\theta_{0} = 13$ and $130$ $\mu$rad, correspondingly. 
Common legend is presented in the right graph.
}
\label{ep-20GeV-0.1A-120period.fig}
\end{figure}

In connection with these experiments, which initially had been planned to be carried out
with both electron and positron beams, the channeling simulations were carried out 
 for 15-35 GeV projectiles \cite{Our-for-SLAC} by means of the \MBNExplorer package.
The simulations of trajectories were supplemented with computation of the spectra of the emitted
radiation for various detector apertures. 
It was recommended to carry out experiments with electrons and with the smallest aperture possible. 
In this case the SASP signal in the spectrum was expected to be the highest. 
Figure \ref{ep-20GeV-0.1A-120period.fig} illustrates theoretical predictions by presenting the
spectral distribution of radiation energy emitted by 20 GeV projectiles in the narrow 
13 $\mu$rad $\approx 1/2\gamma$ (left graph) and wide 13 $\mu$rad $\approx 5/\gamma$ (right graph)
cones along the incoming beam direction. 
The peaks centered around $\hbar\om=7$ GeV are due to the SASP bending. 
The peaks at much lower energy (0.2-0.5 GeV) correspond to ChR.

\subsection{Stack of SASP periodically bent crystals \label{Stack}}

In recent series of experiments at MAMI \cite{Wistisen_etal:PRL_v112_254801_2014} with 600 and 855~MeV 
electrons the effect of the radiation enhancement due to the 
SASP periodic bending has been observed (see discussion in Section \ref{SASP}).
Another set of experiments with thin SASP diamond crystals was planned within the E-212 collaboration at 
the SLAC facility (USA) with 10-20~GeV 
electron beams~\cite{UggerhojWistisen:NIMB_v355_p35_2015}. 

As a case study aimed at producing theoretical benchmarks for the SLAC experiments, a series of numerical 
simulations have been performed 
of the planar channeling of  10-20~GeV electrons and positrons in straight and SASP periodically bent thin 
crystals of silicon and diamond
\cite{Sushko:Thesis_2015,Sushko-EtAl:NTV_v1_p341_2015}.
The crystal thickness $L$ was set to 4~microns, the period of bending $\lamu=0.4$ microns and the bending 
amplitude $a=0.4$ \AA{}, which is lower than half 
of the (110) interplanar distance in both cases. 

\begin{figure}
\centering      
\includegraphics[scale=0.3,clip]{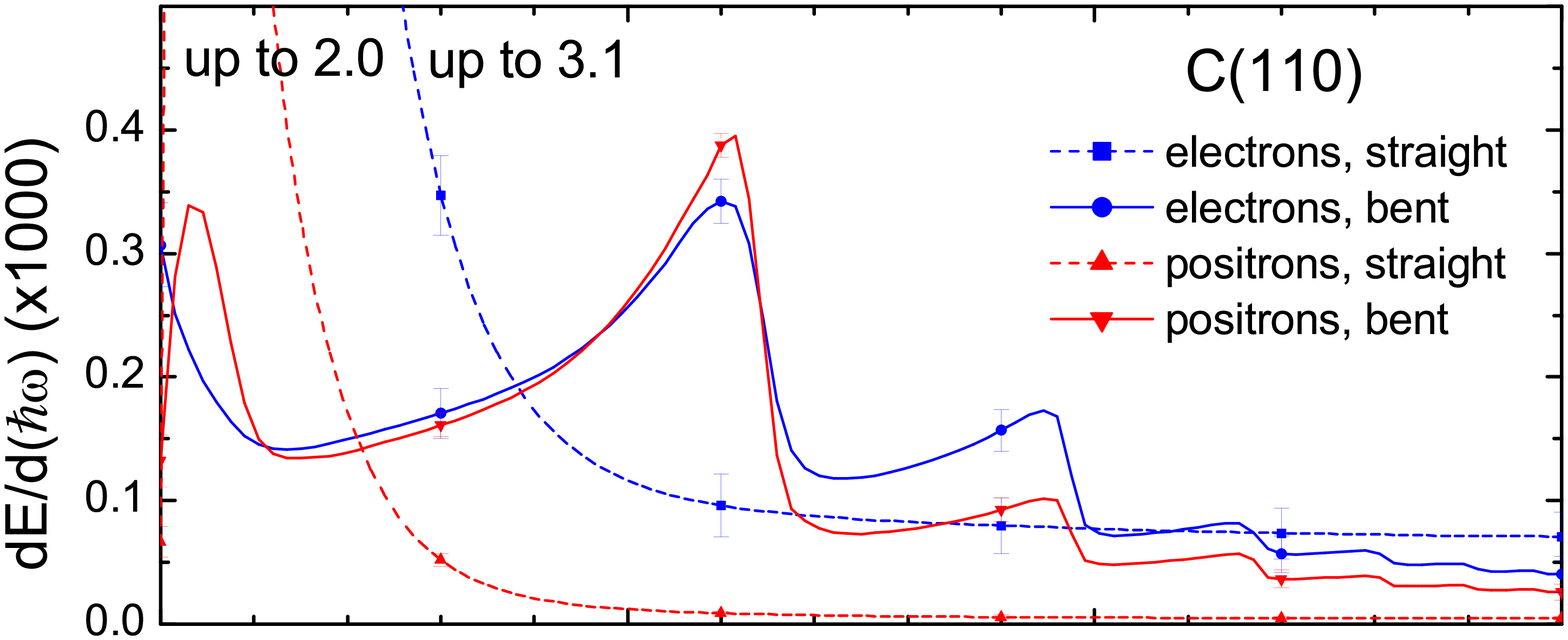}
\\
\includegraphics[scale=0.3,clip]{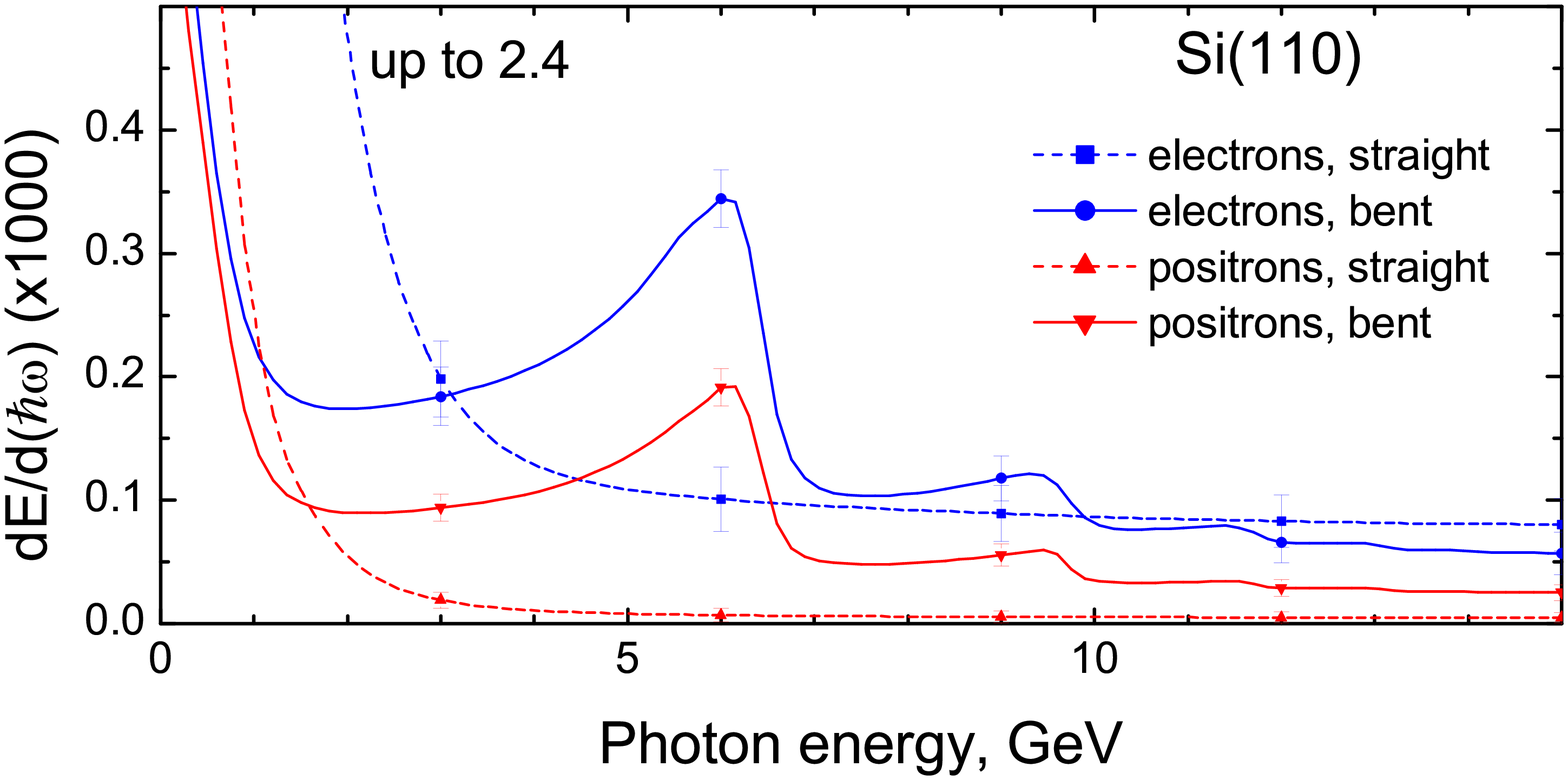}
\caption{
Spectral distribution of radiation emitted by 20 GeV electrons and positrons in straight and SASP periodically bent 
4 microns thick diamond(110) (top) and silicon(110) (bottom) oriented crystals. 
Bending amplitude and period are 0.4~{\AA} and 0.4~$\mu$m, respectively.
Refs. \cite{Sushko:Thesis_2015,Sushko-EtAl:NTV_v1_p341_2015}.
}
\label{diamond-SASP-4microns-spectra.fig}
\end{figure}

In Figure~\ref{diamond-SASP-4microns-spectra.fig} the results of the simulation of radiation of 20~GeV 
projectiles are compared for the cases of straight 
and periodically bent diamond(110) crystals. 
The beam emittance was taken equal to $\psi = 5$~$\mu$rad.
The spectra presented refer to the emission cone $\theta_{0}$ = 150~$\mu$rad, which is 5.8 times higher 
than natural emission angle 
$1/\gamma = 25.6$~$\mu$rad and thus collects virtually all radiation emitted.
In both figures the peaks located below 1~GeV corresponds to the channeling radiation.
For periodically bent targets, the peaks at $\hbar\om \approx 6$ GeV and above are due to the SASP bending.
Note, that bending of a crystal leads to significant suppression of the channeling peak. 
This effect can be explained qualitatively in terms of the continuous potential modification in a SASP 
channel (see Section \ref{Potential-SASP}).
With increase of bending amplitude the depth of the potential well decreases and the width of the potential 
well grows resulting in 
decrease of the number of channeling projectiles and in lowering frequencies of channeling oscillations.

\begin{figure}[!ht]
\centering
\includegraphics[scale=0.3,clip]{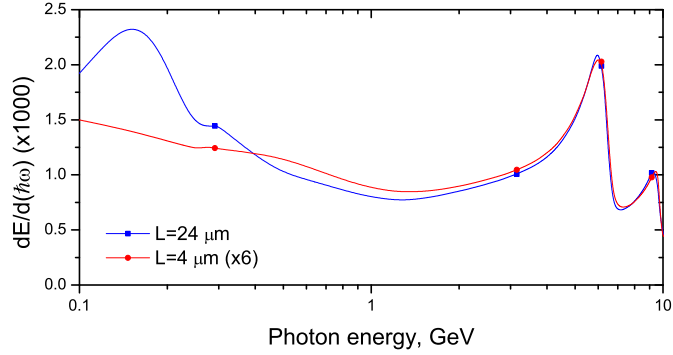}
\caption{
Emission spectra for 20~GeV electrons in SASP bent diamond(110) crystal 
calculated for two different thicknesses, as indicated. 
The value $L=24$~$\mu$m exceeds characteristic channeling oscillations period while $L=4$~$\mu$m is lower than that.
Note the absence of channeling radiation peak around 150~MeV in the latter case. 
For the sake of comparison, the curve for $L=4$~$\mu$m is multiplied by six.
Refs. \cite{Sushko:Thesis_2015,Sushko-EtAl:NTV_v1_p341_2015}.
}
\label{diamond-SASP-thick.fig}
\end{figure}

Another factor that leads to the suppression of the channeling radiation is that for the 20~GeV projectiles 
the characteristic period of channeling 
oscillations, deduced from the simulated trajectories, is about $\langle \lambda_{\rm ch}\rangle \approx10$
microns for both diamond and silicon crystals, so that 
the crystal is too thin to allow for even a single channeling oscillation.
Therefore, the peak of channeling radiation is not that pronounced as in the case of thicker, $L > \lambda_{\rm ch}$.
The emission spectra formed in thick ($L=24$ microns) and thin ($L=4$ microns) crystals are compared in 
Fig.~\ref{diamond-SASP-thick.fig} where
the latter spectrum is multiplied by a factor of 6 for the sake of convenience.
A sharp peak of the channeling radiation at $\hbar\om \approx 150$ MeV is present for the thick crystal whereas 
for the thin one it reduces to a small hump, which is due to the synchrotron-type radiation emitted by 
projectiles moving along the one-ark trajectory.
Remarkable feature, seen in the figure, is that the peaks due to the SASP bending in both curves virtually coincide.

\begin{figure}
\centering
\includegraphics[scale=0.4,clip]{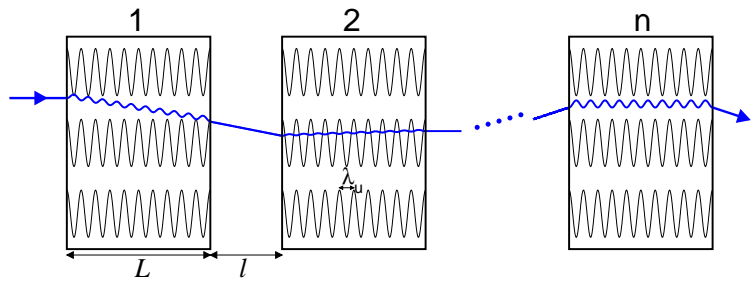}
\caption{
Stack of $n$ SASP bent crystal layers each of the thickness $L$ separated by the gaps $l$.
Periodic bending of the crystalline structure is illustrated by thin cosine curves.
Blue line illustrates a projectile's trajectory which consists of the modulated parts 
(inside the crystals) and of the straight line segments in between the layers.
}
\label{stack.fig}
\end{figure}

The effects of suppression of the channeling radiation but maintaining the level of undulator radiation in thin crystals
can be used to produce intensive radiation at much higher energies corresponding to the SASP bending.
To increase the latter intensity one can increase the crystal thickness $L$.
However, this extensive approach is not optimal taking into account complications related to the technological aspects 
(increase in the time of the crystal growth
as well as in the costs associated, accumulation of the defects in the crystalline structure etc.) 
Alternatively, instead of a single thick crystal, a stack of several aligned thin crystals can be used
\cite{Sushko-EtAl:NTV_v1_p341_2015,Sushko:Thesis_2015}, as illustrated by Fig. \ref{stack.fig}. 
A projectile passes sequentially several layers of periodically bent crystals, which constitute the stack, and 
the radiation produced in each element of the stack adds to the total radiation emitted by
the projectile. 
For SASP undulator the thickness of layers can be taken in the interval
between the bending period $\lamu$ and the characteristic channeling period of projectile.
Such choice of the parameters leads to absence of full channeling oscillation
periods in each channeling segment of trajectory of projectile which results in
suppression of channeling radiation. The effect of undulator radiation in the system
remains and grows with increase of number of layers.
Thickness of each crystal layer can be chosen to be smaller than the period of 
channeling oscillations of the projectile thus suppressing ChR.
The intensity of radiation due to the SASP periodic bending 
increases with the number the stack layers.

\begin{figure}
\centering
\includegraphics[scale=0.4,clip]{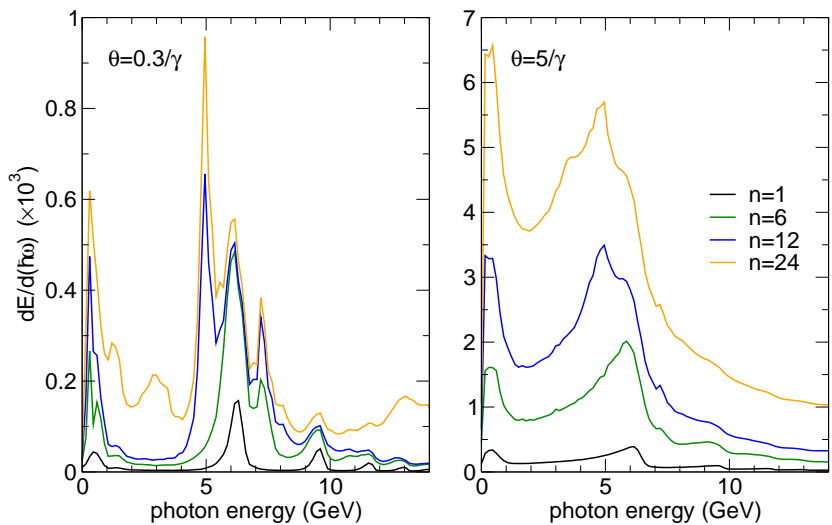}
\caption{
Radiation spectra for the small $\theta_0=0.3/\gamma$ (left panel) 
and large $\theta_0=5/\gamma$ (right) apertures calculated for 
different number $n$ of diamond(110) periodically bent crystals in stack (as indicated in the common legend). 
The data refer to 20~GeV positrons, the bending 
amplitude and period are 0.4~{\AA} and 400~nm, respectively.
}
\label{stack-p-allX.fig}
\end{figure}

To simulate the radiation emission from the stack the following system was modeled 
\cite{Sushko-EtAl:NTV_v1_p341_2015,Sushko:Thesis_2015}. 
A set of several $L=4$ microns thick layers of SASP periodically bent crystals (bending period $\lamu=0.4$ microns) 
separated with $l=4$ microns gaps were generated in the simulation box.
The 20 GeV projectiles (positrons) entered the first layer tangent to the (110) crystallographic plane.
Due to the multiple scattering in the layer, a projectile leaves is at some non-zero angle are with respect 
to the initial direction and
this angle serves as the incident angle at the entrance to the second layer, etc.
The multiple scattering leads to a gradual increase of the angular dispersion of the transverse velocity of 
projectiles and to the decrease of the number of 
channeling particles with the growth of the layer's number.
As a result, for moderate number of layers the destructive effect of the multiple scattering is not too 
pronounced, so that the radiation intensity 
increases being proportional to $n$. 
For sufficiently large $n$ values, the intensity reaches its saturation level and the peak intensity becomes
independent on $n$.

Figure~\ref{stack-p-allX.fig} compares the radiation spectra calculated for different number of layers in stack, 
as indicated. 
For the smaller emission angle ($\theta_0=0.3/\gamma=15.3$ $\mu$rad, left panel) the radiation intensity scales 
linearly with the number of stack 
layers until $n=6$. 
For larger values of $n$ the spread of the projectiles' transverse velocities gets wider so that the intensity
of radiation emitted within a narrow cone
along the initial beam direction saturates. 
For the larger emission angle ($\theta_0=5/\gamma=256$ $\mu$rad) the nearly linear growth of the intensity
continues up to $n=24$.

\begin{figure}
\centering
\includegraphics[scale=0.3,clip]{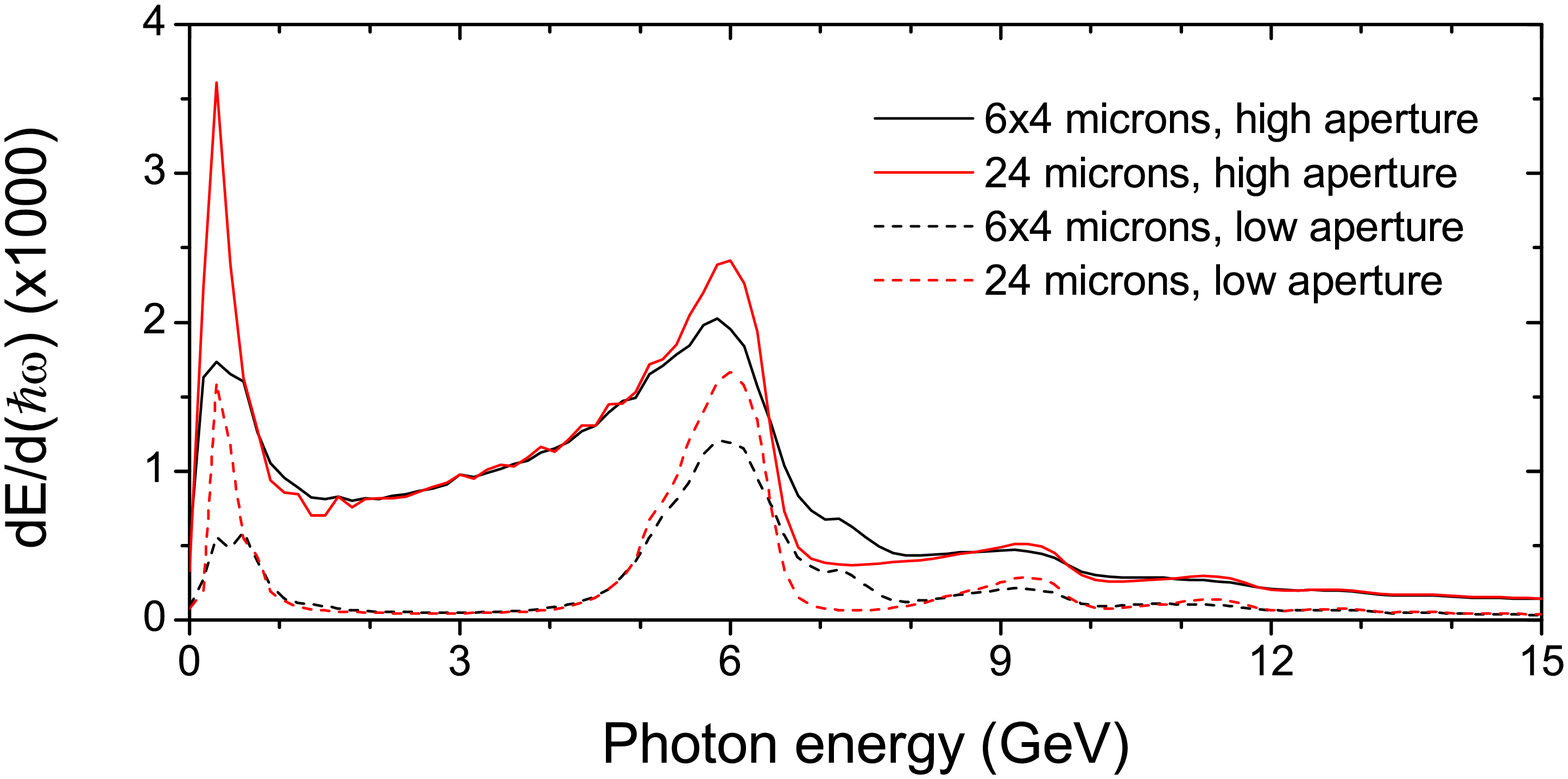}
\includegraphics[scale=0.25,clip]{Figure_26b.eps}
\caption{
\textit{Left panel.}
Comparison of radiation spectra formed by 20 GeV positrons in a $L=24$ microns thick crystal
and in a stack of six $L=4$ microns thick layers. 
Dashed curves present the spectra calculated for the smaller emission angle, $\theta_0=15.3$ $\mu$rad,
solid curves -- for the larger emission angle $\theta_0=256$ $\mu$rad.
Refs. \cite{Sushko:Thesis_2015,Sushko-EtAl:NTV_v1_p341_2015}.
\\
\textit{Right panel.}
Comparison of radiation spectra from 20 GeV positrons (solid curves) and electrons (dashed curves) 
emitted in stacks of diamond(110) SASP periodically bent crystals with different number of layers as indicated.
The data refer to the emission angle $\theta_0=5/\gamma=256$ $\mu$rad.
}
\label{stack-ep-dE.fig}
\end{figure}

Left panel in Fig. \ref{stack-ep-dE.fig} compares the intensities of radiation emitted by 20 GeV positrons traveling 
in a single 24 microns thick SASP diamond(110) crystal and in a stack of six thin ($L=4$ microns) layers. 
It is seen, that for the channeling peak in the spectrum is suppressed in the case of the stack of layers, 
wheres the undulator peaks are of the same intensity for both targets.
Right panel in the figure compares the emission spectra for positrons and electrons of the same energy, 
20 GeV, and in the same target
(stacks of $n=1,2$ and 24 layers. 
The curves presented illustrate weak sensitivity of the spectra formed in SASP periodically bent crystals 
to the sign of a projectile's charge.

Therefore, this regime is favorable for the construction of light sources with the use of intensive electron 
beams which, at present,
are more available than positron beams.

\section{\textcolor{black}{Conclusion} \label{Conclusion}} 

In this paper we have discussed the relativistic molecular dynamic approach implemented in the 
multi-purpose \MBNExplorer software package
for accurate computational modeling of propagation and radiation emission by various ultra-relativistic 
projectiles in crystalline media.
The exemplary case studies presented refer to various straight, bent and periodically bent oriented crystals. 
In cases where the experimental data are available (in particular, the data on the dechanneling length and on 
the emission spectra)
is has been used for the comparison with the results of numerical simulations.

The software package used in the current paper allows for advanced computation exploration, which can be carried out at the 
atomistic level of detail, of a variety of phenomena accompanying propagation of high energy particles in crystals.
Apart from the trajectories simulation and the calculation of the relevant quantities, one can model and 
analyze quantitatively more complex processes.
The latter include: 
\\
(i) structural changes in crystals due to the irradiation and their impact on the projectiles propagation and 
on the emission spectra
(this is important to account for when the target is exposed to highly intensive beams as, for example, 
the FACET beam at SLAC 
\cite{FACET,FACETII_Conceptual_Design_Rep-2015});
\\
(ii) the crystal structure modifications, incl. the defects formation, occurring in the course of fabrication
of bent and periodically bent crystals (see Ref. \cite{HighImpact} for the review of the technologies developed);
\\
(iii) simulation of the projectile' dynamics with account for the radiation damping force  
({\textcolor{black}{this is important when considering propagation of highly energetic electrons and
positrons, $\varepsilon \gtrsim 10^2$ for low-$Z$ crystals and $\varepsilon \gtrsim 10$ for high-$Z$
crystals, when radiative energy losses become noticeable 
\cite{KSG_losses_2000}});\footnote{\textcolor{black}{The radiative damping force can be introduced via different schemes
(see, for example, recent paper \cite{NielsenEtAl-PRD_v102_052004_2020}).
The scheme that has been recently implemented in \MBNExplorer
is based on the formalism due to Landau and Lifshitz \cite{Landau2}.
Previously, it had been implemented in Ref. \cite{Dechan01} within the
continuous potential framework.}}
\\
(iv) account for quantum effects (e.g., ionizing collisions, bond breaking \cite{IDMD-2106}, specific 
quantum features in the
incoherent scattering from the constituent atoms \cite{Tikhomirov:PR-AB_v22_054501_2019}).
which accompany a projectile propagation.
\\
Apart from these, the \MBNExplorer architecture allows one to develop and explore multiscale algorithms for 
simulations of the long-term dynamics of a crystalline medium on time scales significantly exceeding those 
achievable by means of conventional molecular dynamics \cite{MBNExplorer_Book}. 

These important features of \MBNExplorer increase significantly accuracy of its predications, expand its
application areas and go well beyond the limits of molecular dynamics codes which are unable to deal with the multiscale
modeling as well as of the codes based on the continuous potential model for the crystalline field.

The atomistic approaches and the computational algorithms implemented in \MBNExplorer open a broad
range of possibilities for the virtual design of the novel crystal-based LSs, which are mentioned in the 
Introduction section and
discuss in greater detail in Ref. \cite{HighImpact}. 
Thus, the multiscale  all-atom relativistic molecular dynamics simulations of the particle propagation 
and radiation in realistic 
crystals can be carried out. 
Combined with modern numerical algorithms, advanced computational facilities and computing technologies, 
it will bring the predictive power
of the software up to the accuracy level comparable or maybe even higher than achievable experimentally. 
Ultimately, it will turn computational modeling into the instrumental tool that could substitute 
(or become an alternative
to) expensive laboratory experiments, and thus reduce the experimental and technological costs. 
The important outcome of this analysis will enable us to provide the realistic characterization of the novel LSs in
the photon energy range up to GeV region and allow for their
optimization with respect to a particular experimental setup and targeted application.

\ack

The work was supported in part by the DFG Grant (Project No. 413220201).

\appendix

\section{Atomic and interplanar potentials \label{AtomicPotential}}

In this section, for the sake of reference, we compare atomic and interplanar 
potentials
build using the parameterization due to Moli\`{e}re \cite{Moliere}, 
Doyle and Turner \cite{DoyleTurner1968}, and Pacios \cite{Pacios1993}. 

The atomic system of units, $e=m_{e} =\hbar=1$, is used in this Section.

\subsection{Atomic potential parameterization \label{AtomicPotentials}}

Below we summarize the parameterizations 
for atomic potential, $\Uat(r)$, and its Fourier image, $\tUat(q)$:
\begin{eqnarray}
\Uat(r) 
= {4\pi \over (2\pi)^3}
\int_0^{\infty} {\sin(qr) \over qr}\, \tUat(q) q^2 \d q\,.
\label{AtomicPotentials:eq.01}
\end{eqnarray}

\begin{itemize}
\item
The parameterization  due to Moli\`{e}re \cite{Moliere} is based 
on the Thomas-Fermi atomic model:
\begin{eqnarray}
\cases{
U_{\rm at, M}(r) = {Ze\over r}\, \sum_{j=1}^3
a_j \, \exp\left(-{b_j r \over \aTF}\right)
\\
\widetilde{U}_{\rm at, M}(r)
 =
4\pi {Z e}
\sum_{i=1}^3 
{\alpha_j  \over q^2 + \gamma_j^2}
}.
\label{AtomicPotentials:eq.02}
\end{eqnarray}
Here $Z$ is the nucleus charge, $\aTF=0.8853Z^{-1/3}$ is the Thomas-Fermi radius.
The dimensionless Moli\`{e}re's coefficients are:
$a_{1,2,3}=(0.1, 0.55, 0.35)$,  $b_{1,2,3}=(6.0, 1.2, 0.3)$.
The short-hand notation $\gamma_j =\beta_j/\aTF$ is used.

\item
Doyle and Turner \cite{DoyleTurner1968} introduced parametric fits  
to kinematic scattering factors for X-rays and electrons
with the use of relativistic Hartree-Fock approximation 
(see also Ref. \cite{WaasmaierKirfel_ActaCrys_v51_p416_1995}).
From their formulae one derives the following parameterization:
\begin{eqnarray}
\cases{
U_{\rm at, DT}(r) 
=
{(4\pi)^3 \over 4\sqrt{\pi} }  
\sum_{j=1}^4
{a_j  \over b_j^{3/2}}
\exp\left(-{4\pi^2r^2 \over b_j} \right) 
\\
\widetilde{U}_{\rm DT}(q)  
=
2\pi
\sum_{j=1}^4
a_j \exp\left(-{b_j q^2 \over (4\pi)^2}\right)
}.
\label{AtomicPotentials:eq.03}
\end{eqnarray}
For carbon, silicon and germanium atoms the Doyle-Turner 
parameters $a_j$ (in \AA) and $b_j$ (in \AA$^2$) 
are listed in Table \ref{Doyle-Turner.Table}.

 \begin{table}[h]
\hspace*{-1cm}
 \caption{
 Parameters $a_j$ (in \AA) and $b_j$ (in \AA$^2$) for the 
 Doyle-Turner fit for several neutral atoms as indicated.
 }
 \footnotesize\rm\item[]
 \begin{tabular}{@{}lllllllll}
 \br
Atom & $a_1$  & $b_1$   & $a_2$  & $b_2$   & $a_3$  & $b_3$ & $a_4$ & $b_4$    \\
 \br
B    & 0.9446 & 46.4438 & 1.3120 & 14.1778 & 0.4188 & 3.2228 & 0.1159 & 0.3767 \\
C    & 0.7307 & 36.9951 & 1.1951 & 11.2966 & 0.4563 & 2.8139 & 0.1247 & 0.3456 \\
Si   & 2.1293 & 57.7748 & 2.5333 & 16.4756 & 0.8349 & 2.8796 & 0.3216 & 0.3860 \\
Ge   & 2.4467 & 55.8930 & 2.7015 & 14.3930 & 1.6157 & 2.4461 & 0.6009 & 0.3415 \\
 \br
 \end{tabular}
 \label{Doyle-Turner.Table}
 \end{table}

It has been pointed out (see, e.g., Ref. \cite{Dedkov1995})
that the D-T scheme does not provide correct behaviour 
of the potential at small distances since 
$\lim_{r\to 0}U_{\rm a, DT}(r) \neq \infty$.

\item
For atoms from H to Kr, Pacios \cite{Pacios1993} proposed the following parameterization 
based on the Hartree-Fock potentials:
\begin{eqnarray}
\cases{
U_{\rm at, P}(r)
=
{4\pi \over r}
\sum_{j=1}^M
{a_j \over b_j^3}\, \bigl(2 + b_j r\bigr)\exp\left(-b_jr\right),
\\
\widetilde{U}_{\rm at, P}(q)  
=
2(4\pi)^2 
\sum_{j=1}^M
{a_j \over b_j^3}
 \left(
 {1 \over q^2 + b_j^2}
 +
 {b_j^2 \over (q^2 + b_j^2)^2}
 \right)
}
 \label{AtomicPotentials:eq.04}
\end{eqnarray}
Sets of the coefficients $a_j$ and $b_j$ ($j=1,\dots M$) and values of the integer $M$
for several selected atoms are presented in table \ref{Pacios.Table}.
Note that the nucleus charge $Z$ is absent in Eqs. (\ref{AtomicPotentials:eq.04}).
Although it is not indicated either in Ref. \cite{Pacios1993} or in earlier papers
by the author, 
the nucleus charge and the coefficients are related as 
$Z =8\pi \sum_{j=1}^M{a_j/b_j^3}$.

 \begin{table}[h]
\hspace*{-1cm}
 \caption{
 Parameters $a_j$  and $b_j$ (in a.u.)
 of the Pacios potential Eq. (\ref{AtomicPotentials:eq.04}) 
 for several atoms as indicated. 
 }
 \footnotesize\rm\item[]
 \begin{tabular}{@{}llllllllll}
 \br
     & M &  $a_1$    & $a_2$     &  $a_3$    & $a_4$ 
         &  $b_1$    & $b_2$     &  $b_3$    & $b_4$      \\
 \br
 B   & 3 & 72.22775  & -1.021225 & 0.778090  & \dash  
         & 9.828608  & 2.984085  &  1.689647 & \dash    \\
 C   & 4 & 128.0489  & -2.535155 & 2.041774  & \dash  
         & 11.84981  & 3.508196  &  2.099930 & \dash    \\
 Si  & 4 & 1713.363  & 158.9419  & -107.9461 & 1.348130  
         & 29.95277  & 4.305803  &  3.906608 & 1.627379 \\
 Ge  & 4 & 20901.16  & 1399.193  & 169.1339  & 0.991756  
         & 68.65812  & 22.95161  &  5.903443 & 1.541315 \\
 \br
 \end{tabular}
 \label{Pacios.Table}
 \end{table}

\end{itemize}

\begin{figure} [h]
\centering 
\includegraphics[scale=0.5,clip]{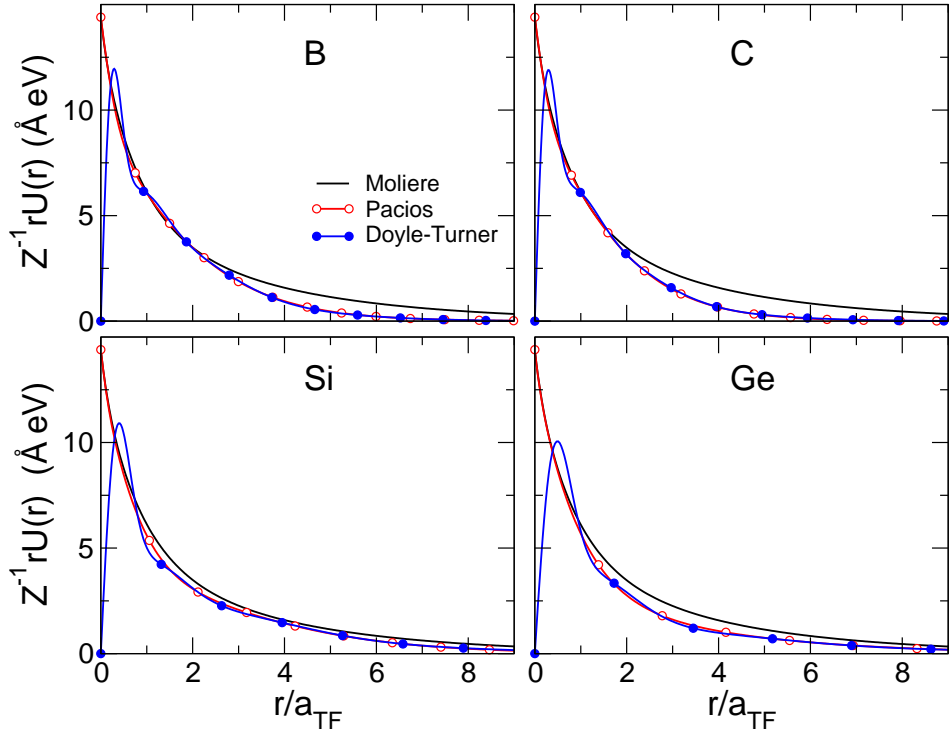}
\caption{
Dependences $rU(r)/Z$ on $r/\aTF$ 
calculated for B, C , Si and Ge atoms within 
the Moli\`{e}re, Pacios and Doyle-Turner approximations.
}
\label{B-C-Si-Ge-pot.fig}
\end{figure} 

Figure \ref{B-C-Si-Ge-pot.fig} compares the dependences $rU_{\rm a}(r)/Z$ on 
$r/\aTF$ calculated for several atoms by means of different parameterizations. 
the Moli\`{e}re, Pacios and Doyle-Turner approximations.
We note that at large distances the Pacios and Doyle-Turner curves practically
coincide whereas the Moli\`{e}re approximation provides larger values for the 
potential.
At small distances, where the Doyle-Turner parameterization fails, both Moli\`{e}re
and Pacios schemes lead to the same result.

These differences in the behaviour of the atomic potentials reveal themselves 
in the scattering process of an ultra-relativistic projectile from an atom.
Within the framework of classical small-angle scattering framework,
the scattering angle $\theta$ is related to the change of the transverse momentum
$\theta \approx c|\Delta \bfp_{\perp} | /\E$.
To calculate $\Delta \bfp_{\perp} $ one assumes, that the projectile moves along 
a straight line (the $z$ direction) with a constant speed $v\approx c$ (see, e.g.,
Ref. \cite{Landau1}). 
As a result, the scattering angle as a function of the impact parameter
$\rho$ is written as follows:
\begin{eqnarray}
\theta(\rho) 
\approx
{2 \over \E}
\left|{\partial \over \partial \rho}
\int_{0}^{\infty} U_{\rm a}(r) \, \d z 
\right|_{r=\sqrt{\rho^2 + z^2}} .
\label{AtomicPotentials:eq.06}
\end{eqnarray}

Using Eqs. (\ref{AtomicPotentials:eq.02})-(\ref{AtomicPotentials:eq.04}) in
(\ref{AtomicPotentials:eq.06}) one derives 
\begin{eqnarray}
\theta(\rho) 
=
{1 \over \E}
\cases{
{2Z \over \aTF}
\sum_{j=1}^3 a_j b_j
K_1\left(b_j \,{\rho\over \aTF}\right)
& Moli\`{e}re approx. \\
{\rho\over 4}
\sum_{j=1}^4
{a_j  \over B_j^2}
\, \exp\left(-{\rho^2 \over 4 B_j}\right)
& Doyle-Turner approx. \\
8\pi \rho
\sum_{j=1}^M
{a_j \over b_j}  \, K_2(b_j\rho)
& Pacios approx.
}.
\label{AtomicPotentials:eq.06a}
\end{eqnarray}
Here $K_1(.)$ and $K_2(.)$ stand for the MacDonald functions of the first and second 
order. respectively, and notation $B_j=bj/(4\pi)^2$ is introduced in the case of 
Doyle-Turner formula.

For small arguments, $K_1(z) \to z^{-1}$ and $K_2(z) \to 2z^{-2}$. 
Using these, one finds that in the limit 
of small impact parameters, $\rho\ll \aTF$, both the Moli\`{e}re and Pacios
formulae reduce to a correct result $2Z/\E \rho$ which is the scattering angle 
in the point Coulomb field $Z/r$.
The Doyle-Turner approximation produces incorrect result, $\theta \propto \rho$, 
in this limit.

Figure \ref{Theta_E855MeV_C_Si.fig} shows the dependences 
$\theta_{\rm M,P,DT}(\rho)$ calculated for a $\E=855$ MeV electron/positron
scattering by carbon and silicon atoms.
For the sake of comparison, the dependence for 
the point Coulomb field is also shown.

\begin{figure} [h]
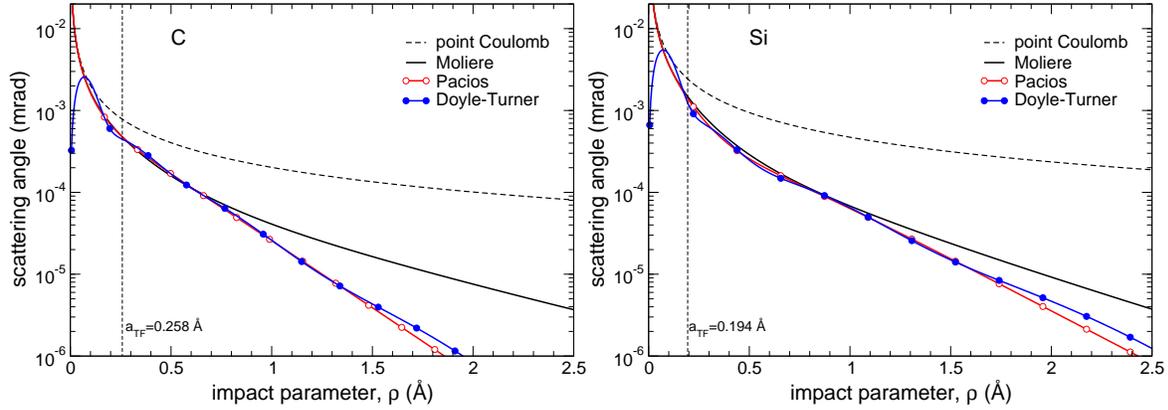

\centering
\includegraphics[scale=0.31,clip]{Figure_A02a.eps}
\includegraphics[scale=0.31,clip]{Figure_A02b.eps}
\caption{
Scattering angle $\theta$ 
versus impact parameter calculated 
for a 855 MeV electron (positron) scattered from a carbon (left) and
silicon (right) atom.
The dependencies obtained within the 
Moli\'{e}re, Pacios and Doyle-Turner approximations as well as for the
point Coulomb field, $Z/r$, are presented.
}
\label{Theta_E855MeV_C_Si.fig}
\end{figure} 

\subsection{Continuous planar potential \label{PlanarPot}}

Continuous potential $\calUpl$ of a plane one obtains
summing up the potentials $\Uat$ of individual atoms
assuming that the latter are distributed uniformly along the plane \cite{Lindhard}.
Directing the $y$-axis perpendicular to the plane one writes:  
\begin{equation}
\calUpl(y) 
= 
\calN
\int w(\Delta)\,
\d^3 \bDelta 
\int\limits_{-\infty}^{\infty}
\int\limits_{-\infty}^{\infty}
 \d z
\d x
\, 
\Uat(|\bfr-\bDelta|)\,.
\label{PlanarPot:eq.01}
\end{equation}
Here $\calN=\langle n \rangle d$ denotes the mean surface density of the atoms 
expressed in terms of mean volume density $\langle n \rangle$ and 
interplanar distance $d$.
Vector $\bDelta$ stands for the displacement of an atom from its 
equilibrium position $\bfr$ due to thermal vibrations,
which are accounted for via the distribution (\ref{PlanarPot:eq.02}).

To transform the r.h.s. of Eq. (\ref{PlanarPot:eq.01}) one expresses 
$\Uat(|\bfr-\bDelta|)$ in terms of its Fourier transform $\tilde{U}_{\rm a}(q)$ 
and carries out the spatial integrals:
\begin{eqnarray}
\calUpl(y)
& = 
{\calN \over \pi}
\int\limits_{0}^{\infty}
\d q 
\,
\ee^{-{q^2u_T^2 \over 2}}\,
\cos(q y) \,\tUat(q)\, .
\label{PlanarPot:eq.03}%
\end{eqnarray}
Using here the Fourier transforms from Eqs. 
(\ref{AtomicPotentials:eq.02})-(\ref{AtomicPotentials:eq.04})
one derives planar potentials within different parameterization schemes.
Below we present the collection of formulae for $\calUpl(y)$.

\begin{itemize}
\item \textit{Moli\'ere approximation.}

\begin{eqnarray}
\calUpl(y)
= 
\pi Z \calN
\sum_{i=1}^3 
{\alpha_j\over \gamma_j}
\ee^{\gamma_j^2u_T^2\over 2}
\!
\left[
F(y; \gamma_j,u_T)
+
F(-y; \gamma_j,u_T)
\right]
\label{PlanarPot:eq.04}
\end{eqnarray}
where 
\begin{eqnarray}
F(\pm y; \gamma_j,u_T)
=
\ee^{\pm\gamma_j y}
{\rm erfc}\!\left({\gamma_j u_T \over \sqrt{2}}\pm{y \over \sqrt{2}u_T }\right)\,.
\label{PlanarPot:eq.04a}
\end{eqnarray}
with ${\rm erfc}(x)$ being the complementary error function. 

These expressions coincide with those presented in Refs. 
\cite{Erginsoy_PRL_v15_360_1965,AppletonEtAl_PR_v161_330_1967}.

\item \textit{Doyle-Turner approximation.}

\begin{eqnarray}
\calUpl(y)
= 
2\pi^{1/2} \calN
\sum_{j=1}^4
{a_j \over \sqrt{4B_j + 2u_T^2 }}
\exp\left(-{y^2 \over 4B_j + 2u_T^2 }\right)
\label{PlanarPot:eq.05}
\end{eqnarray}
This expression coincides with the formulae presented in 
Refs. \cite{Dedkov1995,Moeller:NIMA_v361_p403_1995}.
The seeming deviations are due to different definitions of 
(i) the coefficients $B_j$ (in the cited papers they are defined as
$B_j =b_j/4\pi^2$ whereas here it is four times less),
and (ii) the rms amplitudes of thermal vibrations: in
\cite{Moeller:NIMA_v361_p403_1995} $\rho^2=2u_T^2$ stands for the two-dimensional
rms amplitude.

\item \textit{Pacios approximation.}

\begin{eqnarray}
\fl
\calUpl(y)
 = 
\sqrt{2\pi}\, Z\,\calN
u_T
\ee^{-{y^2 \over 2 u_T^2 }}
\label{PlanarPot:eq.06}\\ 
\fl
\quad
+
4\pi^2 \calN
\sum_{j=1}^M
{a_j \over b_j^4}
\ee^{b_j^2u_T^2\over 2} 
\left[
(3 -b_j^2u_T^2 -b_jy)F(y; b_j,u_T)
+
(3 -b_j^2u_T^2 + b_jy)F(-y; b_j,u_T)
\right]
\nonumber
\end{eqnarray}
with $F(\pm y; b_j,u_T)$ defined in (\ref{PlanarPot:eq.04a}).

\end{itemize}

\subsection{Continuous inter-planar potentials
 \label{InterPlanarPot}} 

The inter-planar potential is obtained by summing the potentials 
$\calUpl(y)$ of individual separate planes. 
For electrons it can be presented in the form 
\begin{equation}
U_{\rm pl}(y) 
= \calUpl(y) + \sum_{n=1}^{N_{\max}} 
            \left[ \calUpl(y+nd) + \calUpl(y-nd) \right]
          + C\,,
\label{InterPlanarPot:eq.01}
\end{equation}
where $y$ is the transverse coordinate with respect to an arbitrary selected 
reference plane, and the sum describes a balanced contribution from the 
neighboring planes. 
The constant term $C$ one chooses to ensure $U_{\rm pl}(y)(y=0)=0$.
The planar potentials (\ref{PlanarPot:eq.04}), (\ref{PlanarPot:eq.05}) and
(\ref{PlanarPot:eq.06}) fall off rapidly with increasing distance from the plane.
Therefore, Eq. (\ref{InterPlanarPot:eq.01}) provides a good approximation for the inter-planar 
potential at already moderate numbers of the terms included in the sum. 
Numerical data presented below refer to $N_{\max}=2$. 
For positrons, the inter-planar potential can be obtained from 
Eq.~(\ref{InterPlanarPot:eq.01}) by reversing the signs of the $\calUpl$ terms and 
selecting the constant $C$ to adjust $U_{\rm pl}(y=\pm d/2)=0$. 
 
Three graphs in Fig. \ref{C-Si-Ge110_PlanarPot.fig} 
compares the Moli\'ere, Pacios and Doyler-Turner electron and positron planar (110) 
potentials in diamond, silicon and germanium crystals.

\begin{figure} [h]
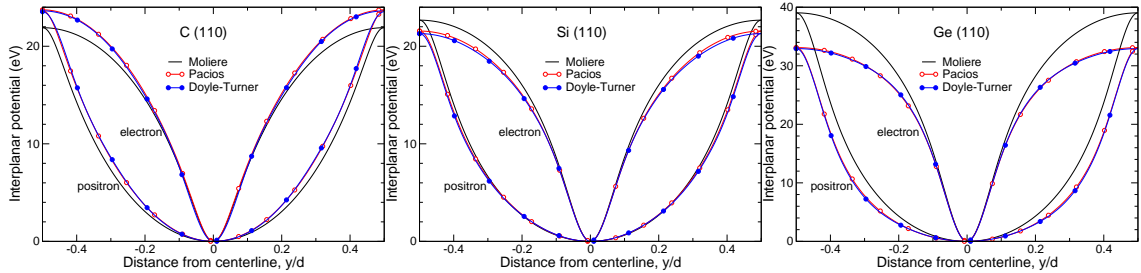

\centering
\includegraphics[scale=0.21,clip]{Figure_A03a.eps}
\includegraphics[scale=0.21,clip]{Figure_A03b.eps}
\includegraphics[scale=0.21,clip]{Figure_A03c.eps}
\caption{
Electron and positron planar (110) potentials in diamond, silicon and germanium crystals
calculated within frameworks of the 
Moli\'{e}re, Pacios and Doyle-Turner approximations.
}
\label{C-Si-Ge110_PlanarPot.fig}
\end{figure} 
 
For the sake of reference we present Fig. \ref{Graphite-C-Si-Ge-W110_PlanarPot.fig}
that compares the Moli\'ere planar potentials in different oriented crystals as indicated.

\begin{figure} [h]
\centering
\includegraphics[scale=0.4,clip]{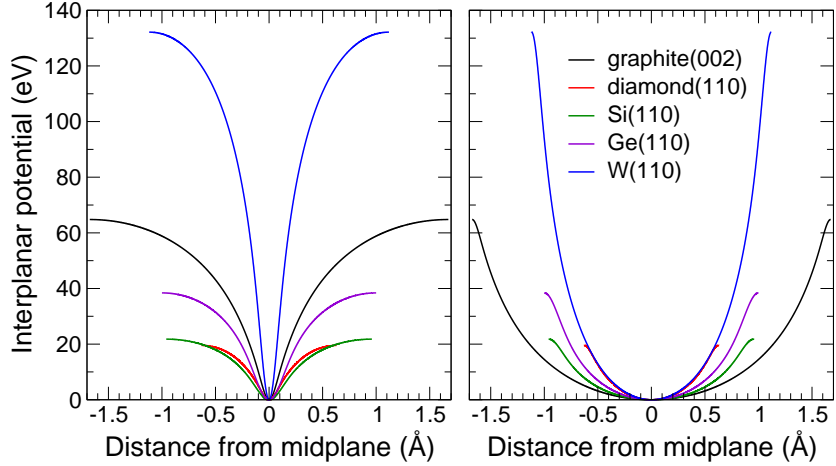}
\caption{
Interplanar potentials for electrons (left) and positrons (right)
in Graphite (002), C(110), Si(110), Ge(110) and W(110) calculated at $T=300^{\circ}$
within the Moli\'ere approximation.
}
\label{Graphite-C-Si-Ge-W110_PlanarPot.fig}
\end{figure}

\section{Continuous potential and transverse motion in a SASP Crystal \label{PotentialMotion-SASP}}

In this supplementary section, explicit formulae are derived that describe 
non-periodic and periodic parts of the 
continuous planar potential in a SASP bent crystal.
The analytical and numerical analysis of the results obtained 
allow us to qualitatively explain the peculiar features in the motion of 
ultra-relativistic projectiles as well as in the radiative spectra. 

\subsection{Continuous potential in a SASP Crystal \label{Potential-SASP}}

Consider a crystallographic plane which coincides with the 
$(xz)$ Cartesian plane.
For the sake of clarity, let us introduce a cosine periodic bending,  
$a\cos(\ku z)$ with $\ku=2\pi/\lamu$, 
of the plane in the transverse $y$ direction.
The bending amplitude $a$ and period $\lamu$ satisfy the SASP bending 
condition
\begin{equation}
a < d \ll \lamu,
\label{Pot-CU:eq.02}
\end{equation}
where $d$ stands for the interplanar distance.

Similar to the procedure used for a straight plane (see Sect. \ref{PlanarPot}), 
the continuous potential of a periodically bent plane one obtains
summing up the potentials of individual atoms
assuming that the latter are distributed uniformly along the plane:
\begin{equation}
\calUpl(y,z) 
= 
\calN
\int w(\Delta)\,
\d \bDelta 
\int\limits_{-\infty}^{\infty}
 \d z^{\prime}
\int\limits_{-\infty}^{\infty}
\d x^{\prime}
\, 
\Uat(|\bfr-\bDelta|)\,.
\label{Pot:straight:eq.01}
\end{equation}
Vector $\bDelta$ stands for the displacement of an atom from its 
equilibrium position, characterized by the 
coordinates $(x^{\prime}, y^{\prime}, z^{\prime})$ with 
with $y^{\prime}=a\cos \ku z^{\prime}$
(see illustrative Fig. \ref{Pot_CU.fig}) due to thermal vibrations.

\begin{figure} [h]
\centering
\includegraphics[scale=0.4,clip]{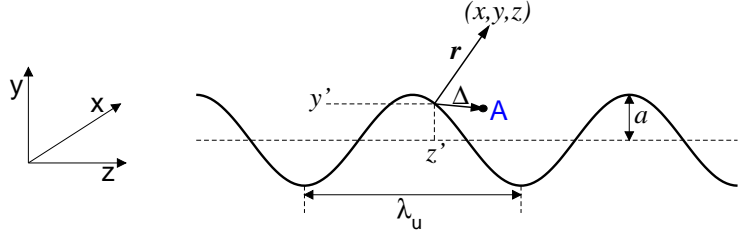}
\caption{
Supplementary figure illustrating the derivation of the 
continuous potential of a periodically bent crystallographic plane 
(the thick curve represents the bending profile).
The atoms are displaced randomly from their equilibrium positions
$(x^{\prime},y^{\prime},z^{\prime})$ due to 
to thermal vibrations (the vector $\bDelta$ shows the position of a displaced atom 
'A'). }
\label{Pot_CU.fig}
\end{figure}

Expressing $\Uat$ in terms of its Fourier transform and 
using (\ref{PlanarPot:eq.02}) one integrates over 
$x^{\prime}, z^{\prime}, \bDelta$ and presents 
the planar potential in the form of a series:
\begin{eqnarray}
\calUpl(y,z)
& = 
V_0(y; a) 
+
\sum_{n=1}^{\infty}
\cos (n\ku z)\, V_n(y;a)
\label{PlanePot_II:eq.06}
\end{eqnarray}
with
\begin{eqnarray}
V_0(y;a)
& = 
{\calN \over \pi}
\int\limits_{0}^{\infty}
\d q 
\,
\ee^{-{q^2u_T^2 \over 2}}\,
\cos(q y) \,J_0(q a)
\tUat(q)
\label{PlanePot_II:eq.07}
\end{eqnarray}
\begin{eqnarray}
V_n(y;a)
= 
{2 \calN \over \pi}
\int\limits_{0}^{\infty}
\d q \,
\ee^{-{Q_n^2 u_T^2 \over 2}}
J_n(q a)\,
\tUat\left(Q_n\right)
\times
\left\{
\begin{array}{l}
\displaystyle
(-1)^{n\over 2} \cos(q y)\\
\displaystyle
 (-1)^{n-1\over 2} \sin(q y)
\end{array}
\right.
\label{PlanePot_II:eq.08}
\end{eqnarray}
where the short-hand notation $Q_n^2=q^2 + (n\ku)^2$ is used.
The upper line stands for even $n$ values, the lower line -- for the odd ones.

In the limit of a straight channel, $a=0$, the right-hand side of Eq. (\ref{PlanePot_II:eq.06}) 
reduces to that in (\ref{PlanarPot:eq.03}).
Indeed, taking into account that $J_0(0)=0$ and $J_n(0)=0$ for $n>0$, one 
notices that this leads to $V_{n}(y;0)\equiv 0$ for all terms defined by (\ref{PlanePot_II:eq.08}) 
whereas the term $V_{0}(y;0)$, Eq. (\ref{PlanePot_II:eq.07}), reduces to (\ref{PlanePot_II:eq.06}). 

To determine the inter-planar potential $U(y,z)$ one uses Eq. (\ref{InterPlanarPot:eq.01}) 
where the potentials $\calUpl(y,z)$ of individual planes are to be inserted. 

 The integrals on the right-hand sides of Eqs. (\ref{PlanePot_II:eq.07})
and (\ref{PlanePot_II:eq.08}) 
can be evaluated explicitly for a number of analytic approximations for
$\Uat$ which can be found in literature 
\cite{Lindhard,Gemmell,Baier,DoyleTurner1968,Dedkov1995,Pacios1993,ChouffaniUberall1999}.
For reference purposes, we present the explicit formulae derived within the framework of the 
Moli\`{e}re approximation (\ref{AtomicPotentials:eq.02}):
\begin{eqnarray}
V_n(y;a)
&=
(1+\delta_{n0})
\calN Z e 
\sum_{i=1}^3
{\alpha_j \over \Gamma_{nj}}\,
\ee^{\gamma_{j}^2 u_T^2\over 2} \,
 \calT_n(y;a,\Gamma_{nj}) 
\label{Terms_Vn:eq.03}
\end{eqnarray}
Here $\delta_{n0}$ is the Kronecker symbol,
$\Gamma_{nj}=\left(\gamma_j^2+(n\ku)^2\right)^{1/2}$,
$\calT_n$ stands for the integral:
\begin{eqnarray}
\fl
\calT_n(y;a,\Gamma)
=
\int\limits_0^{\pi/2} 
 \d \theta \cos(n\theta) 
\Bigl(
\calF(y - a \cos\theta;\Gamma)
+
(-1)^n
\calF(y + a \cos\theta;\Gamma)
\Bigr)
\label{Terms_Vn:eq.04}
\end{eqnarray}
where
\begin{eqnarray}
\calF(Y;\Gamma,u_T)
=
F(Y; \Gamma,u_T) + F(-Y; \Gamma,u_T)
\label{Terms_Vn:eq.05}
\end{eqnarray}
with $F(\pm Y; \Gamma,u_T)$ defined as in (\ref{PlanarPot:eq.04a}).

For $n=0$, Eq. (\ref{Terms_Vn:eq.03}) reproduces the expression derived 
in Ref. \cite{Korol-EtAl:NIMB_v387_p41_2016}.

Similar to the case of a straight crystal, the inter-planar potential $U(y,z)$ 
in a SASP bent crystal is obtained by summing up the potentials (\ref{PlanePot_II:eq.06}) 
of individual planes. 
For the electron channel, the result can be written in the form 
\begin{eqnarray}
U(y,z) =  \sum_{n=0}^{\infty} \cos (n\ku z)\, U_n(y)
\label{PlanarPotential:eq.01}
\end{eqnarray}
where 
\begin{equation}
U_n(y) = V_n(y) 
          + \sum_{k=1}^{K_{\max}} 
            \Bigl( V_n(y+kd) + V_n(y-kd) \Bigr)
          + C_n\,.
\label{ContPot:eq.01}
\end{equation}
Here, $y$ is the transverse coordinate with respect to an arbitrary selected 
reference plane, and the sum describes a balanced contribution from the 
neighboring planes. 
The constants $C_n$ can be chosen to satisfy the condition $U_n(0)=0$.

For positrons, the inter-planar potential can be obtained from 
Eq.~(\ref{ContPot:eq.01}) by reversing the signs of the planar potentials and 
selecting the constants $C_n$ to ensure $V_n(\pm d/2)=0$. 
Similar summation schemes allow one to calculate the charge densities, 
nuclear and electronic, across  the periodically bent channels.

\subsection{Transverse Motion in the SASP Channel
 \label{SASP-Motion}} 

The function $y(t)$ describes the transverse motion of a particle with 
respect to the centerline of the channel.
The equation of motion (EM) reads
\begin{eqnarray}
\ddot{y}  
=
- {1 \over m \gamma }\, {\partial  U(y,z) \over \partial y}\,.
\label{EM:eq.01} 
\end{eqnarray}
In what follows we outline a perturbative solution of the EM.

Assuming the longitudinal coordinate $z$ changes linearly with time,
$z\approx ct$, 
one re-writes the potential (\ref{PlanarPotential:eq.01})
substituting $z$ with $ct$.
Then, the EM is written as follows:
\begin{eqnarray}
\ddot{y}  
=
{1 \over m \gamma }
\left(
f_0(y) 
+ 
\sum_{n=1}^{\infty} \cos (n\Omu t)\, f_n(y)
\right)
\label{EM-SASP:eq.04}
\end{eqnarray}
with
\begin{eqnarray}
\Omu = \ku c = {2\pi c \over  \lamu},
\quad
f_{0}(y) = - {\d U_{0} \over \d y},
\quad
f_{n}(y) = - {\d U_{n} \over \d y}\,.
\label{EM-SASP:eq.05}
\end{eqnarray}

The action of the time-independent force $f_{0}$ results in the channeling 
oscillations.
At the same time, the projectile experiences local small-amplitude oscillations 
(the jitter-like motion) due to the driving forces
$f_{n}\cos(n\Omu t)$ ($n=1,2,\dots$).
In a SASP channel, the frequency $\Omu$ (and, respectively, 
its higher harmonics $n\Omu$)
exceeds greatly the frequency of the channeling oscillations: 
$\Omu  \gg \Om_{\rm ch}$.
As a result, the EM (\ref{EM-SASP:eq.04}) can be integrated following the 
perturbative procedure outlined in  Ref. \cite{Landau1}, Sect. 30, for the 
motion in a rapidly oscillating field.
Namely, $y(t)$ is represented as a sum
\begin{eqnarray}
y(t) = Y(t) + \Xi(t)
\label{EM-SASP:eq.06}
\end{eqnarray}
where $\Xi(t)$ stands for a small (but rapidly oscillating) correction to the
smooth dependence $Y(t)$ which describes the channeling oscillations.
The function $\Xi(t)$ satisfies the equation in which the coordinate 
$Y$ is treated as a parameter:
\begin{eqnarray}
\ddot{\Xi}  
=
\sum_{n=1}^{\infty} \cos (n\Omu t)\, {f_n(Y)\over m \gamma }\,.
\label{EM-SASP:eq.07}
\end{eqnarray}
Its solution reads
\begin{eqnarray}
\Xi(t, Y)
=  
\sum_{n=1}^{\infty} \xi_n(Y) \cos(n \Omu t)\,.
\label{EM-SASP:eq.08}
\end{eqnarray}
where
\begin{eqnarray}
\xi_n(Y)
=  
- {1 \over m \gamma \Om_{\rm u}^2} 
{f_n(Y) \over n^2} 
\label{EM-SASP:eq.09}
\end{eqnarray}
The presence of the term $\Xi(t, Y)$ modifies the EM for $Y(t)$.
In addition to the force $f_0=-\d U_0(Y)/\d Y$ due to the static potential,
a ponderomotive force $f_{\rm pond}$ appears.
It can be calculated as follows
\cite{Landau1} (below, the overline denotes averaging over the
period $2\pi/\Omu$ which is much smaller that the characteristic time of the
channeling motion and, thus, does not affect the value of $Y(t)$):
\begin{eqnarray}
f_{\rm pond}(Y)  
&=
\overline{
\Xi(t, Y)
\sum_{n=1}^{\infty} \cos (n\Omu t)\, {\d f_n(Y) \over \d Y}
}
=
- {\d  U_{\rm pond}\over \d Y}
\label{EM-SASP-add:eq.01}
\end{eqnarray}
The ponderomotive potential, $U_{\rm pond}$, introduced here is defined as
follows:
\begin{eqnarray}
U_{\rm pond}(Y)  
=
{\lambda_{\rm u}^2 \over 16\pi^2 \E}
\sum_{n=1}^{\infty} {f_n^2(Y) \over n^2}
\label{EM-SASP-add:eq.02}
\end{eqnarray}
Note that the ponderomotive correction to the potential
becomes smaller as the energy increases since $U_{\rm pond}\propto 1/\E$.

Therefore, the channeling oscillations $Y=Y(t)$ are described by the EM
\begin{eqnarray}
\ddot{Y}  
=
- {1 \over m \gamma } {\d U_{\rm eff} \over \d Y}
\label{EM-SASP:eq.10}
\end{eqnarray}
where the total effective potential reads
\begin{eqnarray}
 U_{\rm eff}(Y) 
=
U_0(Y) + U_{\rm pond}(Y)\,.
\label{EM-SASP:eq.11}
\end{eqnarray}

\begin{figure} [h]
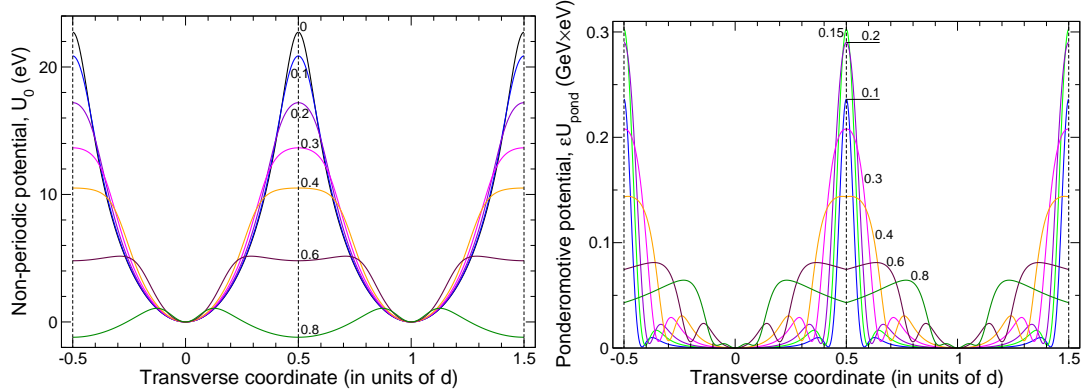

\centering
\includegraphics[width=7cm,clip]{Figure_B02a.eps}
\hspace*{0.1cm}
\includegraphics[width=7cm,clip]{Figure_B02b.eps}
\caption{
\textit{Left.} The non-periodic part $U_0(y)$ of the continuous 
inter-planar potential for positrons in Si(110) 
calculated at different values of bending amplitude indicated in 
\AA{} near the curves ($a=0$ stands for the 
straight crystal).
\textit{Right.} The ponderomotive correction $U_{\rm pond}$ 
multiplied by $\E$ in GeV calculated for various $a$ 
and for fixed bending period $\lamu=308$ microns. 
The potentials shown are evaluated for temperature 300 K 
by using the Moli\`{e}re atomic potentials.
The vertical dashed lines mark the adjacent (110) planes in straight crystal
(the interplanar spacing is $d=1.92$ \AA).
}
\label{U0-Upond.fig}
\end{figure}

The non-periodic potential $U_0$ and the ponderomotive term $U_{\rm pond}$
calculated within the Moli\`ere approximation for {\it positron} SASP Si(110) 
channel are presented in Fig. \ref{U0-Upond.fig}.
The curves correspond to different bending amplitude as indicated.
The right panel shows the dependence of the product $\E U_{\rm pond}$
(with $\E$ measured in GeV) which is independent on the projectile energy.
In both panels the vertical lines mark the (110)-planes in the straight crystal.

The modification of $U_0$ with increase of the bending amplitude is clearly 
seen on the left panel in Fig. \ref{U0-Upond.fig}. 
In detail, this issue was discussed in Ref. \cite{Korol-EtAl:NIMB_v387_p41_2016}.
Here, for the sake of consistency, we mention several features relevant to
the topic of the current paper.
For small and moderate amplitude values, $a \leq 0.4$ \AA, the major
changes in the potential is the decrease of the interplanar potential barrier.
As the $a$ values approach the $0.4 \dots 0.6$ \AA{} range, the volume density 
of atoms becomes more friable leading to flattening of the potential maximum. 
For larger amplitudes, the potential changes in a more dramatic way as 
additional potential well appears.
In the figure, this feature is clearly seen in the behaviour of the
the $U_0$ curve for $a=0.8$ \AA:
in addition to the "regular" channels centered at the midplanes, i.e. at 
$y/d=\dots, -1,0,1,\dots$,
"complementary" channels appear centered at $y/d=\dots, -0.5, 0.5, \dots$.
As a result, a positron can experience channeling oscillations moving
in the channels of the two different types. 
Similar feature characterizes the electron channeling phenomenon at sufficiently large
values of bending amplitude \cite{Korol-EtAl:NIMB_v387_p41_2016}.

\begin{figure} [h]
\centering
\includegraphics[scale=0.4,clip]{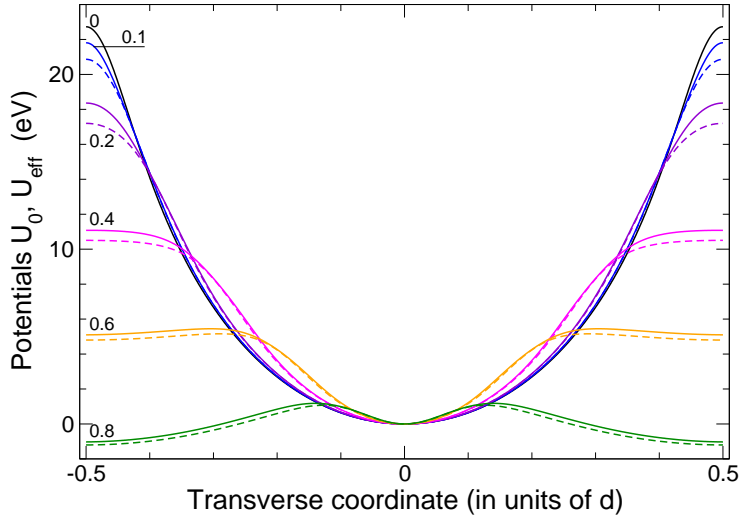}
\caption{
The effective interplanar potential $U_{\rm eff}$, Eq. (\ref{EM-SASP:eq.11}), 
(solid lines) for a 250 MeV positron in Si(110) at different values of bending 
amplitude indicated in 
\AA{} near the curves ($a=0$ stands for the straight crystal).
The dashed curves show the continuous potential $U_{\rm 0}$ without the
ponderomotive corrections.
The bending period is $\lamu=308$ microns. 
}
\label{U0_Ueff_all-a_p-250MeV.fig}
\end{figure}

Comparing the absolute values of the ponderomotive term $U_{\rm pond}$ 
(curves in the right panel of Fig. \ref{U0-Upond.fig} correspond to
the terms at $\E=1$ GeV) 
and those of $U_0$ one can state that for all bending amplitudes $U_{\rm pond}$
is a negligibly small correction to $U_0$ for the projectile energies above 1 GeV. 
For much lower energies, say for a few hundreds-MeV,  the contribution of
$U_{\rm pond}$ becomes more noticeable, reaching the eV range in the regions
$2|y|/d \approx \dots, -1,0,1,\dots$.
Thus, it should be accounted if an accurate integration of the
EM (\ref{EM-SASP:eq.10}) is desired.
Figure \ref{U0_Ueff_all-a_p-250MeV.fig} illustrates the change in the continuous
interplanar Si(110) potential due to the ponderomotive correction.
Solid curves who the corrected potentials, the dashed ones -- the  term $U_0$.
The data refer to 250 MeV positron channeling in Si(110) bent periodically
with the period $\lamu=308$ microns; 
the values of bending amplitude (in \AA{}) are indicated in the figure.

\section*{References}

\end{document}